\documentclass{lmcs} %%% last changed 2014-08-20
\pdfoutput=1

\usepackage{lastpage}
\lmcsdoi{15}{4}{17}
\lmcsheading{}{\pageref{LastPage}}{}{}%
{Mar.~08,~2019}{Dec.~19,~2019}{}

\usepackage{subcaption} %% For complex figures with subfigures/subcaptions

\usepackage{hyperref}
\usepackage{bbold}
\usepackage{color}
\usepackage{extarrows}
\usepackage[bbgreekl]{mathbbol}
\usepackage{microtype}%if unwanted, comment out or use option "draft"
\usepackage{multicol}
\usepackage{multirow}
\usepackage{proof}
\usepackage{stmaryrd}
\usepackage[normalem]{ulem}
\usepackage{xspace}
\usepackage[all]{xy}\CompileMatrices
\usepackage{wrapfig}

\DeclareSymbolFont{bbold}{U}{bbold}{m}{n}
\DeclareSymbolFontAlphabet{\mathbbold}{bbold}

\newcommand{\Nat}{\ensuremath{\mathbb{N}}}
\newcommand{\Natinf}{\ensuremath{\Nat_{\infty}}}
\newcommand{\Bin}{\ensuremath{\mathbb{2}}}

\newcommand{\down}[1]{\ensuremath{\lfloor #1 \rfloor}}

\newcommand{\minp}[1]{\ensuremath{{}^\circ{#1}}}
\newcommand{\maxp}[1]{\ensuremath{#1^\circ}}

\newcommand{\preS}[1]{\ensuremath{{}^\bullet{#1}}}
\newcommand{\postS}[1]{\ensuremath{#1^\bullet}}

\newcommand{\pntrans}{\to}

\newcommand{\BC}[1]{\textsc{bc}(#1)}
\newcommand{\bc}{\mathbb{C}}

\newcommand{\dn}[1]{\textsc{dn}(#1)}

\newcommand{\DT}[1]{\mathbb{T}(#1)}

\newcommand{\scelleq}{\leftrightarrow}

\newcommand{\enc}[1]{\llbracket#1\rrbracket}
\newcommand{\dyntopt}[1]{\llparenthesis#1\rrparenthesis}

\newcommand{\tr}[1]{\xlongrightarrow{#1}}
\newcommand{\Tr}[1]{\xLongrightarrow{#1}}

\newcommand{\co}[1]{\overline{#1}}

\newcommand{\remInitial}[2]{#1\ominus#2}

\definecolor{mygray}{RGB}{120,120,120}

\newcommand{\drawplace}{*[o]=<.9pc,.9pc>{\ }\drop\cir{}}
\newcommand{\drawmarkedplace}{*[o]=<.9pc,.9pc>{\bullet}\drop\cir{}}

\newcommand{\drawpersistentplace}{*[o]=<.9pc,.9pc>[Foo:mygray]{\ }}
\newcommand{\drawmarkedpersistentplace}{*[o]=<.9pc,.9pc>[Foo:mygray]{\color{mygray}\bullet}}

\newcommand{\nameplaceup}[1]{\POS[]+<0pc,.85pc>\drop{\scriptstyle{#1}}}
\newcommand{\nameplacedown}[1]{\POS[]-<0pc,.75pc>\drop{\scriptstyle{#1}}}
\newcommand{\nameplaceright}[1]{\POS[]+<.75pc,0pc>\drop{\scriptstyle{#1}}}
\newcommand{\nameplacerightm}[1]{\POS[]+<1.2pc,-0.2pc>\drop{\scriptstyle{#1}}}
\newcommand{\nameplaceleft}[1]{\POS[]-<.75pc,0pc>\drop{\scriptstyle{#1}}}
\newcommand{\nameplaceleftm}[1]{\POS[]-<1pc,0.2pc>\drop{\scriptstyle{#1}}}

\newcommand{\namepersup}[1]{\POS[]+<0pc,.85pc>\drop{\color{mygray}\co{\scriptstyle{\bf #1}}}}
\newcommand{\namepersdown}[1]{\POS[]-<0pc,1pc>\drop{\color{mygray}\co{\scriptstyle{\bf #1}}}}
\newcommand{\namepersright}[1]{\POS[]+<.9pc,.05pc>\drop{\color{mygray}\co{\scriptstyle{\bf #1}}}}
\newcommand{\namepersrightm}[1]{\POS[]+<1.1pc,.05pc>\drop{\color{mygray}\co{\scriptstyle{\bf #1}}}}
\newcommand{\namepersleft}[1]{\POS[]-<.8pc,-.05pc>\drop{\color{mygray}\co{\scriptstyle{\bf #1}}}}
\newcommand{\namepersleftm}[1]{\POS[]-<1pc,-.05pc>\drop{\color{mygray}\co{\scriptstyle{\bf #1}}}}

\newcommand{\namepersupD}[1]{\POS[]+<0pc,.9pc>\drop{\color{mygray}{\scriptstyle{\bf #1}}}}
\newcommand{\namepersrightD}[1]{\POS[]+<1.1pc,0pc>\drop{\color{mygray}{\scriptstyle{\bf #1}}}}

\newcommand{\drawtrans}[1]{*=<2pc,1pc>{#1}\drop\frm{-}}
\newcommand{\drawtransu}[1]{*=<2pc,1pc>[F:mygray]{#1}}
\newcommand{\drawtransd}[1]{*=<2pc,1pc>[F--]{#1}}
\newcommand{\drawtransud}[1]{*=<2pc,1pc>[F--:mygray]{#1}}

\newcommand{\OneInfSafe}{$1$-$\infty$-safe\xspace}

\newcommand{\Tpos}{T_{\rm pos}}
\newcommand{\Tneg}{T_{\rm neg}}

\newcommand{\events}[1]{{\it ev}(#1)}
\newcommand{\dec}[1]{\lVert#1\rVert}

\newcommand{\changed}[1]{#1}
\theoremstyle{plain} %\crefname{satz}{Satz}{S\"atze}
\begin{document}

\title[Concurrency and Probability: Removing Confusion, Compositionally]{Concurrency and Probability: Removing Confusion, Compositionally}         

%\titlecomment{{\lsuper*}OPTIONAL comment concerning the title, \eg, 
%  if a variant or an extended abstract of the paper has appeared elsewhere.}

\author[R.~Bruni]{Roberto Bruni}	%optional
\address{University of Pisa, Italy}	%optional
\email{bruni@di.unipi.it}  %optional
%\thanks{thanks 2, optional.}	%optional

\author[H.~Melgratti]{Hern\'an Melgratti}	%optional
\address{ICC - Universidad de Buenos Aires - Conicet, Argentina}	%optional
\email{hmelgra@dc.uba.ar}  %optional

\author[U.~Montanari]{Ugo Montanari}	%optional
\address{University of Pisa, Italy}	%optional
\email{ugo@di.unipi.it}  %optional

%% etc.

%% required for running head on odd and even pages, use suitable
%% abbreviations in case of long titles and many authors:

%%%%%%%%%%%%%%%%%%%%%%%%%%%%%%%%%%%%%%%%%%%%%%%%%%%%%%%%%%%%%%%%%%%%%%%%%%%

%% the abstract has to PRECEDE the command \maketitle:
%% be sure not to issue the \maketitle command twice!

\begin{abstract}
Assigning a satisfactory truly concurrent semantics to Petri nets with confusion and distributed decisions is a long standing problem, especially if one wants to resolve decisions by drawing from some probability distribution.
Here we propose a general solution to this problem based on a recursive, static decomposition of (occurrence) nets in loci of decision, called \emph{structural branching cells} (\emph{s-cells}). 
Each s-cell exposes a set of alternatives, called \emph{transactions}.
Our solution  transforms  a given Petri net, possibly with confusion, into another net whose transitions are the transactions of the s-cells and whose places are those of the original net, with some auxiliary nodes for bookkeeping. The resulting net is confusion-free by construction, 
and thus conflicting alternatives can be equipped with probabilistic choices, while nonintersecting alternatives are purely concurrent and their probability distributions are independent.
The validity of the construction is witnessed by a tight correspondence with the recursively stopped configurations of Abbes and Benveniste.  
Some advantages of our approach are that: 
i)~s-cells are defined statically and locally in a compositional way;
ii)~our resulting nets faithfully account for concurrency.%exhibit the complete concurrency property.
\end{abstract}

\maketitle
%% start the paper here:

% !TEX root =  main.tex

\section{Introduction}

Concurrency theory and practice provide a useful abstraction
for the design and use of a variety of systems.
Concurrent computations (also \emph{processes}), as defined in many models, 
are equivalence classes of executions, called \emph{traces},
where the order of concurrent (i.e., independent) events is inessential.
A key notion in  concurrent models is \emph{conflict} (also known as choices or decisions). 
Basically, two events are in conflict when they cannot occur in the same execution. 
The interplay between concurrency and conflicts introduces a phenomenon in 
which the execution of an event can be influenced by the occurrence of 
another concurrent (and hence independent) event. Such situation,
known as \emph{confusion}, naturally arises in concurrent  and distributed systems 
and is intrinsic to problems involving mutual exclusion~\cite{DBLP:journals/tcs/Smith96}. 
When interleaving semantics is considered, the problem is less compelling, 
however it has been recognised and studied from the 
beginning of net research~\cite{DBLP:conf/ac/RozenbergE96}, and to address it in a general and
acceptable way can be considered as a long-standing open problem for concurrency theory.

%
%
%
%Sequences in the same class should be indistinguishable
%for the current purpose of interest.
%
%Probabilistic programming is widespread in fields like
%security, approximate computing, machine learning, quantum computing.
%For example, it is shown in~\cite{DBLP:conf/esop/KaminskiKMO16} that it is possible to equip ordinary programming languages with probabilistic choices in such a way that weakest precondition analysis can be used to prove important properties of probabilistic programs.
%
%However, the interplay between concurrency and choices driven by general 
%probabilistic distributions has proved hard to convey in one computational model,
%as recognised, e.g., in~\cite{DBLP:conf/concur/EisentrautHZ10}.
%

To illustrate confusion, we rely on
Petri nets~\cite{Pet:KMA,DBLP:books/daglib/0032298}, which are a basic, well understood model of
concurrency.
\begin{figure}[t]
\begin{subfigure}[b]{0.4\textwidth} 
$$
 \xymatrix@R=.75pc@C=.8pc{
 &
 \drawmarkedplace\ar[d]\ar[rd]
 \nameplaceright 1
 \\
 & \drawtrans a\ar[d] 
 & \drawtrans d\ar[d] 
 \\
 \drawmarkedplace\ar[d]\ar[dr]
 \nameplaceup 2
 &
 \drawplace\ar[d]
 \nameplaceright 3
 &
 \drawplace
 \nameplaceright 6
 \\
 \drawtrans b\ar[d] 
 &
 \drawtrans c\ar[d] 
 \\
 \drawplace
 \nameplaceright {4}
 &
 \drawplace
 \nameplaceright {5}
 }
$$
$$
    \xymatrix@R=.6pc@C=1.2pc{
      &
      a\ar[d]\ar@{~}[r]
      &
      d
      \\
      b\ar@{~}[r]
      &
      c
    }
$$
%\vskip-8pt
\subcaption{Asymmetric confusion}\label{fig:condconfusion}
\end{subfigure}
\begin{subfigure}[b]{0.4\textwidth} 
$$
 \xymatrix@R=.75pc@C=.8pc{
 &
 \drawmarkedplace\ar[ld]\ar[rd]
 \nameplaceright 1
 \\
 \drawtrans a\ar[d] 
 & & \drawtrans d\ar[d] \ar[rd]
 \\
 \drawplace\ar[d]\ar[dr]
 \nameplaceright 3
 &
 \drawmarkedplace\ar[d] \ar[ld] \ar[rd]
 \nameplaceright 2
 &
 \drawplace
 \nameplaceright 6
 &
 \drawplace \ar[ld]
 \nameplacedown {\neg c}
 \\
 \drawtrans c\ar[d] 
 &
 \drawtrans {b_1}\ar[d] 
 &
 \drawtrans {b_2}\ar[ld] 
 \\
 \drawplace
 \nameplaceright {5}
 &
 \drawplace
 \nameplaceright {4}
 }
$$
$$
    \xymatrix@R=.6pc@C=1.2pc{
      &
      a\ar[ld]\ar[d]\ar@{~}[r]
      &
      d \ar[d]
      \\
      b_1\ar@{~}[r]
      &
      c
      &
      b_2
    }
$$
%\vskip-8pt
\subcaption{Removing confusion}\label{fig:remconfusion}
\end{subfigure}
%\begin{subfigure}[b]{0.3\textwidth} 
%$$
% \xymatrix@R=.8pc@C=.8pc{
% &
% \drawmarkedplace\ar[d]\ar[rd]
% \nameplaceright 1
% \POS[]+<1.1pc,-1pc> *+=<4.2pc,3.6pc>[F--]{}
% \POS[]+<2.5pc,0pc>\drop{\bc_1} 
% \\
% & \drawtrans a\ar[d] 
% & \drawtrans d\ar[d] 
% \\
% &
% \drawplace\ar[dd]
% \nameplaceright 3
% &
% \drawplace
% \nameplaceright 6
% \\
% \drawmarkedplace\ar[d]\ar[dr]
% \nameplaceup 2
% \POS[]+<1.2pc,-0.8pc> *+=<4.8pc,4pc>[F--]{}
% \POS[]+<3pc,0pc>\drop{\bc_2} 
% \POS[]+<0.2pc,-0.85pc> *+=<2.4pc,3.6pc>[F--]{}
% \POS[]+<1pc,0pc>\drop{\bc_3} 
% \\
% \drawtrans b\ar[d] 
% &
% \drawtrans c\ar[d] 
% \\
% \drawplace
% \nameplaceright {4}
% &
% \drawplace
% \nameplaceright {5}
% }
%$$
%\subcaption{S-cells}\label{fig:scellscondconfusion}
%\end{subfigure}
\caption{Some nets (top) and their event structures (bottom)}\label{fig:allconfusion}
\end{figure}
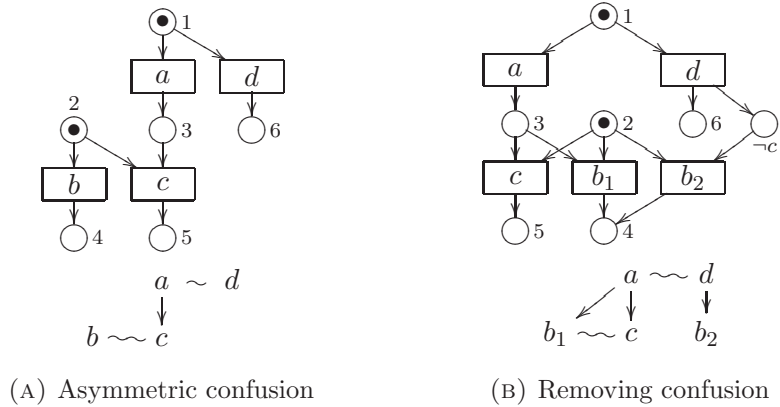
The simplest example of (asymmetric) confusion is the net in Fig.~\ref{fig:condconfusion}.
We assume the reader is familiar with the firing semantics of Petri nets, otherwise see the short summary in Section~\ref{sec:petrinets}.
The net has two traces involving the concurrent 
events $a$ and $b$, namely $\sigma_1 = a;b$ and $\sigma_2 = b; a$. Both 
traces define the same concurrent execution.
% because
%they just differ on the order in which two concurrent events are executed.
%Apparently, transition $b$ is concurrent w.r.t. $a$, but the firing of $a$ enables $c$, which is in conflict with $b$. 
%Differently, the firing of $d$ definitively disables $c$. 
Contrastingly,  $\sigma_1$  and $\sigma_2$  are associated
with completely different behaviours of the system as far as the resolution of choices is concerned.
In fact, the system makes two choices while executing $\sigma_1$: firstly, it chooses $a$ over $d$,  enabling $c$ as an alternative to $b$;
secondly,   $b$ is selected over $c$. Differently, the system makes just one choice in $\sigma_2$: since initially $c$ is not enabled, $b$ is executed without any choice;
after that, the system chooses $a$ over $d$.
As illustrated by this example, the choices made by two different traces of the same concurrent computation may differ
 depending on the order in which concurrent events occur.  
 
 The fundamental problem behind confusion relates to 
  the description of distributed, global choices. Such 
 problem becomes essential when choices are driven by %general 
probabilistic distributions and one wants to assign probabilities to executions, as it is 
the case with probabilistic, concurrent models.
Consider again %the net in 
Fig.~\ref{fig:condconfusion} and assume that 
%$p_a$ is the probability of choosing $a$ over $d$ and $p_b$ is the probability of choosing $b$ over $c$. 
 $a$ is chosen over $d$ with probability $p_a$ while $b$ is chosen over $c$ with probability $p_b$.
When driven by independent choices, the trace $\sigma_1$ 
has probability $p_a \cdot p_b$, while $\sigma_2$ has probability $1 \cdot p_a = p_a$. 
Hence,  two linear representations of the same concurrent computation, which are deemed  equivalent,
would be assigned different probabilities. 

Different solutions have been proposed in the literature for 
%combining 
%concurrency and probabilistic choices and, in particular,  for 
adding probabilities to 
Petri nets~\cite{DBLP:conf/performance/DuganTGN84,DBLP:journals/tocs/MarsanCB84,Molloy:1985:DTS:4101.4110,DBLP:conf/apn/EisentrautHK013,kudlek2005probability,haar2002probabilistic,bouillard2009critical,katoen1993modeling,DBLP:journals/cj/BrinksmaKLL95}.
As a matter of fact, most of them
replace nondeterminism with probability only in part, or take an interleaving semantics approach that disregards concurrency, 
or 
introduce time dependent stochastic distributions, thus giving up the
abstract flavour of untimed truly concurrent models. % (our first item in the list).
%Alternatively, the study is restricted to certain classes of nets that do not exhibit confusion.
%notably {\em free choice} and {\em confusion free} nets. The
%former are defined statically in terms of net structure and the latter in
%terms of net executions.  
Confusion-free probabilistic models have
been studied in~\cite{DBLP:journals/tcs/VaraccaVW06}, but 
this class, which subsumes free-choice nets, is usually considered quite restrictive.
More generally, the distributability of 
%nondeterministic 
decisions has been studied, e.g., in~\cite{DBLP:journals/corr/GlabbeekGS13,DBLP:conf/esop/KatoenP13},
%where loci of choice are determined, 
but while the results in~\cite{DBLP:journals/corr/GlabbeekGS13} apply to some restricted classes of nets,
the approach in~\cite{DBLP:conf/esop/KatoenP13} requires nets to be decorated with agents and produces
distributed models with both nondeterminism and probability, where concurrency  depends on the 
scheduling of agents.

\begin{figure}[t]
\begin{subfigure}[b]{.3\textwidth} 
$$
    \xymatrix@R=.6pc@C=1.2pc{
      &
      a\ar[d]\ar@{~}[r]
      \POS[]-<1.5pc,0pc>\drop{\bc_1} 
      &
      d
      \POS[]-<1pc,0pc> *+=<3.5pc,1pc>[F--]{}
      \\
      b\ar@{~}[r]
      &
      c
    } 
$$
%\vskip-7pt
\subcaption{Initial configuration}\label{fig:AB1a}
\end{subfigure}
\begin{subfigure}[b]{0.3\textwidth} 
$$
    \xymatrix@R=.6pc@C=1.2pc{
      &
      a\ar[d]
      \\
      b\ar@{~}[r]
      \POS[]+<0pc,1pc>\drop{\bc_2} 
      &
      c
      \POS[]-<1pc,0pc> *+=<3.5pc,1pc>[F--]{}
    } 
 $$
 \subcaption{$a$ is chosen}\label{fig:AB1b}
\end{subfigure}
\begin{subfigure}[b]{0.3\textwidth} 
$$
    \xymatrix@R=.6pc@C=1.2pc{
      &
      d
      \\
      b
      \POS[] *+=<1.5pc,1pc>[F--]{}
      \POS[]+<0pc,1pc>\drop{\bc_3} 
    }
$$
\subcaption{$d$ is chosen}\label{fig:AB1c}
\end{subfigure}
\caption{AB's dynamic branching cells for the example in Fig.~\ref{fig:condconfusion}}\label{fig:AB1}
\end{figure}
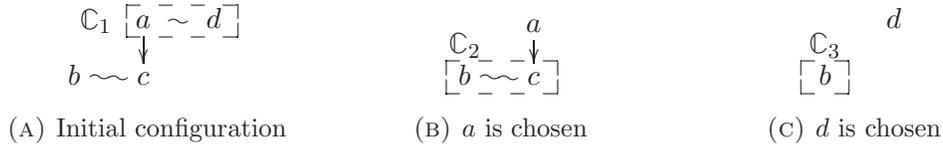
A substantial advance 
%in the study of concurrent, probabilistic models
has been contributed by Abbes and Benveniste
(AB)~\cite{DBLP:conf/fossacs/AbbesB05,DBLP:journals/iandc/AbbesB06,DBLP:journals/tcs/AbbesB08}. They
consider prime event structures and provide a 
 {\em branching cell decomposition}   that 
establishes the order in which choices are resolved (see Section~\ref{sec:ABcells}).
Intuitively, the
%Here branching cells are obtained by transitively closing the prime conflict relation.
event structure in Fig.~\ref{fig:condconfusion} has the three branching cells outlined in 
Fig.~\ref{fig:AB1}. First a decision between $a$ and $d$ must be taken (Fig.~\ref{fig:AB1a}):
if $a$ is executed, then a subsequent branching cell $\{b,c\}$ is enabled (Fig.~\ref{fig:AB1b});
otherwise (i.e., if $d$ is chosen) the trivial branching cell $\{b\}$ is enabled (Fig.~\ref{fig:AB1c}). 
In this approach,  the trace $\sigma_2 = b;a$ is not admissible, because
the branching cell $\{b\}$ does not exist in the original decomposition (Fig.~\ref{fig:AB1a}): it appears  {\em after} the  choice of $d$ over  $a$ has been resolved.
% $b$ depends 
%on the choice between $a$ and $d$. 
%the internal cell around $b$ being activated only if the $a$
%choice has been performed earlier.
%Moreover, there are two other processes: one that comprises $b$ and $d$ as concurrent events (both its traces $b;d$ and $d;b$ have probability $p_d = 1 - p_a$) and one that comprises $a$ and $c$ (with $a$ a cause of $c$), whose unique underlying trace $a;c$ has probability $p_a \cdot p_d = p_a \cdot (1 - p_b)$. 
%From sanity check \#2, we expect the sum of probabilities of all processes to be $1$: this is the case if the process with $a$ and $b$ is assigned probability $p_a \cdot p_b$, i.e. if the trace $\sigma_2$ is not admissible.
%
%AB decomposition provides a straightforward mechanism for assigning probabilities:
Branching cells are equipped with independent probability
distributions and the probability assigned to a concurrent execution is given
by the product of the probabilities assigned by its branching cells. 
Notably,  the sum of the probabilities of
maximal configurations is 1.
Every decomposition of a configuration yields an
execution sequence compatible with that configuration. 
Unfortunately, 
% the converse does not hold in general: 
certain sequences of events, legal w.r.t. the configuration, 
are not executable according to AB.

\subsection*{Problem statement}
The question addressed in this paper is a foundational one: 
\emph{can concurrency and general probabilistic distributions 
coexist in Petri nets?
If so, under which circumstances?}
By \emph{coexistence} we mean that all the following issues must be addressed:
\begin{enumerate}
\item \emph{Time independence}: Truly concurrent semantics usually assumes computation to be independent from the relative speed of processes. In this sense, although truly concurrent models have been extended in the literature with some notion of time such that occurrences of events are studied in terms of stochastic distributions, here we consider the more abstract case of untimed models only.

\item \emph{Schedule independence}: %In connection to the previous item, 
%The probability distribution that drives the choice of a transition 
%%(both its probability and its alternatives) 
%must not be affected by the execution of concurrent events, i.e., 
Concurrent events must be driven by independent probability distributions.
This item is tightly related to the confusion problem, where the set of alternatives, and thus their probability distribution, can be changed by the execution of some concurrent event.

\item \emph{Probabilistic computation}: Nondeterministic choices must be replaceable by probabilistic choices. This means that whenever two transitions are enabled, the choice to fire one instead of the other is either inessential (because they are concurrent) or is driven by some probability distribution.

\item \emph{Complete concurrency}: It must be possible to establish a bijective correspondence between  equivalence classes of firing sequences  and a suitable set of concurrent processes. 
In particular, given a concurrent process it must be possible to recover all its underlying firing sequences.
%Moreover, given a process it must be possible to recover all its corresponding firing sequences. 

%\item \emph{Complete concurrency}: It should be possible to partition the firing sequences in equivalence classes and establish a bijective correspondence between such classes and a suitable set of concurrent, deterministic processes. Moreover, given a process it must be possible to recover all its corresponding firing sequences. 

\item \emph{Sanity check \#1}: All  firing sequences of the same  process carry the same probability, i.e., the probability of a concurrent computation is independent from the order of execution.

\item \emph{Sanity check \#2}: The sum of the probabilities assigned to all possible maximal processes must be $1$.
\end{enumerate}

\noindent
In this paper we provide a positive answer for  finite occurrence nets:
given any such net we show how to define loci of decisions, 
called \emph{structural branching cells} (\emph{s-cells}), 
and construct another net where independent probability distributions can be assigned to concurrent 
events. % At the implementation level 
This means that each s-cell can be assigned to a distributed random agent
and that any concurrent computation is independent from the scheduling of agents.

\subsection*{Overview of the approach}
\begin{figure*}[t]
\begin{subfigure}[b]{0.45\textwidth} 
$$
 \xymatrix@R=.65pc@C=.6pc{
 &
 \drawmarkedplace\ar[d]\ar[rd]
 \nameplaceright 1
 & &
 \drawmarkedplace\ar[d]\ar[rd]
 \nameplaceright 7
 \\
 & \drawtrans a\ar[d] 
 & \drawtrans d\ar[d] 
 & \drawtrans e\ar[d] 
 & \drawtrans f\ar[d] 
 \\
 \drawmarkedplace\ar[d]\ar[dr]
 \nameplaceup 2
 &
 \drawplace\ar[d]
 \nameplaceright 3
 &
 \drawplace
 \nameplaceright 6
 &
 \drawplace\ar[dll]\ar[d]
 \nameplaceright 8
 &
 \drawplace
 \nameplaceright 9
 \\
 \drawtrans b\ar[d] 
 &
 \drawtrans c\ar[d] 
 & &
 \drawtrans g\ar[d] 
 \\
 \drawplace
 \nameplaceright {4}
 &
 \drawplace
 \nameplaceright {5}
 & &
 \drawplace
 \nameplaceright {10}
 }
$$
$$
    \xymatrix@R=.3pc@C=1.2pc{
      &
      a\ar[d]\ar@{~}[r]
      &
      d
      &
      e\ar@{~}[r]\ar[dll]\ar[d]
      &
      f
      \\
      b\ar@{~}[r]
      &
      c\ar@{~}[rr]
      &
      &
      g
    }
$$
\subcaption{Confusion with OR-causes}\label{fig:multcondconfusion}
\end{subfigure}
\begin{subfigure}[b]{0.45\textwidth} 
$$
 \xymatrix@R=.75pc@C=.5pc{
 &
 \drawmarkedplace\ar[d]\ar[rd]
 \nameplaceright 1
 & & &
 \drawmarkedplace\ar[d]\ar[rd]
 \nameplaceright 7
 \\
 & \drawtrans a\ar[d] 
 & \drawtrans d\ar[d] \ar[rd]
 & & \drawtrans e\ar[d] 
 & \drawtrans f\ar[d] \ar[lld]
 \\
 &
 \drawplace\ar[dd]\ar[rrdd]
 \nameplaceright 3
 &
 \drawplace
 \nameplaceright 6
 & 
 \drawplace \ar@{<->}[llldd]\ar@{<->}[rdd]
 \nameplaceup {\neg c}
 &
 \drawplace\ar[ddlll]\ar[dd]\ar[ldd]
 \nameplaceright 8
 &
 \drawplace
 \nameplaceright 9
 \\
 \drawmarkedplace\ar[d]\ar[dr]\ar[rrrd]
 \nameplaceup 2
 \\
 \drawtrans b\ar[d] 
 &
 \drawtrans c\ar[d] 
 & & 
 \drawtrans {bg}\ar[llld]\ar[rd] 
 &
 \drawtrans g\ar[d] 
 \\
 \drawplace
 \nameplaceleft {4}
 &
 \drawplace
 \nameplaceright {5}
 & & &
 \drawplace
 \nameplaceright {10}
 }
$$
\subcaption{An attempt}\label{fig:attmultcondconfusion}
\end{subfigure}
\begin{subfigure}[b]{0.9\textwidth} 
$$
 \xymatrix@R=.75pc@C=.5pc{
 &
 \drawmarkedplace\ar[d]\ar[rd]
 \nameplaceright 1
 & & &
 \drawmarkedplace\ar[d]\ar[rd]
 \nameplaceright 7
 \\
 & \drawtrans a\ar[d] 
 & \drawtrans d\ar[d] \ar[rd]
 & & \drawtrans e\ar[d] 
 & \drawtrans f\ar[d] \ar[lld]
 \\
 &
 \drawplace\ar[dd]\ar[rrdd]
 \nameplaceright 3
 &
 \drawplace
 \nameplaceright 6
 & 
 \drawpersistentplace \ar[llldd]\ar[rdd]
 \nameplaceup {\neg c}
 &
 \drawplace\ar[ddlll]\ar[dd]\ar[ldd]
 \nameplaceright 8
 &
 \drawplace
 \nameplaceright 9
 \\
 \drawmarkedplace\ar[d]\ar[dr]\ar[rrrd]
 \nameplaceup 2
 \\
 \drawtrans b\ar[d] 
 &
 \drawtrans c\ar[d] 
 & & 
 \drawtrans {bg}\ar[llld]\ar[rd] 
 &
 \drawtrans g\ar[d] 
 \\
 \drawplace
 \nameplaceleft {4}
 &
 \drawplace
 \nameplaceright {5}
 & & &
 \drawplace
 \nameplaceright {10}
 }
$$
\subcaption{A solution with persistent places}\label{fig:persistency}
\end{subfigure}
\caption{Running example}\label{fig:orconfusion}
\end{figure*}
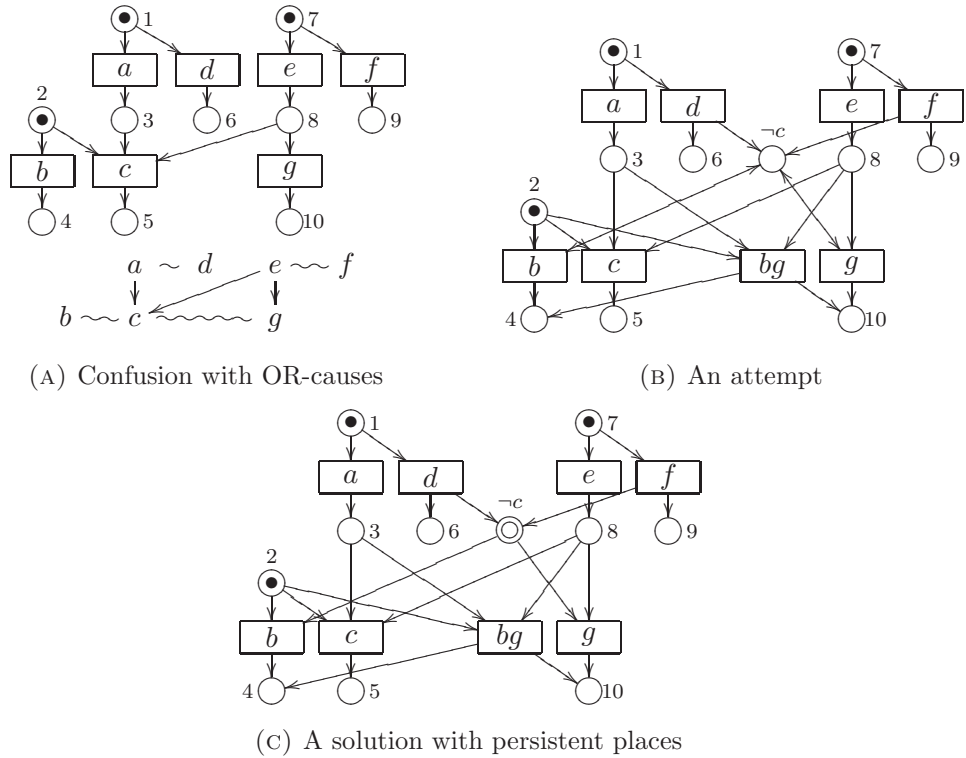

%$$
% \xymatrix@R=.8pc@C=.8pc{
% &
% \drawmarkedplace\ar[ld]\ar[rd]
% \nameplaceright 1
% & & &
% \drawmarkedplace\ar[d]\ar[rd]
% \nameplaceright 7
% \\
% \drawtrans a\ar[d] 
% & & \drawtrans d\ar[d] 
% & & \drawtrans f\ar[d] 
% & \drawtrans e\ar[d] 
% \\
% \drawplace\ar[d]
% \nameplaceright 3
% &
% \drawmarkedplace\ar[d]\ar[ld]
% \nameplaceright 2
% &
% \drawplace
% \nameplaceright 6
% & &
% \drawplace
% \nameplaceright 9
% &
% \drawplace\ar[dlllll]\ar[d]
% \nameplaceright 8
% \\
% \drawtrans c\ar[d] 
% &
% \drawtrans b\ar[d] 
% & & & &
% \drawtrans g\ar[d] 
% \\
% \drawplace
% \nameplaceright {4}
% &
% \drawplace
% \nameplaceright {5}
% & & & &
% \drawplace
% \nameplaceright {10}
% }
%$$

Following the rationale behind AB's approach, a net is transformed into another one that postpones the execution of choices that can be
 affected by pending  decisions.  
%As a general guideline, if the firing of a transition changes the set of alternatives available at some other site of the net,
%then it means that such transition is best executed before the choice at the other site happens, i.e., some causal dependency enforcing a suitable ordering of events must be added.
According to this intuition, the net in Fig.~\ref{fig:condconfusion} is transformed into another one that
delays the execution of $b$ until all its potential alternatives (i.e., $c$)
are  enabled or definitively excluded. In this sense, $b$ should never be executed before the decision between $a$ and $d$ is 
taken, because $c$ could still be enabled (if $a$ is chosen). % or  discarded  (if $d$ is chosen).
As a practical situation, imagine that $a$ and $d$ are the choices of your partner to either come to town ($a$) or go to the sea ($d$) and that 
you can go to the theatre alone ($b$), which is always an option, or go together with him/her ($c$), which is possible only when he/she is in town and accepts the invitation. Of course you better postpone the decision until you know if your partner is in town or not.
This behaviour is faithfully represented, e.g., by the confusion-free net in Fig.~\ref{fig:remconfusion},
where  two variants of $b$ are made explicit:  $b_1$ (your partner is  in town) and $b_2$ (your partner is not in town).
The new place $\neg c$ represents the fact that $c$ will never be enabled.
Now, from the concurrency point of view, there is a single process that 
comprises both $a$ and $b_1$ (with $a$ a cause of $b_1$), whose overall probability is 
the product of the probability of choosing $a$ over $d$ by the probability of choosing $b_1$ over $c$. 
The other two processes comprise, respectively, $d$ and $b_2$ (with $d$ a cause of $b_2$) and  $a$ and $c$ (with $a$ a cause of $c$).
As the net is confusion-free 
%(although not free-choice) 
all criteria in the  desiderata are met.

The general situation is  more involved because: i) there can be several ways to disable the same transition;
ii) resolving a choice may require to execute several transitions at once. Consider the net in Fig.~\ref{fig:multcondconfusion}: 
i) $c$ is discarded as soon as $d$ \emph{or} $f$ fires; and ii) when both $a$ and $e$ are fired we can choose to execute $c$ alone or \emph{both} $b$ and $g$.  
%
%For example we can imagine that this time three persons are involved: Alice would like to play tennis with Carol, but they need Bob as a referee. Alice is already at the tennis court: she would like to play ($c$) but she can also practice alone ($b$); Bob and Carol can choose to go to the tennis court ($a$ and $e$, respectively) or to stay home ($d$ and $f$, respectively); if at the tennis court, Carol can also decide to  practice alone ($g$).
%
Likewise the previous example, %in this second, more general scenario 
we may expect to transform the net as in Fig.~\ref{fig:attmultcondconfusion}.
Again, the place $\neg c$  represents  the permanent disabling of $c$.
This way a probability distribution can drive the choice between $c$ and (the joint execution of) $bg$, whereas  $b$ and $g$ (if enabled) can fire concurrently when $\neg c$ is marked. 

A few things are worth  remarking: i) a token  in $\neg c$ can be needed several times (e.g., to fire $b$ and  $g$), hence tokens should be read but not consumed from $\neg c$ (whence the double headed arcs from $\neg c$ to $b$ and $g$, called \emph{self-loops}); ii) several tokens can appear in the place $\neg c$ (by firing both $d$ and $f$).
These facts have severe repercussions on the concurrent semantics of the net. 
%Suppose $d;f;b;g$ fire in this order: is the firing of $b$ causally dependent on that of $d$ or that of $f$ (or on both)? Moreover, is the firing of $g$ causally dependent on $b$ (due to the self-loop $b$ has on $\neg c$)?
{Suppose the trace $d;f;b$ is observed. 
The firings of $d$ and $f$ produce two tokens in the place $\neg c$:
Does $b$ causally depend on the token generated from $d$ or from $f$ (or from both)?
Moreover, consider the trace $d;e;b;g$, in which $b$ takes and releases a token in $\neg c$. Does $g$ causally depend on $b$ (due to such self-loop)?}
This last question can be solved by replacing self-loops with \emph{read arcs}~\cite{DBLP:journals/acta/MontanariR95}, %(see Fig.~\ref{fig:readmultcondconfusion}), 
so that the firing of $b$ does not alter the content of $\neg c$ and thus no causal dependency  arises between $b$ and $g$.
Nevertheless, if process semantics or event semantics is considered, then we should explode all possible combinations of causal dependencies, thus introducing a new, undesired kind of nondeterminism.
In reality, we should not expect any causal dependency between $b$ and $g$, while both have OR dependencies on $d$ and $f$.

To account for OR dependencies, we exploit the notion of  \emph{persistence}: tokens in a persistent place have infinite weight and are collective. Namely, once a token reaches a persistent place, it cannot be removed and if two tokens reach the same persistent place they are indistinguishable. Such networks are a variant of ordinary P/T nets and have been studied in~\cite{DBLP:journals/entcs/CrazzolaraW05}. In the example, we can declare $\neg c$  to be a persistent place and replace self-loops/read arcs on $\neg c$ with ordinary outgoing arcs (see Fig.~\ref{fig:persistency}).
Nicely we are able to introduce a process semantics for nets with persistent places that satisfies complete concurrency.

The place $\neg c$ in the examples above is just used to sketch the general idea: our transformation  introduces persistent places like $\mathbf{\overline{3}}$ to express that a token will never appear in the regular place $3$.

\subsection*{Contribution.}
In this paper we show how to systematically derive confusion-free nets (with persistency) from any (finite, occurrence) Petri net and equip them with probabilistic distributions and concurrent semantics in the vein of AB's construction. 

Technically, our approach is based on a structurally recursive decomposition of the original net in s-cells.
A simple kind of Asperti-Busi's dynamic nets is used as an intermediate model to structure the coding. While not strictly necessary, 
%we feel that 
the intermediate step 
%makes the construction 
%more understandable: the translation from the given net to the dynamic net 
emphasises the hierarchical nature of the construction.
The second part is a general flattening step independent of our special case.
%
%Of course the general construction is more sophisticated than those in the examples shown above: 
%to keep the approach compositional, some auxiliary transitions and persistent places are associated with each s-cell of the original net.
Our definition is purely local (to s-cells), static and compositional, whereas AB's is dynamic and global 
(i.e., it requires the entire PES).
Using nets with persistency, we compile the execution strategy of nets with confusion in a statically defined, confusion-free,  operational model.
The advantage is that  the concurrency within a process of the obtained p-net is consistent with
execution, i.e., all linearizations of a persistent process are executable.

\subsection*{Structure of the paper}
After fixing notation in Section~\ref{sec:background}, our solution to the confusion problem consists of the following steps:
(i)~we define s-cells in a compositional way (Section~\ref{sec:structural-cells});
(ii)~from s-cells decomposition and the use of dynamic nets, we derive a confusion-free net with persistency (Section~\ref{sec:encscells});
(iii)~we prove the correspondence with AB's approach (Section~\ref{sec:cells}); 
(iv)~we define a new notion of process that accounts for OR causal dependencies and satisfies complete concurrency (Section~\ref{sec:concurrency}); and
(v)~we show how to assign probability distributions to s-cells (Section~\ref{sec:probability}).
For the sake of readability, all proofs of main results can be found in Appendix.

\section{Preliminaries}\label{sec:background}

\subsection{Notation}\label{sec:notation}

%We fix some notation.
We let $\Nat$ be the set of natural numbers, 
$\Natinf = \Nat \cup \{\infty\}$ and 
$\Bin = \{0,1\}$.  
We write $U^S$ for the set of functions from $S$ to $U$:  
hence
%Following this convention 
a subset of $S$ is an element of $\Bin^S$, 
a multiset $m$ over $S$ is an element of $\Nat^S$, 
%where elements of $S$ can be assigned a finite multiplicity
and a bag $b$ over $S$ is an element of $\Natinf^S$. 
%where elements of $S$ can be assigned a finite or infinite multiplicity.  
By overloading the notation, union, difference
and inclusion of sets, multisets and bags are all denoted by the same
symbols: $\cup$, $\setminus$ and $\subseteq$, respectively.  
In the case of bags, the difference $b\setminus m$ is defined only
when the second argument is a multiset, with the convention that $(b\setminus m)(s) =
\infty$ if $b(s)=\infty$. Similarly, $(b\cup b')(s) = \infty$ if
$b(s)=\infty$ or $b'(s)=\infty$. 
A set can be seen as a multiset or a bag whose
elements have unary multiplicity.  
Membership is denoted by  $\in$: 
for a multiset $m$ (or a bag $b$), we write $s \in m$ for $m(s) \neq 0$ ($b(s) \neq 0$).
Given a relation $R \subseteq S \times S$, we let $R^+$
be its transitive closure and $R^*$ be its reflexive and transitive
closure.  We say that $R$ is \emph{acyclic} if $\forall s\in S.~ (s,s)
\not\in R^+$.

\subsection{Petri Nets, confusion and free-choiceness}
\label{sec:petrinets}

%A Petri net is a bipartite graph whose nodes are partitioned into the
%set of places $P$ and transitions $T$.
%
%Intuitively, an automaton is a special case of a Petri net where only
%one token is present at any time and where each transition removes
%one token from one place and add one token to a place.
%
%Formally 
A \emph{net structure} $N$ (also \emph{Petri net})~\cite{Pet:KMA,DBLP:books/daglib/0032298} is a tuple $(P,T,F)$ where: $P$ is the set of
places, $T$ is the set of transitions, and $F \subseteq (P\times
T)\cup (T\times P)$ is the flow relation.  For $x\in P\cup
T$, we denote by $\preS{x} = \{ y \mid (y,x)\in F\}$ and $\postS{x} =
\{ z \mid (x,z)\in F\}$ its \emph{pre-set} and \emph{post-set},
respectively.
We  assume that $P$ and $T$ are disjoint and non-empty
and that  $\preS{t}$ and  $\postS{t}$ are non empty for every
$t\in T$.
We write $t: X \to Y$ for $t\in T$ with $X = \preS{t}$ and $Y =
\postS{t}$.

A \emph{marking} is a multiset $m\in\Nat^P$.
% assigns $m(p)$ tokens to each $p\in P$ and 
We say that  $p$ is \emph{marked} at $m$ if $p\in m$.
% (i.e., $m(p)\neq 0$).
We write  $(N,m)$ for the net $N$ \emph{marked} by $m$.
We write $m_0$ for the initial marking of the net, if any.  

Graphically, a Petri net is a directed graph whose
nodes are the places and transitions and whose set of arcs is $F$.
Places are drawn as circles and transitions as rectangles.  The
 marking $m$ is represented by inserting $m(p)$
tokens in each place $p\in m$ (see Fig.~\ref{fig:allconfusion}).

A transition $t$ is \emph{enabled} at the marking $m$,
written $m\xrightarrow{t}$,
%, if all the places in the pre-set of $t$ are marked at $m$, i.e., 
if $\preS{t}\subseteq m$.  
The execution of a transition $t$ enabled at $m$, called  \emph{firing}, is written 
$m \xrightarrow{t} m'$ with $m' = (m\setminus \preS{t}) \cup \postS{t}$.
%Roughly, the firing of $t$ removes $\preS{t}$ tokens and add $\postS{t}$ tokens.
%
%When the underlying set of transitions is not clear from the
%context, we write a marking as the pair $(T,m)$ and the firing of the
%transition $t$ as $(T,m) \xrightarrow{t} (T,m')$. This notation is
%especially useful in the case of dynamic nets as introduced later.
%where the set of transitions can be extended by a firing.
%
A firing sequence from $m$ to $m'$ is a finite sequence of firings
$m=m_0 \xrightarrow{t_1}  \cdots \xrightarrow{t_n}
m_n=m'$, abbreviated to $m \xrightarrow{t_1 \cdots t_n} m'$
or just $m \rightarrow^* m'$.
%when intermediate states are irrelevant, 
Moreover,  it is  %$m \xrightarrow{t_1 \cdots t_n} m'$ is 
\emph{maximal} if no transition is
enabled at $m'$.
We write $m \xrightarrow{t_1 \cdots t_n}$ if there is 
$m'$ such that $m \xrightarrow{t_1 \cdots t_n} m'$.  
We say that  $m'$ is
\emph{reachable} from $m$ if  $m \rightarrow^* m'$.  
The set of markings reachable from $m$ is written $[m\rangle$. 
%
%For example, the marked net in Fig.~\ref{fig:confusion} has places $\{1,2,3,4,5\}$, transitions $\{a,b,c\}$ and marking $m_0$ such that 
%$m_0(1) = m_0(2) = 1$ and $m_0(p) = 0$ otherwise.
%It has three maximal firing sequences:
%$m_0 \xrightarrow{a\ b}$,   $m_0 \xrightarrow{b\ a}$ and $m_0 \xrightarrow{a\ c}$.
%
A marked net $(N,m)$ is \emph{safe} if each $m'\in [m\rangle$ is a set.
%, i.e., $m(p)\in \Bin$ for all $p\in P$. 
%Safeness guarantees that $[m\rangle$ is finite (if $P$ is finite)
%and that each place will contain at most one token at any time.

Two transition $t,u$ are in \emph{direct conflict} if $\preS{t}\cap \preS{u} \neq
\emptyset$.   A net is called \emph{free-choice} if for all transitions $t,u$ we
have either $\preS{t} = \preS{u}$ or $\preS{t}\cap \preS{u} =
\emptyset$, 
%
%The notion of free-choice is structural, i.e., it is independent from
%the initial marking.
%
%Intuitively, a free-choice net guarantees that when 
i.e., if a transition $t$
is enabled then all its conflicting alternatives are also enabled. 
%So each $\preS{t}$ is a locus of decision.
% (not necessarily a singleton). 
Note that free-choiceness is purely structural. 
Confusion-freeness considers instead  the dynamics of the net.
%
%Confusion free nets provide a similar guarantee by taking into account
%the dynamics of the net.
%
%A marked net $(N,m_0)$ is \emph{confusion free} if it has neither symmetric 
%nor asymmetric confusion, as defined below.
%
%A net has \emph{symmetric confusion} if there are a reachable marking
%$m$ and transitions $t,u,v$ enabled at $m$ such that: 
%(i)~$\preS{t} \cap \preS{u} \neq \emptyset \neq \preS{u} \cap \preS{v}$, and 
%(ii)~$\preS{t} \cap \preS{v} = \emptyset$.
%%
%An example of symmetric confusion is given by transitions $t=b$, $u=c$
%and $v=g$ in Fig.~\ref{fig:multcondconfusion}.
%
%A net has \emph{asymmetric confusion} if there are a reachable marking
%$m$ and transitions $t,u,v$ such that: 
%(i)~$t$ and $u$ are enabled at $m$, 
%(ii)~$v$ is not enabled at $m$ but it becomes enabled after the firing of $t$, and
%(iii)~$\preS{t} \cap \preS{u} = \emptyset$ and $\preS{v} \cap \preS{u} \neq \emptyset$.
%%
%An example of asymmetric confusion is given by transitions $t=a$,
%$u=b$ and $v=c$ in Fig.~\ref{fig:confusion}.
%
%For any free-choice net $N$ and any marking $m$,
%the marked net $(N,m)$ is confusion free.
%
A safe marked net $(N,m_0)$  has {\em confusion} iff there exists  a reachable marking
$m$ and transitions $t,u,v$  such that: 
\begin{enumerate}
 \item 
(i)~$t,u,v$ are enabled at $m$, 
(ii)~$\preS{t} \cap \preS{u} \neq \emptyset \neq \preS{u} \cap \preS{v}$,  
(iii)~$\preS{t} \cap \preS{v} = \emptyset$ ({\em symmetric case}); or
 \item
(i)~$t$ and $v$ are enabled at $m$, 
(ii)~$u$ is not enabled at $m$ but it becomes enabled after the firing of $t$, and
(iii)~$\preS{t} \cap \preS{v} = \emptyset$ and $\preS{v} \cap \preS{u} \neq \emptyset$ ({\em asymmetric case}).
\end{enumerate}
In case 1, $t$ and $v$ are concurrently enabled but the firing of one disables an alternative ($u$) to the other.
In case 2, the firing of $t$ enables an alternative to $u$.
An example of symmetric confusion is given by $m=\{2,3,8\}$, $t=b$, $u=c$
and $v=g$ in Fig.~\ref{fig:multcondconfusion}, 
while for the asymmetric case take $m=\{1,2\}$, $t=a$,
$v=b$ and $u=c$ in Fig.~\ref{fig:condconfusion}.
A net is \emph{confusion-free} when it has no confusion. 

%For any free-choice net $N$ and any marking $m$,
%the marked net $(N,m)$ is confusion free.

%\[
% \xymatrix@R=1.2pc@C=.8pc{
% &
% \drawplace\ar[dl]\ar[dr] 
% \nameplaceup 1
% &&
% &
% \drawplace\ar[dl]\ar[dr] 
% \nameplaceup 2
% &&
% &
%\drawplace\ar[dl]\ar[dr] 
% \nameplaceup 3
% \\
% \drawtrans a\ar[d] 
% &
% &
% \drawtrans b\ar[d] 
% &
% \drawtrans c\ar[d] 
% &
% &
% \drawtrans d\ar[d] 
% &
% \drawtrans e\drop\frm{-}\ar[d] 
% &
% &
% \drawtrans f\ar[d] 
% &
%\\
%\drawplace
% \nameplaceright 4
% &
% &
%\drawplace\ar[d]\ar[drrr]
% \nameplaceright 5
% &
%\drawplace
% \nameplaceright 6
% &
% &
%\drawplace\ar[d] 
% \nameplaceright 7
% &
%\drawplace
% \nameplaceright 8
% &
% &
%\drawplace\ar[d]\ar[dlll]
% \nameplaceright 9
% \\
% &&
% \drawtrans g\ar[d] 
% &&
% &
% \drawtrans h\ar[d] 
% &
% &&
% \drawtrans i\ar[d] 
%\\
% &&
% \drawplace
% \nameplaceright {10}
% &&
% &
% \drawplace
% \nameplaceright {11}
% &&
% &
% \drawplace
% \nameplaceright {12}
% }
%\]

\subsection{Deterministic Nonsequential Processes}\label{sec:Petriproc}

%When tokens are seen as resources that are manipulated by transitions, we can address 
%the issue of establishing when a firing is causally
%dependent on another one (because it removes some tokens 
%produced by the other), or concurrent with it.
%% (because they are independent).
%%
%This allows to consider equivalence classes of firing sequences of $N$
%up to a permutation of the order in which concurrent firings are
%executed.  Such concurrent runs take the name of \emph{deterministic
%  nonsequential processes}~\cite{DBLP:journals/iandc/GoltzR83}.

A \emph{deterministic 
nonsequential 
process} (or just \emph{process})~\cite{DBLP:journals/iandc/GoltzR83}
represents the equivalence class of all  firing sequences of a net that only differ  in the order in which concurrent firings are
executed.  
It is given as a mapping $\pi:\mathcal{D} \to N$ from a
\emph{deterministic occurrence net} $\mathcal{D}$ to $N$ (preserving pre- and post-sets), 
where a deterministic occurrence net is such that:
(1)~the flow relation is acyclic,
(2)~there are no backward conflicts
%i.e., 
($\forall p\in P.~ |\preS{p}|\leq 1$), and
(3)~there are no forward conflicts
%i.e., 
($\forall p\in P.~|\postS{p}|\leq 1$).
We let $\minp{\mathcal{D}} = \{ p\, \mid\, \preS p = \emptyset\}$ and
$\maxp{\mathcal{D}} = \{ p\, \mid\, \postS p = \emptyset\}$ be the
sets of \emph{initial} and \emph{final places} of $\mathcal{D}$,
respectively (with $\pi(\minp{\mathcal{D}})$ be the initial marking of $N$).
When $N$ is an acyclic safe net, the mapping
$\pi:\mathcal{D} \to N$ is just an injective graph homomorphism:
%(preserving pre- and post-sets): 
without loss of generality, we 
name the nodes in $\mathcal{D}$ as their images in $N$ and
let $\pi$ be the identity. 
The  firing sequences of a processes $\mathcal{D}$ are its maximal
firing sequences starting from the marking $\minp{\mathcal{D}}$.
A process of $N$ is
\emph{maximal} if its firing sequences
are maximal in $N$.

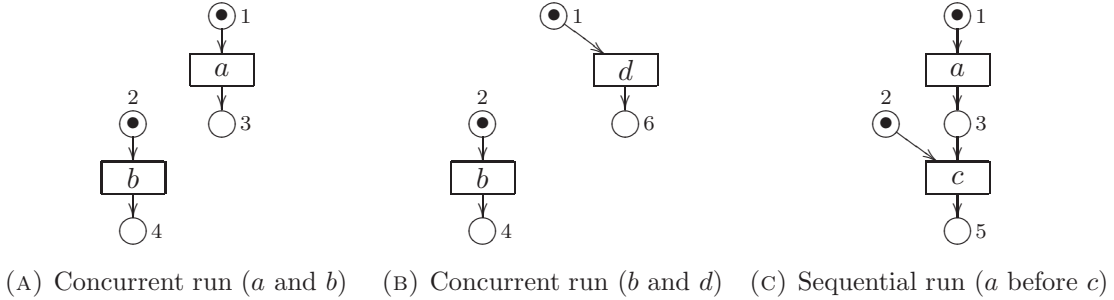
\begin{figure}[t]
\begin{subfigure}[b]{0.32\textwidth} 
$$
 \xymatrix@R=.75pc@C=.8pc{
 &
 \drawmarkedplace\ar[d]
 \nameplaceright 1
 \\
 & \drawtrans a\ar[d] 
 \\
 \drawmarkedplace\ar[d]
 \nameplaceup 2
 &
 \drawplace
 \nameplaceright 3
 \\
 \drawtrans b\ar[d] 
 \\
 \drawplace
 \nameplaceright {4}
 }
$$
%\vskip-8pt
\subcaption{Concurrent run ($a$ and $b$)}\label{fig:ab}
\end{subfigure}
\begin{subfigure}[b]{0.32\textwidth} 
$$
 \xymatrix@R=.75pc@C=.8pc{
 &
 \drawmarkedplace\ar[rd]
 \nameplaceright 1
 \\
 & 
 & \drawtrans d\ar[d] 
 \\
 \drawmarkedplace\ar[d]
 \nameplaceup 2
 &
 &
 \drawplace
 \nameplaceright 6
 \\
 \drawtrans b\ar[d] 
 \\
 \drawplace
 \nameplaceright {4}
 }
$$
%\vskip-8pt
\subcaption{Concurrent run ($b$ and $d$)}\label{fig:bd}
\end{subfigure}
\begin{subfigure}[b]{0.32\textwidth} 
$$
 \xymatrix@R=.75pc@C=.8pc{
 &
 \drawmarkedplace\ar[d]
 \nameplaceright 1
 \\
 & \drawtrans a\ar[d] 
 \\
 \drawmarkedplace\ar[dr]
 \nameplaceup 2
 &
 \drawplace\ar[d]
 \nameplaceright 3
 \\
 &
 \drawtrans c\ar[d] 
 \\
 &
 \drawplace
 \nameplaceright {5}
 }
$$
%\vskip-8pt
\subcaption{Sequential run ($a$ before $c$)}\label{fig:ac}
\end{subfigure}
\caption{Maximal processes for the net in Fig.~\ref{fig:condconfusion}}\label{fig:3proc}
\end{figure}

For example,  take the net in Fig.~\ref{fig:condconfusion}. 
It has three maximal processes that are reported in Fig.~\ref{fig:3proc}.
The equivalence class of the firing sequences 
$m_0 \xrightarrow{a\, b}$ and $m_0 \xrightarrow{b\, a}$ is the maximal process $\mathcal{D}$ in Fig.~\ref{fig:ab} with places $\{ 1, 2, 3, 4 \}$ and transitions
  $\{ a:1 \to 3, b:2\to 4 \}$, where 
  $\minp{\mathcal{D}} = \{ 1, 2\}$ and $\maxp{\mathcal{D}} = \{3,4\}$.
Likewise, the equivalence class of the firing sequences 
$m_0 \xrightarrow{b\, d}$ and $m_0 \xrightarrow{d\, b}$ is the maximal process in Fig.~\ref{fig:bd}.
As $c$ can only be executed after $a$, the corresponding process is in Fig.\ref{fig:ac}.

%\subsection{Prime Event Structures}\label{sec:eventstruct}

%When tokens are seen as resources that are manipulated by transitions,
%we can address the issue of establishing when a firing is causally
%dependent on another one (because it removes some token that was
%produced by the other), or in conflict (because they require the same
%resource) or concurrent (because they are independent from the order
%of execution).

%As shown in~\cite{DBLP:journals/tcs/NielsenPW81}, every (safe) Petri net $N$ can be associated with 
%a prime event structure $\mathcal{E}_N$ through a chain of coreflections.
% As a consequence, to each prime event structure
%$\mathcal{E}$, there is a standard, unique (up to isomorphism)  occurrence net $N_{\mathcal{E}}$
%that yields $\mathcal{E}$, hence we can freely move from one setting to
%the other.

Given an acyclic net we let $\preceq=F^*$ be the (reflexive) \emph{causality} relation 
and say that two transitions $t_1$ and $t_2$
are in \emph{immediate conflict}, written $t_1 \#_0 t_2$ if $t_1 \neq
t_2 \;\wedge\; \preS{t_1} \cap \preS{t_2} \neq \emptyset$.  
%Moreover, 
The \emph{conflict relation}
$\#$ is defined by letting 
$x \# y$ if there are $t_1,t_2\in T$ such that
  $(t_1,x), (t_2,y) \in F^+$ and $t_1 \#_0 t_2$.
Then, a \emph{nondeterministic occurrence net} (or just \emph{occurrence net}) is a net
$\mathcal{O}=(P,T,F)$ such that:
(1)~the flow relation is acyclic,
(2)~there are no backward conflicts ($\forall p\in P.~ |\preS{p}|\leq 1$), 
and
(3)~there are no self-conflicts ($\forall t\in T.~ \neg(t\# t)$).
The \emph{unfolding} $\mathcal{U}(N)$ of a safe Petri net $N$
%=(P,T,F,m_0)$ 
is an occurrence net that accounts
for all (finite and infinite) runs of $N$: its transitions 
%of $\mathcal{U}(N)$ 
model all the possible instances
%(\emph{occurrences}) 
of transitions in $N$ and its places 
%of $\mathcal{U}(N)$ 
model all the tokens that can be created in any
run.
%
%To fix the notation, 
%
Our construction takes a finite occurrence net {as input}, which can be, e.g., the (truncated) unfolding of any safe net.

\subsection{Nets With Persistency}\label{sec:persistentnets}
Nets with persistency (\emph{p-nets})~\cite{DBLP:journals/entcs/CrazzolaraW05}
 partition the set of places into regular
places $P$ (ranged by $p,q,...$) and persistent places $\mathbf{P}$
(ranged by $\mathbf{p}, \mathbf{q},...$).
%, with the assumption that
%once a persistent place is marked it stays marked.
%
We use $s$ to range over $\mathbb{S}=P\cup \mathbf{P}$ and write a p-net
as a tuple $(\mathbb{S},T,F)$.
Intuitively, persistent places  guarantee some sort of
monotonicity about the knowledge of the system.
%%
%For our purposes, persistent places represent the fact that certain
%places will stay empty and thus certain transitions will never become
%enabled, and thus certain alternatives are discarded, and thus certain
%decisions can be taken.
%
Technically, this is realised by letting states be bags of places
$b\in \Natinf^{\mathbb{S}}$ instead of multisets, with the
constraint that $b(p)\in \Nat$ for any regular place $p\in P$ and
$b(\mathbf{p})\in \{0,\infty\}$ for any persistent place
$\mathbf{p}\in \mathbf{P}$.
To guarantee that this property is preserved by firing sequences, we assume
 that the post-set $\postS{t}$ of a transition $t$ is the bag
such that:
$(\postS{t})(p) = 1$ if $(t,p)\in F$ (as usual); 
$(\postS{t})(\mathbf{p}) = \infty$ if $(t,\mathbf{p})\in F$; 
and $(\postS{t})(s) = 0$ if $(t,s)\not\in F$.
We say that a transition $t$ is \emph{persistent} if it is attached to persistent places only 
(i.e. if $\preS{t}\cup\postS{t} \subseteq \mathbf{P}$).

The notions of enabling, firing, firing sequence and reachability extend in the
obvious way to p-nets (when markings are
replaced by bags). 
For example, a transition $t$ is \emph{enabled} at the bag $b$,
written $b\xrightarrow{t}$,
if $\preS{t}\subseteq b$, and the firing of an enabled transition is written 
$b \xrightarrow{t} b'$ with $b' = (b\setminus \preS{t}) \cup \postS{t}$.

A firing sequence is \emph{stuttering} if it has multiple occurrences of a persistent
transition.
Since firing a persistent transition $t$ multiple times is inessential, we consider
non-stuttering firing sequences. (Alternatively, we can add a 
marked regular place $p_t$ to the preset of each persistent transition $t$, so $t$ fires at most once.)

A marked p-net $(N,b_0)$ is \emph{\OneInfSafe} if each reachable bag $b\in[b_0\rangle$ is such
that $b(p)\in\Bin$ for all $p\in P$ and
$b(\mathbf{p})\in \{0,\infty\}$ for all $\mathbf{p}\in \mathbf{P}$.
Note that in \OneInfSafe nets the amount of information conveyed
by any reachable bag is finite, as each place is associated with one
bit of information (marked or unmarked).
Graphically, persistent places are represented by circles with double
border (and they are either empty or
contain a single token). 
%See Fig.~\ref{fig:persistency} for an example.

The notion of confusion extends to p-nets, by checking direct conflicts w.r.t. regular places only.
%a \OneInfSafe  p-net $(N,b_0)$  has {\em confusion} iff there exist 
%$b\in[b_0\rangle$ and $t,u,v\in T$  such that: 
%\begin{itemize}
% \item 
%(1)~$t,u,v$ are enabled at $b$, 
%(2)~$\preS{t} \cap \preS{u}\cap P \neq \emptyset \neq \preS{u} \cap \preS{v} \cap P$,  and
%(3)~$\preS{t} \cap \preS{v} \cap P = \emptyset$; or
%%
% \item
%(1)~$t$ and $u$ are enabled at $b$, 
%(2)~$v$ is not enabled at $b$ but it becomes enabled after the firing of $t$, and
%(3)~$\preS{t} \cap \preS{u}\cap P = \emptyset$ and $\preS{v} \cap \preS{u} \cap P\neq \emptyset$.
%\end{itemize}

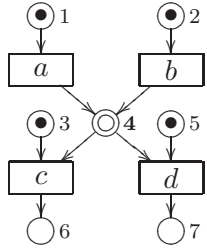
\begin{figure}[t]
$$
 \xymatrix@R=.75pc@C=.6pc{
 \drawmarkedplace\ar[d]
 \nameplaceright 1
 & &
 \drawmarkedplace\ar[d]
 \nameplaceright 2
 \\
 \drawtrans a \ar[dr] 
 & &
 \drawtrans b \ar[dl] 
 \\
 \drawmarkedplace\ar[d] 
 \nameplaceright 3
 &
 \drawpersistentplace\ar[dl] \ar[dr]
 \nameplaceright {{\scriptstyle{\bf 4}}}
 &
 \drawmarkedplace\ar[d] 
 \nameplaceright 5
 \\
 \drawtrans c \ar[d]
 & &
 \drawtrans d \ar[d] 
 \\
 \drawplace
  \nameplaceright 6
 & &
 \drawplace
  \nameplaceright 7
 }
$$
\caption{A marked p-net}\label{fig:persistentplace}
\end{figure}

As an example, consider the \OneInfSafe, confusion-free p-net in Fig.~\ref{fig:persistentplace}.
After firing $a$ and $c$, the firing of $b$ is inessential to enable $d$, because the persistent place ${\bf 4}$ is marked by $\infty$.

% analogously to the ordinary case, by
%restricting to regular places. 
%A \OneInfSafe marked persistent net $(N,m_0)$  is {\em confused} iff there exist  a reachable marking
%$m$ and transitions $t,u,v$  such that: 
%\begin{itemize}
% \item (i)  $t,u,v$ enabled at $m$ (ii)~$\preS{t} \cap \preS{u} \cap P\neq \emptyset \neq \preS{u} \cap \preS{v}\cap P$, and 
%(iii)~$\preS{t} \cap \preS{v} \cap P= \emptyset$; or
%%
% \item
%(i)~$t$ and $u$ are enabled at $m$, 
%(ii)~$v$ is not enabled at $m$ but it becomes enabled after the firing of $t$, and
%(iii)~$\preS{t} \cap \preS{u} \cap P= \emptyset$ and $\preS{v} \cap \preS{u} \cap P\neq \emptyset$.
%\end{itemize}
%
 
%\begin{rem}
%\marginpar{remove?}
%Persistent Petri nets may share similarities with contextual Petri nets~\cite{DBLP:journals/acta/MontanariR95}, 
%in the sense that a marked persistent place can be read concurrently by any number of transitions, as if read arcs
%were used. 
%However, using contextual nets, different persistent tokens would carry different causal dependencies that would be inherited
%after reading~\cite{DBLP:journals/iandc/BaldanCM01}, while using persistent places we disregard the identity of individual persistent tokens and introduce causal OR-dependencies that are needed to achieve complete concurrency (see Section~\ref{sec:concurrency}). 
%\end{rem}

%%% Local Variables:
%%% mode: latex
%%% TeX-master: "main.tex"
%%% End:

% !TEX root =  main.tex

\subsection{Dynamic Nets}\label{sec:dynnets}
Dynamic nets~\cite{DBLP:journals/mscs/AspertiB09} are Petri nets whose sets of places
and transitions may increase dynamically. We focus on a subclass of
persistent dynamic nets that only allows  for changes in the set of
transitions, which is defined as follows.

%In the following we let $\mathbb{S} = P \cup \mathbf{P}$ be a fixed set of regular places $P$ and
%  persistent places $\mathbf{P}$.
%  The next definition introduces a 
%version with persistency for such subclass.

\begin{defi}[Dynamic p-nets]
\label{def:PDN}
The set \emph{$\dn{\mathbb{S}}$} is the {\em
    least set} satisfying the recursive equation:
  \[
  \dn{\mathbb{S}}
  =
  \{(T, b) \;\; |\;\;   
    T \subseteq 2^{\mathbb{S}} \times \dn{\mathbb{S}}\;\;
    \wedge\;\; T \mbox{ finite } 
    \wedge\;\; b\in \Natinf^{\mathbb{S}} \}
  \]
\end{defi}

The   definition  above is a domain equation for the set of
 dynamic p-nets over the set of places $\mathbb{S}$:
the set $\dn{\mathbb{S}}$ is  the least fixed point of
the equation.
The simplest
elements in $\dn{\mathbb{S}}$ are pairs
$(\emptyset, b)$ with bag $b\in \Natinf^\mathbb{S}$
%
%Analogously to persistent Petri nets, we implicitly assume 
(with $b(p)\in \Nat$ for any $p\in P$ and 
$b(\mathbf{p})\in \{0,\infty\}$ for any $\mathbf{p}\in \mathbf{P}$).
Nets $(T, b)$ are defined recursively; indeed any element $t = (S,N)
\in T$ stands for a transition with preset $S$ and postset $N$,
which is another element of $\dn{\mathbb{S}}$.
An ordinary transition from $b$ to $b'$ has thus the form $(b,(\emptyset,b'))$.
%As usual, we denote $S_N\cup T_N$ by $N$, and omit subscript $N$
%whenever no confusion arises.  Moreover, we abbreviate a transition
%$t\in T$ such that $\delta_{0}(t) = P$ and $\delta_{1}(\mathtt{t}) =
%N$ as $P \pntrans N$.
%
We write $S\pntrans N$ for the transition $t= (S,N)$, 
$\preS t=S$ for its
{\it preset}, and $\postS t= N\in \dn{\mathbb{S}}$ for its {\it postset}.
For $N=(T, b)$ we say that $T$ is the set of \emph{top transitions} of $N$. 
All the other transitions are called dynamic.

The firing rule rewrites a  dynamic p-net $(T,b)$ to another one.
%operational semantics of persistent dynamic nets is defined
%by a labelled transition system whose states are persistent dynamic nets $(T,b)$ and
%whose firing rules is defined as follows: 
%In other words, 
The firing of a transition $t = S\pntrans (T', b') \in T$ consumes the preset $S$ and releases both the transitions $T'$ and the tokens in $b'$.
%in its postset $(T', b')$
Formally, 
if $t = S\pntrans (T', b') \in T$ with $S\subseteq b$ 
then $(T, b) \tr t (T\cup T', (b\setminus S)\cup b')$.
    
%It is evident that transitions are handled as persistent resources: they can be created but not erased.
%
The notion of \OneInfSafe dynamic p-net is defined
analogously to p-nets by considering the bags $b$ of reachable states $(T,b)$.

\begin{figure*} 
\begin{subfigure}[b]{0.45\textwidth} 
$$
 \xymatrix@R=.75pc@C=1.5pc{
 \drawmarkedpersistentplace \ar@[mygray][d]
 \namepersleft 3 
 &&
 \drawmarkedplace \ar@{-->}[ld]\ar[rd]
 \nameplaceleft 2
 &
 \drawplace \ar[d]
 \nameplaceleft 3
 \\
 \drawtransu {{t_3}}\ar@[mygray][d]\ar@[mygray][r] 
 & \drawtransd {t_b}\ar@{-->}[d]
 & 
 & \drawtrans {t_c}\ar@[mygray][d]\ar[ld]
 \\
 \drawpersistentplace
 \namepersleft 5
  &
 \drawplace
 \nameplaceleft 4
 &
 \drawplace
 \nameplaceleft 5
 &
 \drawpersistentplace
 \namepersright 4
  &
 &
 &
 }
$$
\subcaption{A dynamic p-net $N$}\label{fig:dyntaste}
\end{subfigure}
\begin{subfigure}[b]{0.5\textwidth} 
$$
 \xymatrix@R=.75pc@C=.6pc{
 \drawmarkedpersistentplace\ar@[mygray][dr]
&
 \drawmarkedpersistentplace \ar@[mygray][d]
 \namepersleft 3 
 &&&&&
&
 \drawmarkedplace \ar[dll]\ar[drr]
 \nameplaceleft 2
 &
 \drawmarkedpersistentplace\ar@[mygray][dr]
&
 \drawplace \ar[d]
 \nameplaceleft 3
 \\
& \drawtransu {\color{mygray}t_3}\ar@[mygray][d]\ar@[mygray][rr] 
&&
\drawpersistentplace \ar@[mygray][rr] 
  \nameplaceup {\color{mygray}p_b}
& & \drawtrans {t_b}\ar[d]
& & 
& & \drawtrans {t_c}\ar@[mygray][d]\ar[lld]
&&&
 \\
&
 \drawpersistentplace
 \namepersleft 5
&  &&&
 \drawplace
 \nameplaceleft 4
& &
 \drawplace
 \nameplaceleft 5
 &&
 \drawpersistentplace
 \namepersright 4
   }
$$
\subcaption{The corresponding p-net $\dyntopt N$} 
\label{fig:flat-taste3}
\end{subfigure}
\caption{A dynamic p-net  encoded as a p-net}
\end{figure*}
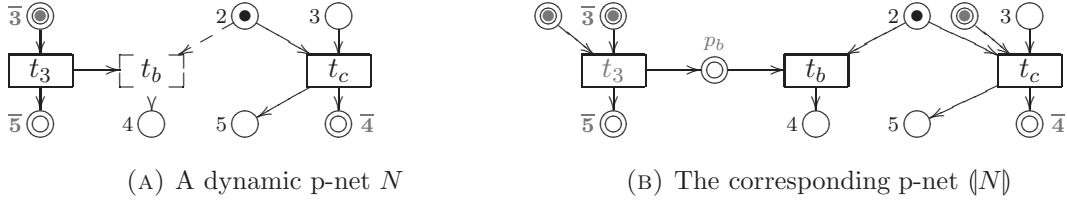
A sample of a dynamic net is shown in Fig.~\ref{fig:dyntaste}, whose only dynamic transition,
which is activated by $t_3$, 
 is depicted with dashed border.  
 The arrow between $t_3$ and $t_b$ denotes the fact that $t_b$ is activated dynamically by 
 the firing of $t_3  : \co {\bf 3} \pntrans (\{ b:  2 \pntrans 4 \}, \{\co{\bf 5}\})$.
%\subsection{Persistent Dynamic Nets as Persistent Nets}\label{sec:dynasflat}

We show that any dynamic p-net can be encoded
as a (flat) p-net. Our encoding resembles the one
in~\cite{DBLP:journals/mscs/AspertiB09}, but it is simpler because we do not need
to handle place creation.  
Intuitively, we release any transition $t$ immediately but we add a persistent place $\mathbf{p}_t$
to its preset, to enable $t$ dynamically ($\mathbf{p}_t$ is initially empty iff $t$ is not a top transition).
%
%the encoding works by
%representing the dynamic enabling of transitions as markings over
%additional persistent places. For a transition $t$, we write ${\bf
%  p}_t$ for the persistent place that represents the enabling of $t$.
%
%
Given a set $T$ of transitions, $b_T$ is the bag such
that $b_{T}({\bf p}_t) = \infty$ if $t\in T$ and $b_{T}(s) = 0$
otherwise. 

For $N = (T, b) \in \dn{\mathbb{S}}$, we let $\DT N = T \cup \bigcup_{t\in T}  \DT {\postS t}$ be the set of
all (possibly nested) transitions appearing in $N$. 
%
%Formally,
%%
%\[ 
%\begin{array}{l@{\ = \ }l}
%  \DT N & T \cup \bigcup_{t\in T}  \DT {\postS t} \\
%\end{array}
%\]
%
From Definition~\ref{def:PDN} it follows that $\DT N$ is finite and well-defined.
%, since the recursive type $\dn{\mathbb{S}}$ has the base case $(\emptyset, b)$ and every set $T$ of transitions is finite.

\begin{defi}[From dynamic to static]
  Given $N = (T, b) \in\dn{\mathbb{S}}$, the corresponding 
  p-net $\dyntopt N$ is defined as $\dyntopt N = (\mathbb{S} \cup {\bf P}_{\DT N}, \DT N,
  F, b\cup b_T)$, where

\begin{itemize}
\item
    ${\bf P}_{\DT N}=  \{ {\bf p}_t \ |\ t\in \DT{N}\}$; and
\item 
  $F$~is such that for any $t=S\pntrans (T', b') \in \DT N$ then $t:\preS t \cup \{{\bf p}_t\} \to b' \cup b_{T'}$.
%    $F = \{ (\preS t \cup \{{\bf p}_t\}, t) \ |\  t \in \DT N\} \
%    \cup\
%    \{(t, b' \cup b_{T'})\ |\ t \in \DT N \mbox{ and } \postS t = (T', b')\}$,
%  \item $b_{0} = b\cup b_T$.
  \end{itemize}
\end{defi}

\noindent
The transitions of $\dyntopt N$ are those from $N$ (set $\DT N$).
%, i.e., the ones in $\DT N$. 
Any place of $N$ is also a place of $\dyntopt N$ (set $\mathbb{S}$).
%(i.e, the ones in $\mathbb{S}$). 
In addition, there is one persistent place
${\bf p}_t$ for each $t\in\DT N$ (set ${\bf P}_{\DT N}$). 
%The flow relation $F$ is defined to preserve the behaviour of $N$ w.r.t. $\mathbb{S}$ 
%but also to handle dynamic transitions, i.e., by acting on ${\bf P}_{\DT N}$.
%Indeed, any pair $(\preS t
%\cup \{{\bf p}_t\}, t)\in F$ indicates that the transition $t$ has the
%place ${\bf p}_t$ as an additional condition to check that $t$ is
%actually enabled. Similarly, a pair $(t, b' \cup b_{T'})$ defining the
%postset of $t$ denotes that the transition $t$ produces not only the
%resources generated by the original transition $(T',b')\in N$ (i.e.,
%in $b'$) but also the persistent resources that simulate the dynamic
%enabling of all the transition in $T'$. 
The initial marking of $\dyntopt N$ is that of $N$ (i.e., $b$) 
together with the
persistent tokens that enable the top transitions of $N$ (i.e., $b_T$).
Adding $b_T$ is convenient for the statement in Proposition~\ref{prop:dynvsflat}, but
we could safely remove ${\bf P}_T\subseteq {\bf P}_{\DT N}$ (and $b_T$) from the flat p-net without any consequence. 
%\marginpar{Not true, we should change the encoding for top transitions}
%This has been done for the flat net in Fig.~\ref{fig:taste3} derived from the dynamic net that combines the ones in Figs.~\ref{fig:taste1}--\ref{fig:taste2}.

\begin{exa} The dynamic p-net $N$  in Fig.~\ref{fig:dyntaste} is encoded as the
 p-net $\dyntopt N$  in Fig.~\ref{fig:flat-taste3}, which has
as many transitions as $N$, but the preset of every transition contains an additional persistent place (depicted in grey) to indicate transition's availability. 
All  the new places but $p_b$  are  marked because the corresponding transitions are initially available. Contrastingly, 
$p_b$  is unmarked because the corresponding transition becomes available after the firing of $t_3$. 
\end{exa}

The following result shows that all computations of a dynamic p-net  can be mimicked by 
the corresponding p-net and vice versa. Hence, the encoding  preserves also 1-safety   over 
regular places. 

\begin{prop}\label{prop:dynvsflat}
  Let $N = (T, b) \in\dn{\mathbb{S}}$. Then,
  \begin{enumerate}
  \item
    $N\tr t N'$ implies $\dyntopt N\tr t \dyntopt {N'}$;
  \item Moreover,  $\dyntopt N \tr t N'$ implies there exists $N''$ such that $N\tr t N''$ and {$N' = \dyntopt {N''}$}.
  \end{enumerate}
\end{prop}

\begin{cor}
  $\dyntopt {N}$ is \OneInfSafe iff  $N$ is   1-safe.
\end{cor}

%%% Local Variables:
%%% mode: latex
%%% TeX-master: "main.tex"
%%% End:

% !TEX root =  main.tex

\section{From Petri Nets to Dynamic P-Nets}
\label{sec:results}

In this section we show that any (finite, acyclic) 
net $N$ can be associated with a confusion-free,  dynamic p-net  $\enc N$
by suitably encoding loci of decision.
% whose
%executions are in one-to-one correspondence with the
%recursively-stopped configurations of $\mathcal{E}_N$. 
The
mapping builds on the structural cell
decomposition introduced below. 

%Then, we recast the notion of branching cell in a
%structural approach, that makes the definition static and
%compositional and, in particular, independent from the past execution
%of the net. This original construction is outlined in
%Section~\ref{sec:structural-cells}.

% !TEX root =  main.tex

\subsection{Structural Branching Cells}
\label{sec:structural-cells}

A structural branching cell represents a statically determined
locus of choice, where the firing of some transitions is considered
against all the possible conflicting alternatives.  
To each transition $t$ we assign an s-cell $[t]$.
This is achieved by taking the equivalence class of $t$ w.r.t. the equivalence relation $\scelleq$
induced by the least preorder $\sqsubseteq$ that
includes immediate conflict $\#_0$ and causality $\preceq$.
%
%As we want s-cells to represent loci of decision, we find it convenient to
For convenience, each s-cell $[t]$ also includes the places in the pre-sets of the transitions in $[t]$, i.e., we let the relation $\mathsf{Pre}^{-1}$ be also included in $\sqsubseteq$, with $\mathsf{Pre}= F \cap (P\times T)$.
This way, if $(p,t)\in F$ then $p \sqsubseteq t$ because $p\preceq t$ and $t\sqsubseteq p$ because $(t,p)\in \mathsf{Pre}^{-1}$.
Formally, we let $\sqsubseteq$ be the transitive closure of the relation 
$\#_0\ \cup \preceq \cup\ \mathsf{Pre}^{-1}$.
Since $\#_0$ is subsumed by the transitive closure of the relation $\preceq \cup\ \mathsf{Pre}^{-1}$,
we  equivalently set $\sqsubseteq\ = (\preceq \cup\ \mathsf{Pre}^{-1})^*$.
Then, we let $\scelleq\ = \{(x,y) \mid x\sqsubseteq y \wedge y \sqsubseteq x\}$.
Intuitively, the choices available in the equivalence class $[t]_{|\scelleq}$ of a transition $t$ must be resolved atomically.

\begin{defi}[S-cells]\label{def:scells}
Let $N=(P,T,F)$ be a finite, nondeterministic occurrence net.
The set $\BC N$ of \emph{s-cells} is 
the set of equivalence classes of $\scelleq$, i.e., $\BC N = \{[t]_{|\scelleq} \mid t\in T\}$.
\end{defi}

\begin{rem}
Exploiting the algebraic structure of monoidal categories, in~\cite{DBLP:journals/corr/abs-1807-06305} we have given an alternative characterization of s-cells as those nets that can be decomposed neither in parallel nor in sequence.
The alternative definition is maybe more intuitive, but its formalization requires some technical machinery which we prefer to leave out of the scope of the present paper.
\end{rem}

We let $\bc$ range over s-cells.  
By definition it follows that for all $\bc,\bc'\in \BC N$, if $\bc \cap \bc'\neq \emptyset$ then $\bc = \bc'$.
%
%Branching cells partitions places and transitions of the net.
%
%\begin{lem}
%  For all $\bc,\bc'\in \BC N$, if $\bc \cap \bc'\neq \emptyset$ then $\bc = \bc'$.
%\end{lem}
%
%\begin{defi}
  For any s-cell $\bc$, we denote by $N_{\bc}$ the
  subnet of $N$ whose elements are in 
  $\bc \cup \bigcup_{t\in \bc} \postS{t}$.
%\end{defi}
%  Note that while $p\in \minp N_{\bc}$ implies that $p\in\bc$, $p\in
%  \maxp N_{\bc}$ actually implies that $p\not\in\bc$.
%
Abusing the notation, we denote by $\minp \bc$ the set of all the
initial places in $N_{\bc}$ and by $\maxp \bc$ the set of all the
final places in $N_{\bc}$.
%Abusing the notation, we denote by $\minp \bc$ the set of all the
%places $p\in \bc$ such that the unique transition $t$ with $\preS{p} =
%\{t\}$ is not in $\bc$ and by $\maxp \bc$ the set of all places $p
%\not\in \bc$ such that there exists $t\in \bc$ with $p\in \postS{t}$.
%We call $\min \bc$ the \emph{premises} of $\bc$.

\begin{defi}[Transactions]
  Let $\bc\in \BC N$.  Then, a
  \emph{transaction} $\theta$ of $\bc$, written $\theta:\bc$, is a maximal (deterministic) process of $N_{\bc}$.
\end{defi}

Since the set of transitions in a transaction $\theta$ uniquely
determines the corresponding process in $N_{\bc}$, we write a
transaction $\theta$ simply as the set of its transitions.

  \begin{figure}[t]
%    \begin{floatrow}
%      \ffigbox[0.54\textwidth]{
        \begin{subfigure}[b]{0.45\textwidth} 
          $$
          \xymatrix@R=.8pc@C=.8pc{
            &
            \drawmarkedplace\ar[d]\ar[rd]
            \nameplaceright 1
            \POS[]+<1.5pc,-1pc> *+=<5.5pc,4pc>[F--]{}
            \POS[]+<3pc,.5pc>\drop{\bc_1} 
            & &
            \drawmarkedplace\ar[d]\ar[rd]
            \nameplaceright 7
            \POS[]+<1.5pc,-1pc> *+=<5.5pc,4pc>[F--]{}
            \POS[]+<3.5pc,.5pc>\drop{\bc_2} 
            \\
            & \drawtrans a\ar[dd] 
            & \drawtrans d\ar[d] 
            & \drawtrans e\ar[dd] 
            & \drawtrans f\ar[d] 
            \\
            &
            &
            \drawplace
            \nameplaceright 6
            &
            &
            \drawplace
            \nameplaceright 9
            \\
            \drawmarkedplace\ar[d]\ar[dr]
            \nameplaceright 2
            &
            \drawplace\ar[d]
            \nameplaceright 3
            \POS[]+<2pc,-1pc> *+=<13.25pc,4pc>[F--]{}
            \POS[]+<8pc,.5pc>\drop{\bc_3} 
            &
            &
            \drawplace\ar[dll]\ar[d]
            \nameplaceright 8
            &
            \\
            \drawtrans b\ar[d] 
            &
            \drawtrans c\ar[d] 
            &
            &
            \drawtrans g\ar[d] 
            \\
            \drawplace
            \nameplaceright {4}
            &
            \drawplace
            \nameplaceright {5}
            &
            &
            \drawplace
            \nameplaceright{\ 10}
          }
          $$
%\vspace{-10pt}
          \subcaption{Structural branching cells}\label{fig:net-and-bc}
        \end{subfigure}
        \begin{subfigure}[b]{0.3\textwidth} 
          $$
          \begin{array}{ll@{ = } l}
            \multicolumn{3}{l}{\bc_1:}
            \\
            & \theta_a  & \{a\}
            \\
            & \theta_d  & \{d\}
            \\[5pt]
            \multicolumn{3}{l}{\bc_2:}
            \\
            & \theta_e  & \{e\}
            \\
            & \theta_f  & \{f\}
            \\[5pt]
            \multicolumn{3}{l}{\bc_3:}
            \\
            & \theta_c  & \{c\}
            \\
            & \theta_{bg}  & \{b,g\}
          \end{array}
          $$
%\vspace{-10pt}
          \subcaption{Transactions}\label{fig:transactions-N}
        \end{subfigure}
%      }
%      {
%        \caption{S-cells and transactions}\label{fig:net-and-bc}
%      }
%      \ffigbox[0.46\textwidth]{
        \begin{subfigure}[b]{0.3\textwidth} 
          $$
          \xymatrix@R=.8pc@C=.8pc{
            \drawplace
            \nameplaceright 3
            &
            \drawplace
            \drawplace\ar[d]
            \nameplaceright 8
            \\
            &
            \drawtrans g\ar[d] 
            \\
            &
            \drawplace
            \nameplaceright{\ 10}
          }
          $$
%\vspace{-4pt}
          \subcaption{$\remInitial{N_{\bc_3}} 2$}\label{fig:C3-2}
        \end{subfigure}
        \hspace{0.1\textwidth}
        \begin{subfigure}[b]{.3\textwidth} 
          $$
          \xymatrix@R=.8pc@C=1.8pc{
            \drawplace\ar[d]
            \nameplaceright 2
            &
            \drawplace\ar[d]
            \nameplaceright 8
            \\
            \drawtrans b\ar[d] 
            \POS[]+<-.5pc,.5pc> *+=<3.4pc,3.6pc>[F--]{}
            \POS[]+<-1.3pc,1.7pc>\drop{\bc_b} 
            &
            \drawtrans g\ar[d] 
            \POS[]+<-.5pc,.5pc> *+=<3.4pc,3.6pc>[F--]{}
            \POS[]+<-1.3pc,1.7pc>\drop{\bc_g} 
            \\
            \drawplace
            \nameplaceright {4}
            &
            \drawplace
            \nameplaceright{\ 10}
          }
          $$
%\vspace{-4pt}
          \subcaption{$\remInitial{N_{\bc_3}} 3$}\label{fig:C3-3}
        \end{subfigure}
        \hspace{0.1\textwidth}
        %\par
        \begin{subfigure}[b]{0.3\textwidth} 
          $$
          \xymatrix@R=.8pc@C=.8pc{
            \drawplace\ar[d]
            \nameplaceright 2
&
            \drawplace
            \nameplaceright 3
             \\
            \drawtrans b\ar[d] 
            \\
            \drawplace
            \nameplaceright {4}
          }
          $$
%\vspace{-4pt}
          \subcaption{$\remInitial{N_{\bc_3}} 8$}\label{fig:C3-8}
        \end{subfigure}
%      }{
%        \caption{Place removal}
%        \label{fig:place-removal}
%      }
%    \end{floatrow}
\caption{Structural branching cells (running example)}
\label{fig:s-cells}
  \end{figure}
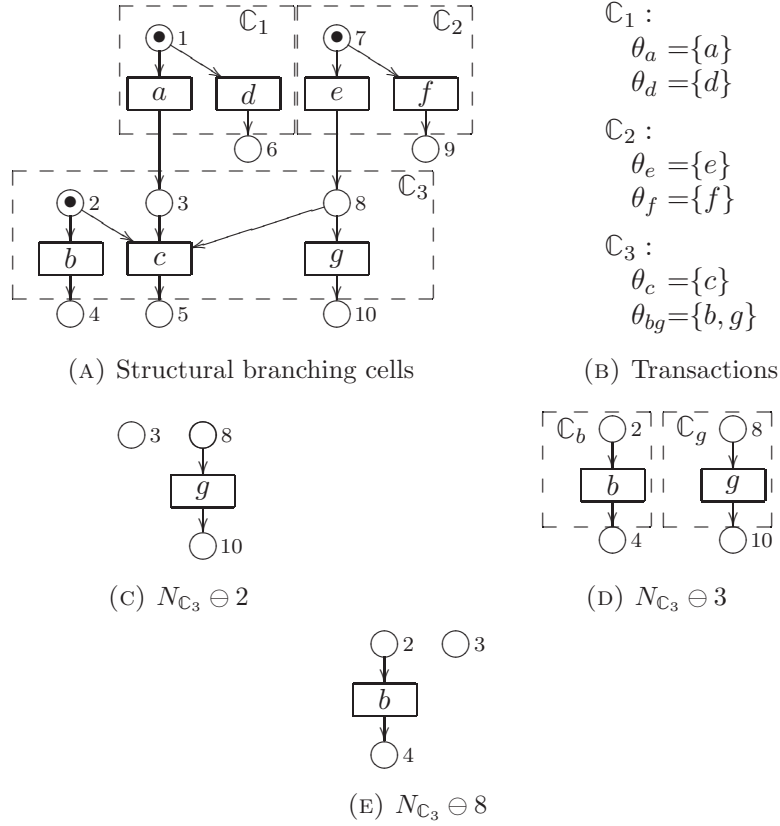
  
\begin{exa}
  The net $N$ in Fig.~\ref{fig:multcondconfusion} has the three s-cells shown in Fig.~\ref{fig:net-and-bc}, whose 
  transactions are listed in Fig.~\ref{fig:transactions-N}. For
  $\bc_1$ and $\bc_2$, each transition defines a transaction; $\bc_3$
  has one transaction associated with 
  $c$ and  one with (the concurrent firing of)
  $b$ and $g$.  
\end{exa}

%\begin{defi}
%(Structural Branching Cells)(nostra) Let \eventStructure{E}{\evSet} be an event structure , then $B\subseteq \evSet$ is a call \emph{structural branching cell}  if:
%\begin{itemize}
%\item $B$ is closed with respect to $\#_\mu$
%\item if $B^\prime\subseteq \evSet$ is a set closed  with respect to $\#_\mu$ then: $B \preceq B^\prime ~\land~B^\prime \preceq  B \Rightarrow B=B^\prime$
%\item $B$ is minimal (i.e. the only structural branching cell in $B$ is $\emptyset$)
%\end{itemize}
%We will indicate with $\setSBC$ the set of structural branching cells of $E$.
%\end{defi}

%Under the assumption that some place $p$
%in $\minp \bc$ will stay empty, the net $N_{\bc}$
%can be recursively decomposed in s-cells (by s-cells decomposition of the net $\remInitial {N_{\bc}} p$ obtained by canceling from $N_{\bc}$
%the nodes that depend on $p$). This fact will be
%exploited  in Section~\ref{sec:results}.

%%% Local Variables:
%%% mode: latex
%%% TeX-master: "main.tex"
%%% End:

%Intuitively, an s-cell $\bc$ stands for a locus of choice in which 
% alternatives are selected  when all places in $\minp \bc$  are marked.
%However, at some point of the computation it can be the case that some of the places in $\minp \bc$ will never receive a token.
%In that case we may recursively decompose the s-cell $\bc$ in smaller s-cells, by erasing the alternatives that are permanently disabled.

The following operation $\ominus$ is instrumental for the definition of our 
encoding and stands for the removal of a minimal place of a
net and all the elements that causally depend on it. Formally, $\remInitial N p$ is the least set that
satisfies the rules (where $\minp{\_}$ has higher precedence over set difference):
\[
\infer{q \in \remInitial N p}{q\in\minp N\setminus\{p\}}
\qquad
\infer{t \in \remInitial N p}{t \in N \qquad \preS t\subseteq \remInitial N p}
\qquad 
\infer{q \in \remInitial N p}{t \in \remInitial N p \qquad q \in \postS t}
\]
%With abuse of notation, we will use $\ominus$ also for branching cells.

%  \begin{figure}[t]
%    \begin{subfigure}[b]{0.12\textwidth} 
%      $$
%      \xymatrix@R=1.2pc@C=.8pc{
%        \drawplace\ar[d]
%        \nameplaceright 8
%        \\
%        \drawtrans g\ar[d] 
%        \\
%        \drawplace
%        \nameplaceright{\ 10}
%      }
%      $$
%      \subcaption{$\remInitial{N_{\bc_3}} 2$}\label{fig:C3-2}
%    \end{subfigure}
%    \hspace{.2cm}
%    \begin{subfigure}[b]{0.14\textwidth} 
%      $$
%      \xymatrix@R=1.2pc@C=.8pc{
%        \drawplace\ar[d]
%        \nameplaceright 2
%        &
%        \drawplace\ar[d]
%        \nameplaceright 8
%        \\
%        \drawtrans b\ar[d] 
%        &
%        \drawtrans g\ar[d] 
%        \\
%        \drawplace
%        \nameplaceright {4}
%        &
%        \drawplace
%        \nameplaceright{\ 10}
%      }
%      $$
%      \subcaption{$\remInitial{N_{\bc_3}} 3$}\label{fig:C3-3}
%    \end{subfigure}
%    \hspace{.1cm}
%    \begin{subfigure}[b]{0.12\textwidth} 
%      $$
%      \xymatrix@R=1.2pc@C=.8pc{
%        \drawplace\ar[d]
%        \nameplaceright 2
%        \\
%        \drawtrans b\ar[d] 
%        \\
%        \drawplace
%        \nameplaceright {4}
%      }
%      $$
%      \subcaption{$\remInitial{N_{\bc_3}} 8$}\label{fig:C3-8}
%    \end{subfigure}
%    \caption{Removal of minimal places of $N$} 
%  \end{figure}

\begin{figure}[t]
$$
 \xymatrix@R=.8pc@C=.8pc{
 &
 \drawmarkedplace\ar[d]\ar[rd]
 \nameplaceright 1
 \POS[]+<1.4pc,-1pc> *+=<5.5pc,4pc>[F--]{}
 \POS[]+<3.5pc,.5pc>\drop{\bc_1} 
 \\
 & \drawtrans a\ar[d] 
 & \drawtrans d\ar[d] 
 \\
 &
 \drawplace\ar[dd]
 \nameplaceright 3
 &
 \drawplace
 \nameplaceright 6
 \\
 \drawmarkedplace\ar[d]\ar[dr]
 \nameplaceleft 2
 \POS[]+<1.8pc,-1pc> *+=<6.8pc,4pc>[F--]{}
 \POS[]+<4.4pc,.30pc>\drop{\bc_2} 
 \POS[]+<0.2pc,-1pc> *+=<2.8pc,3.6pc>[F--]{}
 \POS[]+<1pc,0pc>\drop{\bc_3} 
 \\
 \drawtrans b\ar[d] 
 &
 \drawtrans c\ar[d] 
 \\
 \drawplace
 \nameplaceright {4}
 &
 \drawplace
 \nameplaceright {5}
 }
$$
\caption{S-cells for the net in Fig.~\ref{fig:condconfusion}}\label{fig:scellscondconfusion}
\end{figure}
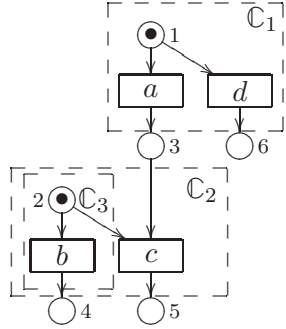

\begin{exa}
Consider the net in Fig.~\ref{fig:condconfusion}. 
There are two main s-cells: $\bc_1$ associated with $\{a,d\}$, and $\bc_2$ with $\{b,c\}$.
There is also a nested s-cell $\bc_3$ that arises from the decomposition of the subnet $\remInitial {N_{\bc_2}} 3$.
All the above s-cells are shown in Fig.~\ref{fig:scellscondconfusion}.
\end{exa}

\begin{exa}
  Consider the s-cells in  Fig.~\ref{fig:net-and-bc}. 
  The net $\remInitial {N_{\bc_1}} 1$
  is empty because every node in $N_{\bc_1}$ causally depends on
  $1$. Similarly, $\remInitial {N_{\bc_2}} 7$ is empty. The cases
  for $\bc_3$ are in Figs.~\ref{fig:C3-2}--\ref{fig:C3-8}.
\end{exa}

\subsection{Encoding s-cells as confusion-free dynamic nets}\label{sec:encscells}
%Under the assumption that some place $p$
%in $\minp \bc$ will stay empty, the net $N_{\bc}$
%is recursively decomposed in s-cells (by s-cells decomposition of 
%the net $\remInitial {N_{\bc}} p$ obtained by canceling from $N_{\bc}$
%the nodes that depend on $p$).
Intuitively, the proposed encoding works by explicitly representing
 the fact that a place 
will not be marked in a  computation.  We denote with $\co {\bf
  p}$ the place that models such ``negative'' information about the regular place
$p$ and let $\co {\bf P} = \{{\bf \co p}\ |\ p\in P\}$.\footnote{The notation $\co{P}$ denotes just a set of places whose names are decorated with a bar; it should not be confused with usual set complement.}
%Note that an arc from $\co {\bf p}$ to a transition $t$ is  
%very different from an inhibitor arc from $p$ to $t$: the former checks that $p$ will stay empty,
%while the latter that $p$ is currently empty.
%
The encoding uses negative information to recursively decompose s-cells 
under the assumption that some of their minimal places will stay empty.

\begin{defi}[From s-cells to dynamic p-nets] 
  \label{def:BC-in-DN}
  Let $N = (P,T,F,m)$ be a marked occurrence net. Its dynamic p-net $\enc N \in \dn {P \cup
    \co {\bf P}}$ is defined as $\enc N = (\Tpos \cup \Tneg, m)$, where:
  \[
  \begin{array}{l@{\ }l@{\ }l} 
  \Tpos &= & \{\
  \minp \bc 
  \pntrans 
  (\emptyset, \maxp\theta \cup \co{\maxp \bc\setminus\maxp \theta})
  \ \mid\ \bc \in \BC N \ \mbox{and }
    \theta : \bc\ \}
    \\ % and
    \Tneg &=& \{\ 
  \co{\bf p} 
  \pntrans 
  (T', \co{\maxp \bc\setminus\maxp {(\remInitial {N_{\bc}} p)}}) 
  \ \mid\; \bc \in \BC N \ \mbox{and }  
  p \in \minp {\bc} \ \quad \\
  &&
  \phantom{\{\ \co{\bf p} \pntrans (T', \co{\maxp \bc\setminus\maxp {(\remInitial {N_{\bc}} p)}}) \ \mid\;}\;
  \mbox{and } (T', b) = \enc{\remInitial {N_{\bc}}
      p}\ \}
   \end{array}
   \]
\end{defi}

%The encoding of a net $N$ is defined in terms of the structural branching cells $\bc$ of $N$.  
For any s-cell $\bc$ of $N$ and
 transaction $\theta : \bc$, the encoding generates a transition
$t_{\theta,\bc} = (\minp \bc \pntrans (\emptyset, \maxp\theta \cup \co{\maxp
    \bc\setminus\maxp \theta})) \in \Tpos$ to mimic the atomic
execution of $\theta$. 
%
%For simplicity, each transaction is encoded as a single transition but we could 
%consider instead copies of the deterministic process associated with the transaction.
%
Despite $\minp \theta$ may be strictly included in $\minp \bc$, 
we define $\minp \bc$ as the preset of $t_{\theta,\bc}$ to ensure that the execution of $\theta$ only
starts when the whole s-cell $\bc$ is enabled.
Each transition $t_{\theta,\bc} \in \Tpos$ is a transition of an ordinary Petri net because
%its firing does not introduce new transitions, i.e., 
its postset consists of (i) the final places of $\theta$
%, i.e., the tokens generated by the execution of $\theta$, 
and (ii) the negative versions of the places in $ \maxp \bc\setminus\maxp \theta$.
A token in $\co{\bf{p}} \in \co{\maxp \bc\setminus\maxp \theta}$ represents the fact that the
corresponding ordinary place $p\in \maxp \bc$ will not be marked because it depends on 
discarded transitions (not in $\theta$).

Negative information is propagated by the transitions in $\Tneg$.  
For each cell $\bc$ and place $p\in \minp \bc$, there exists one dynamic transition 
$t_{p,\bc} =   \co{\bf p} 
  \pntrans 
  (T', \co{\maxp \bc\setminus\maxp {(\remInitial {N_{\bc}} p)}})$
whose preset is just $\co{\bf p}$ and whose postset %propagates the negative
%information about $p$.
%
%Each transition in $t_{p,\bc}\in \Tneg$ 
is defined in terms of the subnet
$\remInitial {N_{\bc}} p$.
%, i.e., the net obtained from $N_{\bc}$
%(i.e., the subnet associated with $\bc$) 
%by removing all the elements
%that causally depend on $p$.
%
The postset of $t_{p,\bc}$   accounts for two effects of
propagation: (i) the generation of the negative tokens for all maximal
places of $\bc$ that causally depend on $p$, i.e., for the negative
places associated with the ones in $\maxp\bc$ that are not in $\maxp
{(\remInitial {N_{\bc}} p)}$; and (ii) the activation of all
transitions $T'$ obtained by encoding $\remInitial {N_{\bc}} p$, i.e., 
the behaviour of the branching cell $\bc$ after the token
in the minimal place $p$ is excluded.  We remark that the bag
$b$ in $(T', b) = \enc{\remInitial {N_{\bc}} p}$ is always
empty,
because i)  $N_{\bc}$ is unmarked
%has an empty initial marking 
and, consequently,  $\remInitial {N_{\bc}} p$ is unmarked,
%has an empty initial marking 
%for any $p$, 
and ii)
the initial marking of $\enc N$ corresponds to the initial
marking of $N$. 
% i.e., there is one transition for the negate place of each minimal place of a structural branching cell;  

%Lemmas: never black and red together, monotony, nested rules do not
%collide

%\begin{lem}[Transitions in $T_neg$ are persitent] Let $(T,b) = \enc
%N$. If $t \in T$ and $\preS t \subseteq \co S$ then t is persistent,
%i.e., $\postS t = (T', b')$ and $b' \subseteq \co{S}$.
%\end{lem}

\begin{figure}[t]
\begin{subfigure}[b]{0.25\textwidth} 
$$
 \xymatrix@R=1.2pc@C=.6pc{
 \drawmarkedplace\ar[d]\ar[rrd]
  \nameplaceup 1
 &
 \drawpersistentplace\ar@[mygray][d]
 \namepersup 1
 \\
 \drawtrans {t_a}\ar[d]\ar@[mygray][rd]
 & \drawtransu {}\ar@[mygray][d]\ar@[mygray][rd] 
 & \drawtrans {t_d}\ar@[mygray][d]\ar[rd]
 \\
 \drawplace
 \nameplaceright 3 
 &
 \drawpersistentplace
 \namepersright 6
 &
 \drawpersistentplace
 \namepersright 3 
 &
 \drawplace
 \nameplaceright 6 
 }
$$
%\vspace{-10pt}
\subcaption{S-cell $\bc_1$}\label{fig:taste1}
\end{subfigure}
\begin{subfigure}[b]{0.38\textwidth} 
$$
 \xymatrix@R=1.2pc@C=.9pc{
 \drawpersistentplace \ar@[mygray][d] 
 \namepersup 2
 &
 \drawpersistentplace \ar@[mygray][d]
 \namepersup 3 
 &&
 \drawmarkedplace \ar@{-->}[ld]\ar[d]\ar[rd]
 \nameplaceup 2
 &
 \drawplace \ar[ld]\ar[d]
 \nameplaceup 3
 \\
 \drawtransu {}\ar@[mygray][rd]\ar@[mygray]@(dr,ul)[rrrrd]
 & \drawtransu {}\ar@[mygray][d]\ar@[mygray][r] 
 & \drawtransd {t_b}\ar@{-->}[d]
 & \drawtrans {t'_b}\ar@[mygray][lld]\ar[ld]
 & \drawtrans {t_c}\ar@[mygray][d]\ar[ld]
 \\
 &
 \drawpersistentplace
 \namepersright 5
  &
 \drawplace
 \nameplaceright 4
 &
 \drawplace
 \nameplaceright 5
 &
 \drawpersistentplace
 \namepersright 4
  &
 &
 &
 }
$$
%\vspace{-10pt}
\subcaption{S-cell $\bc_2$ and its sub s-cell $\bc_3$}\label{fig:taste2}
\end{subfigure}
\begin{subfigure}[b]{0.32\textwidth} 
$$
 \xymatrix@R=.8pc@C=.6pc{
 &
 &
 \drawmarkedplace \ar[ld]\ar[rd]
 \nameplaceright 1 
 \\
 &
 \drawtrans {t_d}\ar@[mygray][dl]\ar[d]
 && \drawtrans {t_a}\ar@[mygray][dl]\ar[rd]
 \\
 \drawpersistentplace \ar@[mygray][d]
 \namepersright 3
 &
 \drawplace 
 \nameplaceright 6
 &
 \drawpersistentplace 
 \namepersup 6
 &
 \drawmarkedplace \ar[ld]\ar[d]\ar[rd]
 \nameplaceup 2
 &
 \drawplace \ar[ld]\ar[d]
 \nameplaceup 3
 \\
 \drawtransu {}\ar@[mygray][d]\ar@[mygray][r] 
 &
 \drawpersistentplace \ar@[mygray][r]
 \nameplaceup {\color{mygray}p_b}
 & \drawtransu {t_b}\ar[d]
 & \drawtrans {t'_b}\ar@[mygray]@(dl,ur)[llld]\ar[ld]
 & \drawtrans {t_c}\ar@[mygray][d]\ar[ld]
 \\
 \drawpersistentplace
 \namepersright 5
 &&
 \drawplace
 \nameplaceright 4
 &
 \drawplace
 \nameplaceright 5
 &
 \drawpersistentplace
 \namepersright 4
 }
$$
%\vspace{-10pt}
\subcaption{Flat net (pruned)}\label{fig:taste3}
\end{subfigure}
%\caption{Structural cells as dynamic nets and their composed, flattened version}\label{fig:taste}
\caption{S-cells as dynamic nets and their composed, flattened version (for the net in Fig.~\ref{fig:condconfusion})}
\end{figure}
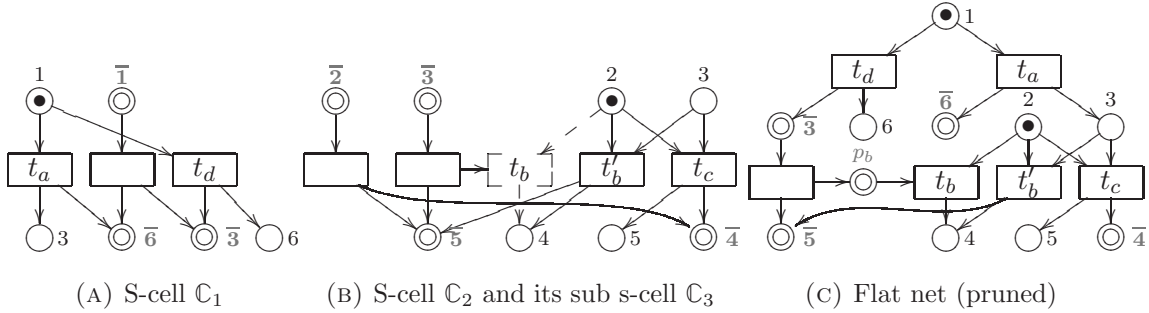

\begin{exa}
We sketch the main ideas over the net in Fig.~\ref{fig:condconfusion}.
We recall that it has two main s-cells ($\bc_1$ associated with $\{a,d\}$, and $\bc_2$ with $\{b,c\}$) and 
a nested one ($\bc_3$): see Fig.~\ref{fig:scellscondconfusion}.
Their dynamic nets are in Figs.~\ref{fig:taste1}--\ref{fig:taste2},
where auxiliary transitions are in grey and unlabeled. Places $\mathbf{\overline{1}}$ and $\mathbf{\overline{2}}$
(and their transitions) are irrelevant, because the places $1$ and $2$ are already marked. 
However, our cells being static, we need to introduce auxiliary places in all cases.
Note that in Fig.~\ref{fig:taste2} there is an arc between two transitions.
As explained before, this is because the target transition is dynamically created when the other is executed (hence the dashed border).
Also note that there are two transitions with the same subscript $b$: one ($t'_b$) is associated with the s-cell $\bc_2$, the other ($t_b$) with the unique s-cell $\bc_3$ of $\remInitial {N_{\bc_2}} 3$ and is released when the place $\mathbf{\overline{3}}$ becomes marked.

After the s-cells are assembled and flattened we get the p-net in Fig.~\ref{fig:taste3} (where irrelevant nodes are pruned).
Initially, $t_a$ and $t_d$ are enabled. Firing $t_a$ leads to the marking $\{2,3,\mathbf{\overline{6}}\}$ where $t'_b:\{2,3\}\to \{4,\mathbf{\overline{5}}\}$ and $t_c:\{2,3\}\to \{\mathbf{\overline{4}},5\}$ are enabled (and in conflict). 
Firing $t_d$ instead leads to the marking $\{2,\mathbf{\overline{3}},6\}$ where only the auxiliary transition can be fired, 
%leading to $\{2,\mathbf{\overline{3}},\mathbf{\overline{5}},6,\mathbf{p}_b\}$ and 
enabling $t_b:\{2,\mathbf{p}_b\} \to 4$. The net is confusion-free, as every conflict involves transitions with the same preset. For example, as the places $3$ and $\mathbf{\overline{3}}$ (and thus $\mathbf{\overline{p}}_b$) are never marked in a same run,
the transitions $t'_b:\{2,3\}\to \{4,\mathbf{\overline{5}}\}$ and $t_b:\{2,\mathbf{p}_b\} \to 4$ will never compete for the token in $2$.
\end{exa}

\begin{exa}
\label{ex:enc-bc-into-dyn}
 
\makeatletter
\newcommand{\vast}{\bBigg@{4}}
\newcommand{\Vast}{\bBigg@{5}}
\makeatother

  \begin{figure}[t]
      $$
      \begin{array}{l@{}l}
      \begin{array}{l@{\ : \ }l@{}l@{\ }l}
        t_a
        & 1 \pntrans (\emptyset, \{3, \co {\bf 6}\})
        && \mbox{for}\ \theta_a %   \{a\}
        \\
        t_d
        & 1 \pntrans (\emptyset, \{6, \co {\bf 3}\})
        && \mbox{for}\ \theta_d  %= \{d\}
        \\[3pt]
        t_1
        & \co {\bf 1} \pntrans (\emptyset, \{\co{\bf 3}, \co{\bf 6}\})
        \\[2pt]
        t_e
        & 7 \pntrans (\emptyset, \{8, \co{\bf 9}\})
        && \mbox{for}\ \theta_e %=  \{e\}
        \\
        t_f
        & 7  \pntrans (\emptyset, \{9, \co{\bf 8}\})
        && \mbox{for}\ \theta_f  %=  \{f\}
        \\[3pt]
        t_7
        & \co {\bf 7} \pntrans (\emptyset, \{\co{\bf 8}, \co{\bf 9}\})
        \\[2pt]
        t_{bg}
        & 2,3,8 \pntrans (\emptyset, \{4, 10, \co {\bf 5}\})
        && \mbox{for}\ \theta_{bg} %=  \{b,g\}
        \\
        t_c
        & 2,3,8 \pntrans (\emptyset, \{5, \co {\bf 4}, \co {\bf 10}\})
        && \mbox{for}\ \theta_c %=  \{c\}
        \end{array}
&
       \begin{array}{l@{\ : \ }l@{}l@{}l}  
        t_2
        & \co {\bf 2} \pntrans (\{t_g, t'_8\}, \{\co {\bf 4}, \co {\bf 5}\})
         \\
        t_3
        & \co {\bf 3}  \pntrans ( \{t_b, t'_2, t_g, t'_8\}, \{\co {\bf 5}\})
        & 
   \\
        t_8
        & \co {\bf 8}  \pntrans (\{t_b, t'_2\} \{\co {\bf 5}, \co {\bf 10}\})
        &
        \\[10pt]
         \multicolumn{2}{l}{\it where}
         \\
         \multicolumn{1}{c}{}&
        t_b : 2 \pntrans (\emptyset, \{4\})
        \\
         \multicolumn{1}{c}{}&
        t'_2 : \co{\bf 2} \pntrans (\emptyset, \{\co{\bf 4}\})
        \\
          \multicolumn{1}{c}{}&
        t_g : 8 \pntrans (\emptyset, \{10\})
        \\
         \multicolumn{1}{c}{}&
        t'_8 : \co{\bf 8} \pntrans (\emptyset, \{\co{\bf 10}\})
        %t_{12}  & 2 \pntrans (\emptyset, \{4\})\\
%t_{13}  & \co{\bf 2} \pntrans (\emptyset, \{\co{\bf 4}\})
      \end{array}
      \end{array}
      $$
%            \vspace{-5pt}
     \caption{Encoding of branching cells (running example)}
     \label{fig:enc-bc-into-dyn}
  \end{figure}

  Consider the net $N$ and its s-cells in Fig.~\ref{fig:net-and-bc}.  Then, 
  $\enc{N} = (T, b)$ is defined such that $b$ is the initial marking of $N$, 
  i.e., $b= \{1,2,7\}$, and $T$ has the transitions shown in 
  Fig.~\ref{fig:enc-bc-into-dyn}.  
%  We now comment on the transitions generated by the encoding.  

  First consider the s-cell
  $\bc_1$. 
   $\Tpos$ contains one transition for each transaction in 
  $\bc_1$,  namely $t_a$ (for $\theta_a : \bc_1$) and $t_d$ (for $\theta_d : \bc_1$).
%$\bc_1$ has two transactions .
    Both $t_a$ and $t_d$
  have $\minp {\bc_1}= \{1\}$ as preset.  By definition of $\Tpos$,
  both transitions have  empty sets of transitions in their postsets. Additionally,
  $\postS{t_a}$ produces  tokens in $\maxp {\theta_a} = \{3\}$ (positive) and 
   $\co{\maxp \bc_1\setminus\maxp {\theta_a}} = \co{\{3, 6\}
    \setminus\{3 \}} = \{\co {\bf 6}\}$ (negative), while $\postS{t_d}$ produces tokens in $\maxp {\theta_d} =
  \{6\}$ and $\co{\maxp \bc_1\setminus\maxp \theta_d} = 
  %\co{\{3, 6\} \setminus\{6 \}} = 
  \{\co {\bf 3}\}$.
  Finally, $t_1\in \Tneg$ propagates negative tokens for the unique place in $\minp {\bc_1} = \{1\}$.  
  Since $\remInitial {N_{\bc_1}} 1$ is the
  empty net, $\enc{\remInitial {N_{\bc_1}} 1} = (\emptyset,
  \emptyset)$. Hence,  $t_1$  produces negative tokens
  for all maximal places of $\bc_1$, i.e., $\{\co{\bf 3}, \co{\bf
    6}\}$.
%
  %The branching cell $\bc_2$ is isomorphic to $\bc_1$; consequently,
%
  For the s-cell $\bc_2$ we  analogously obtain the transitions $t_e$, $t_f$ and $t_7$.

  The s-cell $\bc_3$ has %$\minp {\bc_3} = \{2,3,8\}$, $\maxp {\bc_3} = \{4,5,10\}$ and the
  two transactions $\theta_{bg}$ and $\theta_{c}$.
  Hence, $\enc{N}$ has two transitions $t_{bg},t_c\in \Tpos$.
% are associated with the transactions of $\bc_3$. 
  Despite $\theta_{bg}$ mimics the firing of $b$ and
  $g$, which are disconnected from the place $3$, {it is included in} the preset of
  $t_{bg}$ 
  to postpone the firing of $t_{bg}$ until $\bc_1$ is executed.
  Transitions $t_2,t_3,t_8\in \Tneg$ propagate the negative
  information for the places in $\minp \bc_3 = \{2, 3, 8\}$.
  The transition $t_3$ has  $\preS{t_3}=\{\co {\bf 3}\}$ as its preset and its postset is
  obtained from  $\remInitial {N_{\bc_3}} 3$,
  which has two (sub) s-cells ${\bc_b}$ and ${\bc_g}$ (see Fig.~\ref{fig:C3-3}).  
  The transitions $t_b$ 
  %= 2 \pntrans (\emptyset, 4)$ 
  and $t'_2$
  % = \co {\bf 2} \pntrans (\emptyset, \co {\bf 4})$ 
  arise from ${\bc_b}$, and $t_g$
  % = 8 \pntrans (\emptyset, 10)$ 
  and $t'_8$
  % = \co {\bf 8} \pntrans (\emptyset,\co {\bf 10})$ 
  %are obtained 
  from ${\bc_g}$.  Hence, ${\postS{t_{3}}} = (\{t_b, t'_2,t_g, t'_8\},\{\co{\bf 5}\})$ 
  because
  $\enc{\remInitial {N_{\bc_3}} 3} = 
  (\{t_b, t'_2,t_g, t'_8\},\emptyset)$
  and 
   $\co{\maxp {\bc_3}\setminus\maxp {(\remInitial {N_{\bc_3}} 3)}} = \{\co{\bf 5}\}$.
  Similarly, we derive $t_2$ from $\remInitial {N_{\bc_3}} 2$ and $t_8$ from $\remInitial
  {N_{\bc_3}} 8$.
  
  We now highlight some features of the encoded net.  
  First, the set of top transitions is free-choice: positive and negative transitions have disjoint presets
  and the presets of any two positive transitions either coincide (if they arise from the same s-cell) or
  are disjoint. 
%  For any pair of transitions $t$, $t'$ in $T$ it holds that either $\preS t = \preS
%  t'$ or $\preS t \cap \preS t = \emptyset$, i.e., the net is
%  free-choice wrt the initially enabled transitions.  
  Recursively, this property holds at any level of nesting.
%  , e.g., the transitions in $\{t_b, t'_2, t_g, t'_8\}$ are also free choice.
  Hence, the only source of potential confusion is due to the
  combination of  top transitions and those activated dynamically, e.g., $t_b$ and either $t_{bg}$ or
  $t_c$. However, $t_b$ is activated only when either $\co{\bf 3}$ or
  $\co{\bf 8}$ are marked, while $\preS{t_{bg}} = \preS{t_c} = \{2, 3,8\}$. Then, confusion is avoided if 
  $p$ and $\co{\bf p}$ can never be marked in the same execution
  (Lemma~\ref{lemma:negpos}).

  \begin{figure}[t]
  \vskip-3pt
$$
 \xymatrix@R=1pc@C=.5pc{
&
&
& 
\drawmarkedplace\ar[dl]\ar[dr] 
\nameplaceright 1
 &
 &
 &
 &
 &
 \drawmarkedplace\ar[dl]\ar[dr]
 \nameplaceright 7
 \\
&
 & 
 \drawtrans {t_d}\ar[dl]\ar@[mygray][dr] 
 &
 &
  \drawtrans {t_a}\ar@[mygray][dr]\ar[dddr] 
  &
  &
  &
  \drawtrans {t_e}\ar@[mygray][dl]\ar@/^3ex/[dddllll] 
  &
  &
   \drawtrans {t_f}\ar@[mygray][dl]\ar[dr] 
 \\
&
 \drawplace
 \nameplaceup 6
 &
 &
\drawpersistentplace\ar@[mygray][d] 
 \namepersup{ 3}
 &
 &
 \drawpersistentplace
 \namepersup{ 6}
 &
 \drawpersistentplace
 \namepersup{ 9}
 &
 &
\drawpersistentplace\ar@[mygray][d]\ar@{-->}@[mygray]@/^2ex/[ddr] 
 \namepersup{ 8}
 &
 &
 \drawplace
 \nameplaceright 9
  \\
 &&&
 \drawtransu {\color{mygray}t_3}\ar@[mygray]@/_2ex/[dddlll] \ar@[mygray]@/^6ex/[ddrrrrr]\ar@[mygray]@/^3ex/[drrrrrr]\ar@[mygray]@/_1ex/[ddl]
 &&&&&
 \drawtransu {\color{mygray}t_8}\ar@[mygray][dd]\ar@[mygray]@/^4ex/[dddrrr] \ar@[mygray]@/_7ex/[dddllllllll] 
 \\
% &&&&&&&&&&
%   \\
 &
&&
 \drawplace\ar[dr]\ar[drrr]\ar@{-->}[dl]
 \nameplaceleft {8}
 &
 &
 \drawplace\ar[dl]\ar[dr]
 \nameplaceright {3}
&
&
 \drawmarkedplace\ar[dl]\ar@{-->}[dr]\ar[dlll]
 \nameplaceleft {2}
 &
 &
 \drawtransud {\color{mygray}t'_8}\ar@[mygray]@{-->}[ddrr]  
 &
  \\
 &
 &
 \drawtransd {t_g}\ar@{-->}[dr] 
 &
 &
 \drawtrans {t_{bg}}\ar[drrrr]\ar@[mygray]@(dl,ur)[dllll]\ar[dl]
 &
 &
 \drawtrans {t_c}\ar[d]\ar@[mygray][drrrrr]\ar@[mygray][dll]
  &
  &
 \drawtransd {t_b}\ar@{-->}[d]
  &
 &
 \\
 \drawpersistentplace
 \namepersright{5}
 &
 &
 &
 \drawplace
 \nameplaceright {10}
 &
  \drawpersistentplace
 \namepersright{4}
 &&
 \drawplace
 \nameplaceright {5}
&
&
 \drawplace
 \nameplaceright {4}
 &
 &
 &
  \drawpersistentplace
 \namepersright{10}
 &&
  }
$$
%\vskip-2pt
        \caption{Dynamic net $\enc{N}$ (running example)}
        \label{fig:graphical-encoded-N}   
\end{figure}

  The net $\enc N$ is shown in Fig.~\ref{fig:graphical-encoded-N}, where the places 
  $\{\co{\bf 1},\co{\bf 2},\co{\bf 7} \}$ and the transitions $\{t_1,t_7,t_2,t'_2\}$ are omitted
  because superseded by the initial marking $\{1,2,7\}$.

  We remark that the same dynamic transition can be
  released by the firing of different transitions (e.g., $t_b$ by
  $t_3$ and $t_8$) and possibly several times
  in the same computation. 
  Similarly, the same negative information can be generated multiple times.
  However this duplication has no effect, since we handle persistent tokens.
  For instance, the firing sequence $t_d ; t_f ; t_3 ; t_8$ 
  releases two copies of $t_b$ and marks $\co{\bf 5}$ twice. 
  This is inessential for reachability, but has interesting consequences 
  w.r.t. causal dependencies
%  We remark that the number of
%  times that a transition is activated, and hence the order in which
%  it is activated, is inessential. 
  (see Section~\ref{sec:concurrency}).
  %  In Section~\ref{sec:concurrency} we
%  study concurrent computations of persistent nets.
%  in which persistent places associated with dynamic transitions and with negative information.
%  play a central role.
%  for identifying computations that differ on the order in which dynamic transitions are activated.
\end{exa}

%The remaining of this section is devoted to show that the encoding generates 
%confusion-free nets. 
We now show that the encoding generates 
confusion-free nets. We start by stating a useful property of the encoding that ensures
that an
execution cannot generate tokens in both $p$ and $\co{\bf p}$.  

%\begin{lem}[Negative and positive tokens are mutually exclusive]\label{lemma:negpos} 
%  If $\enc N \rightarrow^* (T,b)$ and $p\in b$ then for all $(T',b')$
%  such that $(T,b) \rightarrow^* (T',b')$ we have $\co {\bf p}\not\in b'$.\\
%  Dually, if $\enc N \rightarrow^* (T,b)$ and $\co {\bf p}\in b$ then $(T,b) \rightarrow^* (T',b')$ implies that $p\not\in b'$.
%\end{lem}

\begin{lem}[Negative and positive tokens are in exclusion]\label{lemma:negpos} 
  If $\enc N \rightarrow^* (T,b)$ and $\co {\bf p}\in b$ then $(T,b) \rightarrow^* (T',b')$ implies that $p\not\in b'$.
\end{lem}

%Since $\co {\bf p}$ is persistent, 
%The above lemma ensures that no
%execution can generate tokens in both $p$ and $\co{\bf p}$.  
We now observe from
Def.~\ref{def:BC-in-DN} that for any transition $t\in\enc N \in
\dn{P\cup\co{\bf  P}}$ it holds that either $\preS t \subseteq P$ or $\preS
t\subseteq \co {\bf P}$.
%
%The next result says that a transition $t$ is enabled only when
%any other transition $t'$ with $\preS t
%\subset \preS t'$ has been discarded, i.e., a negative token has been
%generated for some place in $\preS t'$.
%
%\begin{lem}[Nested rules do not collide]
%  Let $\enc N \in \dn{P\cup\co P}$. If $\enc N \rightarrow^* (T,b) \tr
%  t $ and $\preS t \subseteq P$ then for all $t'\in T$ such that
%  $\preS t \subset \preS t'$, it holds that $\co{\preS t' }\cap b \neq
%  \emptyset $.
%\end{lem}
%
The next result says that whenever there exist two transitions $t$ and 
$t'$ that have different but overlapping presets,  
at least one of them is disabled by the presence of a negative token
in the marking $b$.
\begin{lem}[Nested rules do not collide]
\label{lemma:nested-disjoint-presets}
Let $\enc N \in \dn{P\cup\co {\bf P}}$. If $\enc N \rightarrow^* (T,b)$ then for all $t,t'\in T$ s.t.
$\preS t \neq \preS{t'}$ and  $\preS t \cap \preS{t'} \cap P\neq \emptyset$  it holds that 
there is $p\in P \cap (\preS t \cup \preS{t'})$ such that $\co{\bf p} \in b$.
\end{lem}

The main result states that $\enc{\cdot}$ generates confusion-free nets.

\begin{thm}\label{th:confusion-free}
  Let $\enc N \in \dn{P\cup\co{\bf P}}$. If $\enc N \rightarrow^* (T,b) \tr
  t $ and $(T,b) \tr {t'} $ then either $\preS t = \preS {t'}$ or
  $\preS t \cap \preS {t'} = \emptyset$.
\end{thm}

\begin{cor}%[Confusion-free \#1]
Any net $\enc N \in \dn{P\cup\co{\bf P}}$ is confusion-free.
\end{cor}

Finally, we can combine the encoding $\enc \cdot$ with $\dyntopt \cdot$ (from Section~\ref{sec:dynnets}) to obtain a (flat) \OneInfSafe,  confusion-free,  p-net
$\dyntopt {\enc N}$, that we call the \emph{uniformed net} of $N$.
By Proposition~\ref{prop:dynvsflat} we get that the uniformed net  $\dyntopt {\enc N}$ is also confusion-free by construction.

\begin{cor}%[Confusion-free \#2]
Any p-net $\dyntopt {\enc N}$ is confusion-free.
\end{cor}

%%% Local Variables:
%%% mode: latex
%%% TeX-master: "main.tex"
%%% End:

% !TEX root =  main.tex

\section{Static vs Dynamic cell decomposition}\label{sec:cells}

%As explained in the introduction, 
%The detection of the loci of decisions 
%between mutually exclusive alternatives is important, e.g., for
%distributability~\cite{DBLP:journals/corr/GlabbeekGS13} and for trading
%nondeterminism with a probability distribution over the alternatives.
%The case of free-choice nets is the easiest to handle, as all the alternatives are
%enabled at the same time.
%Varacca and Winskel have focused on probabilistic
%confusion-free nets~\cite{DBLP:journals/tcs/VaraccaVW06}.
%% to accommodate for the interplay between
%%concurrency and probability
%Abbes and Benveniste have proposed a more general approach starting
%from event structures and introducing the notion of a \emph{branching
%  cell}~\cite{DBLP:journals/iandc/AbbesB06,DBLP:journals/tcs/AbbesB08}.

As mentioned in the Introduction, Abbes and Benveniste
%~\cite{DBLP:conf/fossacs/AbbesB05,DBLP:journals/iandc/AbbesB06,DBLP:journals/tcs/AbbesB08}
proposed a way to remove confusion by  dynamically decomposing prime event structures. 
In Sections~\ref{sec:eventstruct} and~\ref{sec:ABcells} we recall the basics of the 
AB's approach
% the main notation and terminology needed to
%define prime event structures and branching cells, 
as introduced in~\cite{DBLP:conf/fossacs/AbbesB05,DBLP:journals/iandc/AbbesB06,DBLP:journals/tcs/AbbesB08}.
Then, we show that there is an  operational correspondence between  
AB decomposition and  s-cells introduced in Section~\ref{sec:structural-cells}.

%We start by recalling the basics of the branching cell approach, which 
% relies on the notion of prime event structures.

%We observe that the definition has some drawbacks:
%(i)~it is hardly accessible to non specialists and requires some ingenuity,
%(ii)~it is not defined statically,
%(iii)~it is not defined in a compositional way but over the whole event structure;
%(iv)~it requires a specialized execution strategy different from the usual computation of nets
%and event structures. 
%Moreover, the notion of a branching cell is dynamic in nature and thus a bit elusive for the analysis.
%
%Our first contribution is to recast the notion of branching cell in a
%structural approach, that makes the definition static and
%compositional and, in particular, independent from the past execution
%of the net. This original construction is outlined in
%Section~\ref{sec:structural-cells}.
%

% !TEX root =  main.tex

\subsection{Prime Event Structures}
\label{sec:eventstruct}
A \emph{prime event structure} (also \emph{PES})
~\cite{DBLP:journals/tcs/NielsenPW81,DBLP:conf/ac/Winskel86} is a triple
$\mathcal{E}=(E,\preceq,\#)$ where: $E$ is the set of \emph{events};
the \emph{causality
  relation} $\preceq$ is a partial order on events; the \emph{conflict relation} $\#$ is a symmetric, irreflexive relation on events
 such that conflicts are inherited by
causality, i.e., $\forall e_1,e_2,e_3\in E.~ e_1\# e_2 \preceq e_3
\Rightarrow e_1 \# e_3$. 

 The %construction of the 
 PES
$\mathcal{E}_N$ associated with a net
$N$ can be formalised using category theory as a chain of universal
constructions, called coreflections.
%, with the category of prime event structures be equivalent to the category of prime algebraic domains. 
Hence, for each PES
$\mathcal{E}$, there is a standard, unique (up to isomorphism) nondeterministic occurrence net $N_{\mathcal{E}}$
that yields $\mathcal{E}$ and thus we can freely move from one setting to
the other.

%\begin{exa}
  Consider the nets in Figs.~\ref{fig:condconfusion} and~\ref{fig:multcondconfusion}. The
  corresponding PESs  are shown
  below each net. 
  Events are in bijective correspondence with the transitions of the nets.  
  Strict causality is depicted by arrows and immediate conflict by curly lines.  
%\qed
%\end{exa}

%The notion of confusion can be extended to event structures.
Given an event $e$, its \emph{downward closure} $\down{e} = \{ e'\in E \mid e'\preceq e\}$ is the set of causes of $e$.
As usual, we assume that $\down{e}$ is finite for any $e$.
Given $B\subseteq E$, we say that $B$ is \emph{downward closed} if
$\forall e\in B.~ \down{e} \subseteq B$ and that $B$ is \emph{conflict-free} if $\forall e,e'\in B.~ \neg (e \# e')$.
We let the \emph{immediate conflict} relation $\#_0$ be
defined on events by letting $e \#_0 e'$ iff $(\down{e} \times \down{e'}) \cap
\# = \{(e,e')\}$, i.e., two events are in immediate conflict if they
are in conflict but their causes are compatible.
%%
%Then, an event structure is \emph{confusion free} if its maximal set of events that are pairwise in immediate conflict and that have the same set of causal predecessors are closed under immediate conflict.

%%% Local Variables:
%%% mode: latex
%%% TeX-master: "main.tex"
%%% End:

\subsection{Abbes and Benveniste's Branching Cells}\label{sec:ABcells}

%In order to define AB's branching cells, some terminology must be introduced first.
%We start by introducing some terminology. 
In the following we assume that a (finite) PES $\mathcal{E}=(E,\preceq,\#)$ is given.
A \emph{prefix} $B\subseteq E$ is any downward-closed set of events
(possibly with conflicts). Any prefix $B$  induces an event structure $\mathcal{E}_B
=(B, \preceq_B, \#_B)$ where $\preceq_B$ and $\#_B$ are
the restrictions of $\preceq$ and $\#$ to the events in $B$.
%
%It is immediate to check that any (downward-closed) subset $B\subseteq
%E$ implicitly defines a \mbox{(sub-)}event structure
%$B_\mathcal{E}=(B,\preceq_B,\#_B)$, where $\preceq_B$ and $\#_B$ are
%the obvious restrictions of $\preceq$ and $\#$ to events in $B$.
%%, i.e., $\preceq_B = \preceq \cap (B \times B)$ and $\#_B = \# \cap (B \times B)$. 
%This fact is important, because the definition of the branching
%cells that are active at some point of the computation takes into
%account the sub-event structure without all the events that are in
%conflict with the events executed so far.
%
A \emph{stopping prefix} is a prefix $B$ that is closed under
immediate conflicts, i.e., $\forall e\in B, e'\in E.~ e\#_0 e'
\Rightarrow e'\in B$.  Intuitively, a stopping prefix is a prefix
whose (immediate) choices are all available.  
It is \emph{initial} if the only stopping prefix strictly
included in $B$ is $\emptyset$.  We assume that any $e\in
E$ is contained in a finite stopping prefix.
% (\emph{local finiteness}).

A \emph{configuration} $v \subseteq \mathcal{E}$ is any set of events
that is downward closed and conflict-free.  Intuitively, a
configuration represents (the state reached after executing) a
concurrent but deterministic computation of $\mathcal{E}$.
Configurations are ordered by inclusion and we denote by
$\mathcal{V}_{\mathcal{E}}$ the poset of
finite configurations of $\mathcal{E}$ and by $\Omega_{\mathcal{E}}$ the poset of maximal
configurations of $\mathcal{E}$.  
%For $B \subseteq E$ we write 
%$\mathcal{V}_{B}$ and $\Omega_{B}$ for  
%$\mathcal{V}_{B_\mathcal{E}}$ and $\Omega_{B_\mathcal{E}}$, respectively.

%Suppose the configuration $v$ has been executed.  
The \emph{future} of a configuration $v$, written $E^v$, is the set of events that can be executed after $v$, i.e.,
$E^v = \{e\in E\setminus v\ \mid\ \forall e'\in v. \neg(e \# e')\}$.
We write $\mathcal{E}^v$ for the event structure induced by $E^v$.  
We assume that any finite configuration enables only
finitely many events, i.e.,  the set of minimal elements in $E^v$
w.r.t. $\preceq$ is finite  for any $v\in
\mathcal{V}_{\mathcal{E}}$.

A configuration $v$ is \emph{stopped } if there is a stopping prefix $B$ with $v\in \Omega_B$.
%
%A configuration
and $v$ is \emph{recursively stopped} if there is a finite sequence of
configurations $\emptyset= v_0 \subset \ldots \subset v_n=v$ such that for any $i\in [0,n)$ the set $v_{i+1} \setminus v_i$ is a
  finite stopped configuration of  $\mathcal{E}^{v_i}$ for
  $v_i$ in $\mathcal{E}$.
%
%We denote by $\mathcal{W}_{\mathcal{E}}$ the set of all finite,
%recursively stopped configurations.  In other terms,
%$\mathcal{W}_{\mathcal{E}}$ is the smallest class of configurations
%that contains all stopped configurations and that is closed under
%concatenation~\cite{DBLP:conf/fossacs/AbbesB05}.

A \emph{branching cell} is any initial stopping prefix of the future
$\mathcal{E}^v$ of a finite recursively stopped configuration $v$.
%\in\mathcal{W}_{\mathcal{E}}$.  
Intuitively, a branching cell is a
minimal subset of events closed under immediate conflict.  
%We denote by $\mathcal{X}_{\mathcal{E}}$ the set of all branching cells.
%
We remark that branching cells are determined by
considering the whole (future of the) event structure $\mathcal{E}$ and
they are recursively computed as $\mathcal{E}$ is executed. Remarkably, 
every maximal configuration has a branching cell decomposition.

\begin{exa}
  \label{ex:bc-decomposition}

  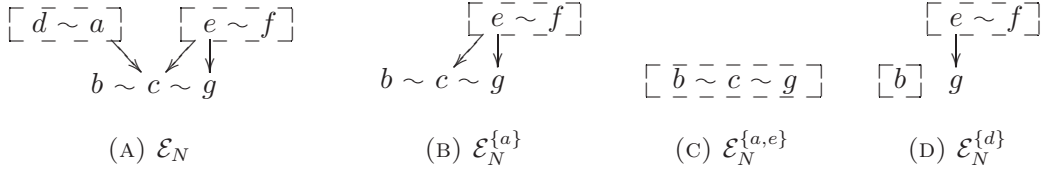
\begin{figure*}[t]
    \begin{subfigure}[b]{0.30\textwidth} 
      $$
      \xymatrix@R=.6pc@C=.8pc{
        d\ar@{~}[r]
        &
        a\ar[dr]
        \POS[]-<1pc,0pc> *+=<3.5pc,1pc>[F--]{}
        &
        &
        e\ar@{~}[r]\ar[dl]\ar[d]
        &
        f
        \POS[]-<1pc,0pc> *+=<3.5pc,1pc>[F--]{}
        \\
        &
        b\ar@{~}[r]
        &
        c\ar@{~}[r]
        &
        g      }
      $$
%\vspace{-10pt}
      \subcaption{$\mathcal{E}_N$}\label{fig:sp-1}
    \end{subfigure}
    \hspace{.1cm}
    \begin{subfigure}[b]{0.22\textwidth} 
      $$
      \xymatrix@R=.6pc@C=.8pc{
        &
        &
        e\ar@{~}[r]\ar[dl]\ar[d]
        &
        f
        \POS[]-<1pc,0pc> *+=<3.5pc,1pc>[F--]{}
        \\
        b\ar@{~}[r]
        &
        c\ar@{~}[r]
        &
        g      }
      $$
%\vspace{-10pt}
      \subcaption{$\mathcal{E}_N^{\{a\}}$}\label{fig:sp-2}
    \end{subfigure}
    \hspace{.2cm}
    \begin{subfigure}[b]{0.17\textwidth} 
      $$
      \xymatrix@R=.6pc@C=.8pc{
        \\
        b\ar@{~}[r]
        &
        c\ar@{~}[r]
        \POS[]-<0pc,0pc> *+=<5.5pc,1pc>[F--]{}
        &
        g      }
      $$
%\vspace{-10pt}
      \subcaption{$\mathcal{E}_N^{\{a,e\}}$}\label{fig:sp-3}
    \end{subfigure}
    \hspace{.3cm}
    \begin{subfigure}[b]{0.15\textwidth} 
      $$
      \xymatrix@R=.6pc@C=.8pc{
        &
        e\ar@{~}[r]\ar[d]
        &
        f
        \POS[]-<1pc,0pc> *+=<3.5pc,1pc>[F--]{}
        \\
        b
        \POS[]-<0pc,0pc> *+=<1.3pc,1pc>[F--]{}
        &
        g
        %\POS[]-<0pc,0pc> *+=<1.3pc,1.7pc>[F--]{}
      }
      $$
%\vspace{-10pt}
      \subcaption{$\mathcal{E}_N^{\{d\}}$}\label{fig:sp-4}
    \end{subfigure}
    
    \caption{AB's branching cell decomposition (running example)}\label{fig:sp-all}
  \end{figure*}

  Consider the PES $\mathcal{E}_N$ in
  Fig.~\ref{fig:multcondconfusion} and its maximal configuration $v =
  \{a,e,b,g\}$. We show that $v$ is recursively stopped by exhibiting a branching cell decomposition. The initial
  stopping prefixes of $\mathcal{E}_N = \mathcal{E}_N^\emptyset$ are
  shown in Fig.~\ref{fig:sp-1}. There are two possibilities for
  choosing $v_1\subseteq v$ and $v_1$ recursively stopped: either $v_1 =
  \{a\}$ or $v_1=\{e\}$. When $v_1 = \{a\}$, the choices for $v_2$ are
  determined by the stopping prefixes of $\mathcal{E}_N^{\{a\}}$
  (see Fig.~\ref{fig:sp-2}) and the only
  possibility is $v_2 = \{a, e\}$. From $\mathcal{E}_N^{\{a,e\}}$
  in Fig.~\ref{fig:sp-3}, we take $v_3 =
  v$. Note that  $\{a,e,b\}$ is not recursively
  stopped because $\{b\}$ is not maximal
  in the stopping prefix of $\mathcal{E}_N^{\{a,e\}}$ (see Fig.~\ref{fig:sp-3}).
Finally, note that the branching cells $\mathcal{E}_N^{\{a\}}$ (Fig.~\ref{fig:sp-2}) and $\mathcal{E}_N^{\{d\}}$ (Fig.~\ref{fig:sp-4}) 
correspond to different choices in $\mathcal{E}_N^\emptyset$ and thus have different stopping prefixes.
%
%  The choices made in a branching cell influence the branching cells available next.
%%  
%%  Stopping prefixes and branching cells change dynamically as a
%%  consequence of the choices made in other branching cells. 
%%  
%  Take $v' =
%  \{d,e,b,g\}$ and start its decomposition from
%%   with the stopping prefixes of 
%  $\mathcal{E}_N = \mathcal{E}_N^\emptyset$ (Fig.~\ref{fig:sp-1}). 
%  Fix $v_1' = \{d\}$, and get
%  $\mathcal{E}_N^{\{d\}}$ as in Fig.~\ref{fig:sp-4}. 
%  Observe that $\mathcal{E}_N^{\{a\}}$(Fig.~\ref{fig:sp-2}) and
%  $\mathcal{E}_N^{\{d\}}$(Fig.~\ref{fig:sp-4}) have different stopping prefixes  
%  because reached after different choices in $\mathcal{E}_N^\emptyset$ ($v_1=\{a\}$
%  and $v_1'=\{d\}$).
\end{exa}

\subsection{Relating s-cells and AB's decomposition}

%We now establish the correspondence
%between  s-cell decomposition of  a  net $N$ and  the
%recursively stopped configurations of (the event structure of) $N$
%(i.e., of $\mathcal{E}_N$). 
%
The recursively stopped configurations of a net $N$ characterise 
all the allowed executions of $N$. Hence, we formally link the recursively stopped configurations of $\mathcal{E}_N$ 
with the computations of the uniformed net $\dyntopt {\enc N}$.
For technical convenience, we first show that the recursively stopped configurations of   $\mathcal{E}_N$ are in 
one-to-one correspondence with the computations of  the dynamic net 
$\enc N$.
%, which is more convenient to handle in the proofs than $\dyntopt {\enc N}$. 
Then, the desired correspondence is obtained by using  Proposition~\ref{prop:dynvsflat} to 
relate the computations of a dynamic net and its associated p-net.

We rely on the  auxiliary map  $\dec {-}$ that  links transitions in $\enc N$ with events in $\mathcal{E}_N$.  
Specifically, 
 $\dec {-}$  associates each transition $t$ of $\enc N$
with the set $\dec t$ of transitions of $N$ (also events in $\mathcal{E}_N$) that are encoded by $t$. Formally,
%
%Given a transaction $\theta$, we write $\events \theta$ for the set of
%transitions of $\theta$.   Formally,
%$\dec t  = \events \theta$ if $t = t_{\theta,\bc} \in \Tpos$ for some s-cell $\bc$ and transaction $\theta:\bc$.
%Otherwise, $\dec t  =  \emptyset$ if $t \in \Tneg$.
%
\[
\dec t  = 
\begin{cases}
  \events \theta & \mbox{if } t = t_{\theta,\bc} \in \Tpos %\quad \mbox{where}\ \events \theta\ \mbox{is the set of transitions in }\ \theta
  %  (\minp \bc, (\emptyset, \maxp\theta \cup \co{\maxp \bc\setminus\maxp \theta}))
  \\
  \emptyset & \mbox{if } t \in \Tneg
\end{cases}
\]
where $\events \theta$ is the set of transitions in $\theta$. 

%If $t\in \Tpos$, i.e., $t$ is defined in terms of a transaction $\theta$, then 
%$\dec t$ denotes the set $\events \theta$ of transitions of $\theta$. On the contrary, 
%$\dec t = \emptyset$ when  $t \in \Tneg$ because $t$ does not 
%encode any transition  of the original net $N$.

\begin{exa}
  Consider the net $N$ in Fig.~\ref{fig:net-and-bc} which is encoded as the dynamic p-net 
  in Fig.~\ref{fig:enc-bc-into-dyn}. The auxiliary mapping $\dec{\_}$ is  as follows
  \[
  \begin{array}{l@{\qquad}l@{\qquad}l@{\qquad}l}
  \dec{t_a} = \{a\} 
& 
  \dec{t_d} = \{d\} 
&
  \dec{t_e} = \{e\} 
&
  \dec{t_f} = \{f\} 
  \\ 
  \dec{t_{bg}} = \{b, g\} 
&
  \dec{t_c} = \{c\} 
&
  \dec{t_b} = \{b\} 
&
  \dec{t_g} = \{g\}   
\\
\multicolumn{4}{l}{\dec{t} = \emptyset \ \mbox{if}\ t\in\{t_1, t_7, t_2, t_3, t_8, t'_2, t'_8\}}
  \end{array}
  \]
   A transition $t_{\theta,\bc}$ of $\dyntopt N$ associated with a transaction $\theta:\bc$ of $N$ is mapped to the  
   transitions of $\theta$. For instance,   $t_a$ is mapped to $\{a\}$, which is 
   the only transition in $\theta_a$. Differently, transitions that propagate negative information, i.e., 
   $t\in\{t_1, t_7, t_2, t_3, t_8, t'_2, t'_8\}$, are mapped to $\emptyset$ because they
    do not  encode any transition of $N$. 
\end{exa}

In what follows we write $M \Tr {} M'$ for a possibly empty firing sequence $M \tr {t_1\cdots t_n} M'$ such that 
$\dec {t_i} = \emptyset$ for all $i\in[1,n]$.  If $\dec {t} \neq \emptyset$, we
write $M \Tr {t} M'$ if $M \Tr {} M_0\Tr{t} M_1 \Tr{} M'$ for some $M_0,M_1$.
Moreover, we write $M \Tr {t_1\cdots t_n}$ if there exist $M_1,...,M_n$ such that
$M \Tr {t_1} M_1 \Tr {t_2} \cdots \Tr {t_n} M_n$.

The following result states that the computations of any dynamic p-net produced by  
$\enc{\_}$ are in
one-to-one correspondence with the recursively stopped configurations
of Abbes and Benveniste.

\begin{lem}%[Correspondence between $\enc{\_}$ and AB]
\label{theo:AB}
 Let $N$ be an occurrence net.  
  \begin{enumerate}
    
  \item
    If $\enc N \Tr {t_1\cdots t_n}$, then $v = \bigcup_{1\leq
      i\leq n} \dec {t_i}$ is recursively stopped in $\mathcal{E}_N$
    and $(\dec {t_i})_{1\leq i\leq n}$ is a valid decomposition of
    $v$.
  \item
    If $v$ is recursively stopped in $\mathcal{E}_N$, then for any
    valid decomposition $(v_i)_{1\leq i\leq n}$ there exists $\enc N
    \Tr {t_1 \cdots t_n}$ such that $\dec {t_i} = v_i$.
  \end{enumerate}
\end{lem}

\begin{exa}
  Consider the branching cell decomposition for  $v = \{a,e,b,g\} \in\mathcal{E}_v$ discussed in Ex.~\ref{ex:bc-decomposition}.
 Then, the  net $\enc N$ in Ex.~\ref{ex:enc-bc-into-dyn} can mimic that decomposition with the following computation
 $$
 \begin{array}{l}
 (T,\{1,2,7\}) \tr{t_a} 
 (T,\{2,3,7,\co{\bf 6}\}) \tr{t_e} 
 (T,\{2,3,8,\co{\bf 6}, \co{\bf 9}\})  %\qquad\qquad
% \\
% \hfill
 \tr{t_{bg}}
 (T,\{4,10,\co{\bf 5},\co{\bf 6}, \co{\bf 9}\})
 \end{array}
 $$
  with
$v_1 = \dec{t_a}=\{a\}$, 
$v_2 = \dec{t_e}=\{e\}$, and 
$v_3 =  \dec{t_{bg}}=\{b, g\}$. 
\end{exa}

%Theorem: enabled transitions of the dynamic nets correspond to branching cells

%The correspondence between the uniformed net and the recursively stopped configurations
%associated with a net $N$ is obtained by combining the previous result with Proposition~\ref{prop:dynvsflat}.
%

%By combining the previous result with Proposition~\ref{prop:dynvsflat} we can obtain the following result.

From Lemma~\ref{theo:AB} and Proposition~\ref{prop:dynvsflat} we obtain the next result.

\begin{thm}[Correspondence]% [Correspondence between $\dyntopt{\enc{\_}}$ and AB]
\label{theo:corr-ab}
 Let $N$ be an occurrence net. 
   \begin{enumerate}
    
  \item
    If $\dyntopt {\enc N} \Tr {t_1\cdots t_n}$, then $v = \bigcup_{1\leq
      i\leq n} \dec {t_i}$ is recursively stopped in $\mathcal{E}_N$
    and $(\dec {t_i})_{1\leq i\leq n}$ is a valid decomposition of
    $v$.
  \item
    If $v$ is recursively stopped in $\mathcal{E}_N$, then for any
    valid decomposition $(v_i)_{1\leq i\leq n}$ there exists $\dyntopt {\enc N}
    \Tr {t_1 \cdots t_n}$ such that $\dec {t_i} = v_i$.
  \end{enumerate}
% For every maximal configuration $v$ of $N$ 
% there exists a firing sequence $\dyntopt {\enc N} \Tr {t_0 \cdots t_n}$ such that $v = \bigcup_{1\leq
%      i\leq n} \dec {t_i}$.
\end{thm}

By (1) above,  any computation of 
$\dyntopt {\enc N}$ corresponds to a (recursively stopped) 
configuration of $\mathcal{E}_N$, i.e., a process  
of $N$. By  
(2),  every  execution of $N$ that can be decomposed in terms of AB's branching cells is preserved
 by $\dyntopt {\enc N}$, because any recursively stopped configuration of $\mathcal{E}_N$ is 
mimicked by $\dyntopt {\enc N}$. 

%Theorem: probabilities distributions by Varacca et al. on the confusion free net coincide with AB’s on the given net.

%%% Local Variables:
%%% mode: latex
%%% TeX-master: "main.tex"
%%% End:

% !TEX root =  main.tex

\section{%Assessing the 
Concurrency of the Uniformed Net}
\label{sec:concurrency}

%In the previous sections we saw how, given a net, possibly with confusion, we can derive a corresponding uniformed, confusion free net. It is easy to see that every firing sequence of the uniformed net corresponds to a firing sequence of the original net (Theorem/Corollary~\ref{}).  The converse does not hold in general: certain firing sequences, legal in the original configuration, are forbidden, namely the new notion of concurrency compatible with lack of confusion is more restrictive.

In this section we study the amount of concurrency still present in the uniformed net $\dyntopt {\enc N}$. 
%
%Concurrency in the original net $N$ is well understood: a deterministic process or an event structure configuration define precisely a concurrent computation corresponding to an equivalence class of firing sequences. 
%However, in the presence of confusion, certain firing sequences that are legal in the configuration should not be executable.
%
%, should be forbidden.
%Instead, a corresponding notion, we believe, is not fully understood for $1-\infty$ safe nets, as uniformed nets actually are.
%
Here, we extend the notion of a process to the case of \OneInfSafe  p-nets and we show that all the \emph{legal} firing sequences
of a process of the uniformed net $\dyntopt {\enc N}$ are executable.
%characterize the corresponding equivalence classes of firing sequences. 
%Finally, we instantiate the resulting notion of concurrency to uniformed nets.

%\subsection{Deterministic Nonsequential Processes of a \OneInfSafe Safe Net}

%For an ordinary Petri net $N$,  a process is a deterministic occurrence net together with a mapping to $N$.
%, where occurrence nets have no conflicts and the preset of every place contains at most a transition. 
%
%For \OneInfSafe nets, a process is still an occurrence net with a mapping,
%: the occurrence net,  equipped with causal and concurrency relations, represents a single deterministic computation as an equivalence class of firing sequences. 
The notion of deterministic occurrence net is extended to p-nets by slightly changing the definitions of conflict and causal dependency: (i)~two transitions  are {\em not} in conflict when all shared places are persistent, (ii)~a persistent place can have more than one immediate cause in its preset, which  introduces OR-dependencies.

\begin{defi}[Persistent process]
 An {\em occurrence p-net} $O=(P\cup \mathbf{P},T,F)$ is an acyclic p-net 
such that $|\postS{p}| \leq 1$ and $|\preS{p}| \leq 1$ for any $p\in P$ (but not necessarily for those in $\mathbf{P}$).

A {\em persistent process} for $N$ is  an occurrence p-net $O$ together with a net morphism $\pi: O \rightarrow N$ that preserves presets and postsets  and the distinction between regular and persistent places.
Without loss of generality, when $N$ is acyclic, we assume that $O$ is  a subnet of $N$ (with the same initial marking) and  $\pi$ is the identity.
\end{defi}

In an ordinary occurrence net, the causes of an item $x$ are all its predecessors.
% $y \prec x$ in the partial order relation $\prec$. 
%In p-nets, we want to model alternative sets of causes
%% for transitions and places 
%and thus the causes of an item $x$ are represented as a formula $\Phi(x)$ of the propositional calculus without negation, where the basic propositions are the transitions of the occurrence net. If we represent such a formula as a sum of products, it corresponds to a set of {\em collections}, i.e. a set of sets of transitions. 
In p-nets, the alternative sets of causes of an item $x$ are given by a formula $\Phi(x)$ of the propositional calculus without negation, where the basic propositions are the transitions of the occurrence net.
If we represent such a formula as a sum of products, it corresponds to a set of {\em collections}, i.e. a set of sets of transitions. 
Different collections correspond to alternative causal dependencies, while transitions within a collection are all the causes of that alternative
and $true$ represents the empty collection.
%Notice that the positive sum of products representation is not unique: for instance dominated (i.e. larger as sets) collections can be omitted.
Such a formula $\Phi(x)$ represents a monotone boolean function, which expresses, as a function of the occurrences of past transitions, if $x$ has enough causes. It is known that such formulas, based on positive literals only, have a unique DNF (sum of products) form, given by the set of prime implicants. In fact, every prime implicant is also essential~\cite{wegener1987complexity}.
%
%We define $\Phi(x)$ as in the ordinary case, in terms of the predecessors $y$ of $x$, written $y \prec x$, with $\prec=F^+$.  
%Alternatively, one could record in the formula just the immediate causes, obtaining a smaller formula $\Psi(x)$.
We define $\Phi(x)$
% and $\Psi(x)$ 
by well-founded recursion:
%, letting $\mathbb{S} = P \cup \mathbf{P}$.
%
\[
%\begin{array}{lr}
 \Phi(x) = 
  \left\{
  \begin{array}{ll}
  \mathit{true} & \mbox{if $x\in P \cup \mathbf{P} \wedge \preS{x}= \emptyset$} \\
  \bigvee_{t\in\preS{x}} (t \wedge \Phi(t)) & \mbox{if $x\in P \cup \mathbf{P} \wedge \preS{x}\neq \emptyset$} \\
  \bigwedge_{s\in\preS{x}} \Phi(s)  & \mbox{if $x\in T$}
  \end{array}
  \right.
%&
% \Psi(x) = 
%  \left\{
%  \begin{array}{ll}
%  \mathit{true} & \mbox{if $x\in \mathbb{S} \wedge \preS{x}= \emptyset$} \\
%  \bigvee_{t\in\preS{x}} t & \mbox{if $x\in \mathbb{S} \wedge \preS{x}\neq \emptyset$} \\
%  \bigwedge_{s\in\preS{x}} \Psi(s)  & \mbox{if $x\in T$}
%  \end{array}
%  \right.
%\end{array}
\]

The boolean formulas above remind the notion of causal automata~\cite{DBLP:journals/tcs/Gunawardena92}, where a set of (labelled) events $E$ is accompanied by an enabling function that assigns a formula in the free Boolean algebra generated by $E$ to each event in $E$. Here note that all formulas are determined by the structure of the p-net and that they are expressed using only true, AND and OR, i.e. they would define a $\{\wedge,\vee\}$-automata according to the terminology in~\cite{DBLP:journals/tcs/Gunawardena92}. 
 
Ordinary deterministic processes satisfy {\em complete concurrency}: each process determines a partial ordering of its transitions, such that the executable sequences of transitions are exactly the linearizations of the partial order. More formally, after executing any firing sequence $\sigma$ of the process, a transition $t$ is enabled if and only if all its predecessors in the partial order (namely its causes) already appear in $\sigma$.
In the present setting a similar property holds.

\begin{defi}[Legal firing sequence]
A sequence  of transitions $t_1 ; \cdots ; t_n$ of a persistent process is \emph{legal}  if for all $k\in [1,n]$ we have that $\bigwedge_{i=1}^{k-1} t_i$ implies $\Phi(t_k)$.
\end{defi}

It is immediate to notice that if the set of persistent places is empty ($\mathbf{P} = \emptyset$) then the notion of persistent process is the ordinary one, $\Phi(x)$ is just the conjunction of the causes of $x$ and a sequence is legal iff it is a linearization of the process.
%In this sense, persistent processes are a conservative extension of ordinary ones.

\begin{thm}
[Complete Concurrency]\label{theo:concur}
Let $\sigma= t_1 ; \cdots ; t_n$ with $n\geq 0$ be a, possibly empty, firing sequence of a persistent process, and $t$ a transition not in $\sigma$.
The following conditions are all equivalent:
(i)~%the transition 
$t$ is enabled after $\sigma$;
(ii)~there is a collection of causes of $t$ which appears in $\sigma$; 
(iii)~$\bigwedge_{i=1}^n t_i$ implies $\Phi(t)$.
%(iv)~$\bigwedge_{i=1}^n t_i$ implies $\Psi(t)$.
\end{thm}

%Notice the analogy with concurrent constraints: assuming the memory constraint $\alpha$ we could write $\mathit{ask}\ \Phi(t): t$ to enable the execution of $t$.

\begin{cor}\label{cor:adm}
Given a persistent  process, a sequence is legal iff it is a firing sequence.
\end{cor}

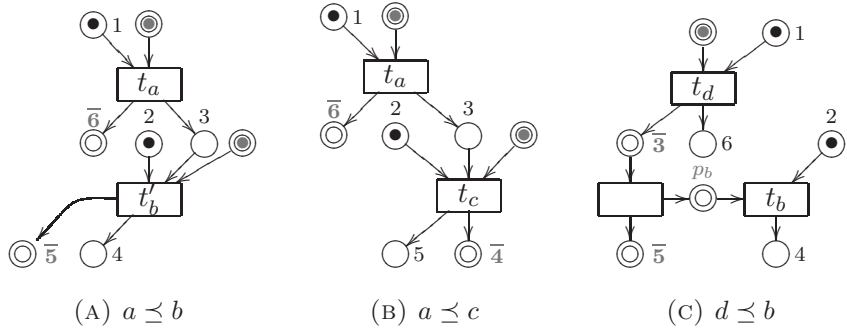
\begin{figure}[t]
\begin{subfigure}[b]{0.25\textwidth} 
$$
 \xymatrix@R=.8pc@C=.3pc{
 &&
 \drawmarkedplace \ar[rd]
 \nameplaceright 1
 &
 \drawmarkedpersistentplace \ar@[mygray][d]
 \\
 {\phantom{t_d}} &&& \drawtrans {t_a}\ar@[mygray][dl]\ar[rd]
 \\
 &
 &
 \drawpersistentplace 
  \namepersup 6
 &
 \drawmarkedplace \ar[d]
 \nameplaceup 2
 &
 \drawplace \ar[ld]
 \nameplaceup 3
 &
 \drawmarkedpersistentplace \ar@[mygray][lld]
 \\
 &
 & 
 & \drawtrans {t'_b}\ar@[mygray]@(l,ur)[llld]\ar[ld]
 & 
 \\
 \drawpersistentplace
 \namepersright 5 &&
 \drawplace
 \nameplaceright 4
 &
 &
 }
$$
%\vspace{-10pt}
\subcaption{$a\preceq b$}\label{fig:tasteproc1}
\end{subfigure}
\begin{subfigure}[b]{0.25\textwidth} 
$$
\xymatrix@R=.8pc@C=.3pc{
 \drawmarkedplace \ar[rd]
 \nameplaceright 1
 &
 \drawmarkedpersistentplace \ar@[mygray][d]
 \\
  {\phantom{t_d}} & \drawtrans {t_a}\ar@[mygray][dl]\ar[rd]
 \\
 \drawpersistentplace 
 \namepersup 6
 &
 \drawmarkedplace \ar[rd]
 \nameplaceup 2
 &
 \drawplace \ar[d]
 \nameplaceup 3
 &
 \drawmarkedpersistentplace \ar@[mygray][ld]
 \\
  {\phantom{t_b}} && 
 \drawtrans {t_c}\ar@[mygray][d]\ar[ld]
 \\
 &
 \drawplace
 \nameplaceright 5
 &
 \drawpersistentplace
 \namepersright 4
 }
$$
%\vspace{-10pt}
\subcaption{ $a\preceq c$}\label{fig:tasteproc2}
\end{subfigure}
\begin{subfigure}[b]{0.25\textwidth} 
$$
 \xymatrix@R=.8pc@C=.3pc{
 &
 \drawmarkedpersistentplace \ar@[mygray][d]
 &
 \drawmarkedplace \ar[ld]
 \nameplaceright 1
 \\
 &
 \drawtrans {t_d}\ar@[mygray][dl]\ar[d]
 \\
 \drawpersistentplace \ar@[mygray][d]
 \namepersright 3
 &
 \drawplace 
 \nameplaceright 6
 &
 &
 \drawmarkedplace \ar[ld]
 \nameplaceup 2
 \\
 \drawtransu {}\ar@[mygray][d]\ar@[mygray][r] 
 &
 \drawpersistentplace \ar@[mygray][r]
 \nameplaceup{\color{mygray}p_b}
 & \drawtrans {t_b}\ar[d]
 \\
 \drawpersistentplace
 \namepersright 5
 &&
 \drawplace
 \nameplaceright 4
 }
$$
%\vspace{-10pt}
\subcaption{ $d\preceq b$}\label{fig:tasteproc3}
\end{subfigure}
%\vspace{-5pt}
\caption{Processes for the net in Fig.~\ref{fig:taste3}}\label{fig:tasteproc}
\end{figure}

\begin{exa}
Figs.~\ref{fig:tasteproc1}--\ref{fig:tasteproc3} show the  maximal processes of the net in Fig.~\ref{fig:taste3}.
%that represent the possible executions of the net.
%We note that, as expected, no concurrent execution are allowed, because the choice between $a$ and $d$ drives the decision in the other cells. 
It is evident that all executions are serialized.
\end{exa}

%
%\begin{figure}
%$$
% \xymatrix@R=.7pc@C=.1pc{
%&
%&
%\drawmarkedpersistentplace\ar@[mygray][d]
% \namepersupD{p_{t_{d}}}
%& 
%\drawmarkedplace\ar[dl]
%\nameplaceup 1
% &
% &
% &
% &
% &
% \drawmarkedplace\ar[dr]
% \nameplaceup 7
% &\drawmarkedpersistentplace\ar@[mygray][d]
% \namepersupD{p_{t_{f}}}
% \\
%&
% & 
% \drawtrans {t_d}\ar[dl]\ar@[mygray][dr] 
% &
% &
%  &
%  &
%  &
%  &
%  &
%   \drawtrans {t_f}\ar@[mygray][dl]\ar[dr] 
% \\
%&
% \drawplace
% \nameplaceup 6
% &
% &
%\drawpersistentplace\ar@[mygray][d] 
% \namepersup{ 3}
% &
% &
% &
% &
% &
%\drawpersistentplace\ar@[mygray][d]
% \namepersup{ 8}
% &
% &
% \drawplace
% \nameplaceright 9
%  \\
% &&&
% \drawtransu {\color{mygray}t_4}\ar@[mygray]@/_2ex/[ddddlll] \ar@[mygray]@/^/[ddrrrrr]\ar@[mygray]@/^2ex/[ddrrrrrrr]\ar@[mygray][dd]
% &&&&&
% \drawtransu {\color{mygray}t_5}\ar@[mygray][dd]\ar@[mygray]@/^5ex/[ddddrrrrr] \ar@[mygray]@/_5ex/[ddddllllllll] 
% \\
% &&&&&&&&&&
%   \\
% &
%&
%&
%  \drawpersistentplace 
% \namepersrightD{p_{t_g}\ }
% &
% &
%&
% \drawmarkedplace\ar[drr]
% \nameplaceright {2}
% &
% &
%  \drawpersistentplace\ar@[mygray][d] 
% \namepersrightD{p_{t_b}}
% &&
% \drawpersistentplace
% \namepersupD{p_{t_{\co g}}}
%  \\
% &
% &
% &
% &
% &
% &
%  &
%  &
% \drawtrans {t_b}\ar[d]
%  &
% &
% &&&&
% \\
% \drawpersistentplace
% \namepersup{5}
% &
% &
% &
%  &
% &
% &
%&
%&
% \drawplace
% \nameplaceright {4}
% &
% &
% &&&
%  \drawpersistentplace
% \namepersright{10}
%  }
%$$
%\caption{A persistent deterministic process for the net $\dyntopt {\enc N}$}
%\label{fig:determ-persistent-process}
%\end{figure}

\begin{exa}  
% !TEX root =  main.tex

\begin{figure}[t]
%%% LR
% \begin{subfigure}[b]{0.45\textwidth}
$$
 \xymatrix@R=1pc@C=.5pc{
&
\drawmarkedpersistentplace\ar@[mygray][d]
 \namepersleftm{p_{t_{d}}}
& 
\drawmarkedplace\ar[dl]
\nameplaceright 1
 &
 &
 &
 &
 &
 \drawmarkedplace\ar[dr]
 \nameplaceleft 7
 &\drawmarkedpersistentplace\ar@[mygray][d]
 \namepersrightm{p_{t_{f}}}
 \\
 & 
 \drawtrans {t_d}\ar[dl]\ar@[mygray][dr] 
 &
 &
  &
  &
  &
  &
  &
   \drawtrans {t_f}\ar@[mygray][dl]\ar[dr] 
 \\
 \drawplace
 \nameplaceup 6
 &
 &
\drawpersistentplace\ar@[mygray][d] 
 \namepersup{ 3}
 &
 &
 &
 &
 &
\drawpersistentplace\ar@[mygray][d]\ar@[mygray]@/^5ex/[dddrrrrrr] 
 \namepersleft{ 8}
 &
 &
 \drawplace
 \nameplaceright 9
  \\
&&
 \drawtransu {\color{mygray}t_3}\ar@[mygray]@/_2ex/[dddll] \ar@[mygray]@/^3ex/[ddrrrrr]\ar@[mygray]@/^1ex/[drrrrrrr]\ar@[mygray][d]
 &&&&&
 \drawtransu {\color{mygray}t_8}\ar@[mygray][d]\ar@[mygray]@/^5ex/[dddrrrrr] \ar@[mygray]@/_5.7ex/[dddlllllll]
\\
&
&
  \drawpersistentplace 
 \namepersrightD{p_{t_g}\ }
 &
 &
&
 \drawmarkedplace\ar[drr]
 \nameplaceright {2}
 &
 &
  \drawpersistentplace\ar@[mygray][d] 
 \namepersrightD{p_{t_b}}
 &&
 \drawpersistentplace\ar@[mygray]@/^2ex/[drrrr]
 \namepersdown{p_{t'_8}}  
  \\
 &
 &
 &
 &
 &
  &
  &
 \drawtrans {t_b}\ar[d]
  &
 &
 &&
 &&
  \drawtransu {\color{mygray}t'_{8}}\ar@[mygray][dl]
 \\
 \drawpersistentplace
 \namepersright{5}
 &
 &
  &
 &
 &
&
&
 \drawplace
 \nameplaceright {4}
 &
 &
 &&&
  \drawpersistentplace
 \namepersright{10}
  }
$$
%\subcaption{}\label{fig:process2}
%\end{subfigure}
%$$
% \xymatrix@R=.7pc@C=.01pc{
%&
%\drawmarkedpersistentplace\ar@[mygray][d]
% \namepersupD{p_{t_{d}}}
%& 
%\drawmarkedplace\ar[dl]
%\nameplaceup 1
% &
% &
% &
% &
% &
% \drawmarkedplace\ar[dr]
% \nameplaceup 7
% &\drawmarkedpersistentplace\ar@[mygray][d]
% \namepersupD{p_{t_{f}}}
% \\
% & 
% \drawtrans {t_d}\ar[dl]\ar@[mygray][dr] 
% &
% &
%  &
%  &
%  &
%  &
%  &
%   \drawtrans {t_f}\ar@[mygray][dl]\ar[dr] 
% \\
% \drawplace
% \nameplaceup 6
% &
% &
%\drawpersistentplace\ar@[mygray][d] 
% \namepersup{ 3}
% &
% &
% &
% &
% &
%\drawpersistentplace\ar@[mygray][d]
% \namepersup{ 8}
% &
% &
% \drawplace
% \nameplaceright 9
%  \\
%&&
% \drawtransu {\color{mygray}t_3}\ar@[mygray]@/_2ex/[ddddll] \ar@[mygray]@/^/[ddrrrrr]\ar@[mygray]@/^2ex/[ddrrrrrrr]\ar@[mygray][dd]
% &&&&&
% \drawtransu {\color{mygray}t_8}\ar@[mygray][dd]\ar@[mygray]@/^5ex/[ddddrrrrr] \ar@[mygray]@/_5ex/[ddddlllllll] 
% \\
% &&&&&&
%   \\
%&
%&
%  \drawpersistentplace 
% \namepersrightD{p_{t_g}\ }
% &
% &
%&
% \drawmarkedplace\ar[drr]
% \nameplaceright {2}
% &
% &
%  \drawpersistentplace\ar@[mygray][d] 
% \namepersrightD{p_{t_b}}
% &&
% \drawpersistentplace
% \namepersupD{p_{t'_8}}
%  \\
% &
% &
% &
% &
% &
%  &
%  &
% \drawtrans {t_b}\ar[d]
%  &
% &
% &&&&
% \\
% \drawpersistentplace
% \namepersright{5}
% &
% &
%  &
% &
% &
%&
%&
% \drawplace
% \nameplaceright {4}
% &
% &
% &&&
%  \drawpersistentplace
% \namepersright{10}
%  }
%$$
\caption{A process for $\dyntopt {\enc N}$ (running example)}
\label{fig:determ-persistent-process}   
\end{figure}

Fig.~\ref{fig:determ-persistent-process} shows a  process for the net $\dyntopt {\enc N}$ of our running example (see $N$ in Fig.~\ref{fig:multcondconfusion} and  $\enc N$ in Fig.~\ref{fig:graphical-encoded-N}). 
The process accounts for the firing of the transitions $d$, $f$, $b$ in $N$. Despite they 
  look as concurrent events in $N$, the persistent 
place $\bf p_{t_b}$ introduces some causal dependencies. In fact, 
we have: $\Phi(t_d) = \Phi(t_f) = \mathit{true}$, $\Phi(t_3) = t_d$, $\Phi(t_8)=t_f$ and $\Phi(t_b) = (t_3 \wedge t_d) \vee (t_8 \wedge t_f)$, thus
 $t_b$ can be fired only after either $t_d$ or $t_f$ (or both).
 The other maximal processes are reported in Appendix~\ref{app:e}.
\end{exa}

\begin{figure}
\begin{subfigure}[b]{0.4\textwidth} 
$$
 \xymatrix@R=.7pc@C=.6pc{
&
& 
\drawmarkedplace\ar[dr] 
\nameplaceup 1
 &
 \drawmarkedpersistentplace\ar@[mygray][d]
 &
 &
 &
 \drawmarkedpersistentplace\ar@[mygray][d]
 &
 \drawmarkedplace\ar[dl]
 \nameplaceup 7
 \\
 & 
 &
 &
  \drawtrans {t_a}\ar@[mygray][dr]\ar[ddddr] 
  &
  &
  &
  \drawtrans {t_e}\ar@[mygray][dl]\ar@/^3ex/[ddddllll] 
  &
  &
 \\
 &
 &
 &
 &
 \drawpersistentplace
 \namepersup{ 6}
 &
 \drawpersistentplace
 \namepersup{ 9}
 &
 &
 &
 &
  \\
 &&
 &&&&&
 \\
 &&&&&&&&&
   \\
&
&
 \drawplace\ar[dr]
  \nameplaceleft {8}
 &
 \drawmarkedpersistentplace\ar@[mygray][d]
 &
 \drawplace\ar[dl]
 \nameplaceright {3}
&
&
 \drawmarkedplace\ar[dlll]
 \nameplaceright {2}
 &
  \\
 &
 &
 &
 \drawtrans {t_{bg}}\ar[drrr]\ar@[mygray][dlll]\ar[dl]
 &
 &
  &
  &
  &
 &
 \\
  \drawpersistentplace
 \namepersright{5}
 &
 &
 \drawplace
 \nameplaceright {10}
 &
 &
 &
 &
 \drawplace
 \nameplaceright {4}
 &
 &
 \\
 \\
 \\
    }
$$
\subcaption{}\label{fig:bgatomic}
\end{subfigure}
\begin{subfigure}[b]{0.4\textwidth} 
$$
 \xymatrix@R=.7pc@C=.6pc{
&
& 
\drawmarkedplace\ar[dr] 
\nameplaceup 1
 &
 \drawmarkedpersistentplace\ar@[mygray][d]
 &
 &
 &
 \drawmarkedpersistentplace\ar@[mygray][d]
 &
 \drawmarkedplace\ar[dl]
 \nameplaceup 7
 \\
 & 
 &
 &
  \drawtrans {t_a}\ar@[mygray][dr]\ar[ddddr] 
  &
  &
  &
  \drawtrans {t_e}\ar@[mygray][dl]\ar@/^3ex/[ddddllll] 
  &
  &
 \\
 &
 &
 &
 &
 \drawpersistentplace
 \namepersup{ 6}
 &
 \drawpersistentplace
 \namepersup{ 9}
 &
 &
 &
 &
  \\
 &&
 &&&&&
 \\
 &&&&&&&&&
   \\
&
&
 \drawplace\ar[dr]
  \nameplaceleft {8}
 &
 \drawmarkedpersistentplace\ar@[mygray][d]
 &
 \drawplace\ar[dl]
 \nameplaceright {3}
&
&
 \drawmarkedplace\ar[dlll]
 \nameplaceright {2}
 &
  \\
 &
 &
 &
 \drawtrans {t_{bg}}\ar@[mygray]@/_4ex/[dddlll]\ar[dl]\ar[drrr]
 &
 &
  &
  &
  &
 &
 \\
&
&
 \drawplace\ar[d]
  \nameplacerightm {8_{t_{bg}}}
 &
 &
&
&
 \drawplace\ar[d]
 \nameplaceleftm {2_{t_{bg}}}
 &
 \\
 &
 &
 \drawtrans {g_{t_{bg}}}\ar[d]
 &
 &
&
&
 \drawtrans {b_{t_{bg}}}\ar[d]
 \\
  \drawpersistentplace
 \namepersright{5}
 &
 &
 \drawplace
 \nameplaceright {10}
 &
 &
 &
 &
 \drawplace
 \nameplaceright {4}
 &
 &
    }
$$
\subcaption{}\label{fig:bgexpanded}
\end{subfigure}
\caption{Expanding transactions I}\label{fig:bg}
\end{figure}
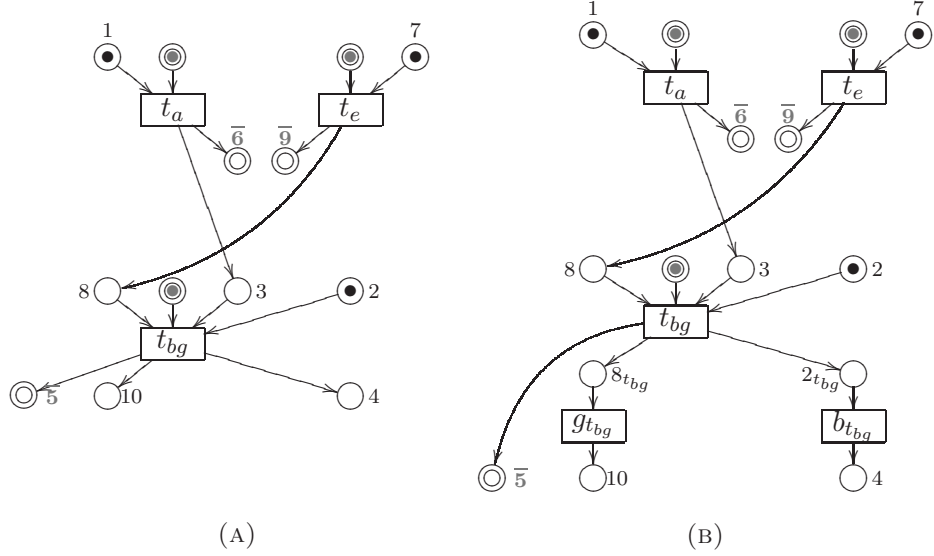

Still referring to the net $\dyntopt {\enc N}$ of our running example, a more interesting case to consider is the process in Fig.~\ref{fig:bgatomic} whose transition $t_{bg}$ stands for the transaction where $b$ and $g$ are executed simultaneously. One may argue that having $t_{bg}$ as an atomic action can reduce the overall concurrency of the system. However, $t_{bg}$ can be expanded with (a fresh copy of) its underlying process as shown in Fig.~\ref{fig:bgexpanded}. We use $t_{bg}$ as a subscript for the new nodes of the process to guarantee they are fresh. The preset of the transition $t_{bg}$ is left unchanged. Its postset takes care of the propagation of negative information and of enabling the initial places of the underlying process. The final places of the underlying process are the (positive) places in the postset of the original transition $t_{bg}$. This transformation has the side effect to separate the choice of the transaction from its execution, but it increases the amount of concurrency, as $b_{t_{bg}}$ and $g_{t_{bg}}$ can now be executed in any order.
While it might be possible in some cases to avoid the additional choice event, the general construction would look cumbersome.

\begin{figure}
\begin{subfigure}[b]{0.33\textwidth} 
          $$
          \xymatrix@R=.8pc@C=.8pc{
            \drawmarkedplace\ar[d]\ar@/^1.1ex/[rddd]
            \nameplaceright 1
            \POS[]+<1pc,-2.9pc> *+=<7.5pc,7.4pc>[F--]{}
            \POS[]-<2pc,0pc>\drop{\bc_1} 
            & 
            \drawmarkedplace\ar[d]\ar[ld]
            \nameplaceright 2
            & &
            \drawmarkedplace\ar[d]
            \nameplaceright 3
            \POS[]+<.9pc,-1.1pc> *+=<4.4pc,3.9pc>[F--]{}
            \POS[]+<2.4pc,0pc>\drop{\bc_2} 
            \\
            \drawtrans a\ar[ddd] 
            & \drawtrans b\ar[d] 
            & 
            & \drawtrans d\ar[d] 
            \\
            &
            \drawplace\ar[d]
            \nameplaceright 4
            &
            &
            \drawplace
            \nameplaceright 5
            \\
            & \drawtrans c\ar[d] 
            \\
            \drawplace
            \nameplaceright 6
            &
            \drawplace
            \nameplaceright 7
          }
          $$
\subcaption{}\label{fig:bcseq}
\end{subfigure}
\begin{subfigure}[b]{0.33\textwidth} 
          $$
          \xymatrix@R=.8pc@C=.8pc{
            \drawmarkedplace\ar[rd]
            \nameplaceright 1
            & 
            \drawmarkedplace\ar[d]
            \nameplaceright 2
            & 
            \drawmarkedpersistentplace\ar@[mygray][ld]
            &
            \drawmarkedplace\ar[d]
            \nameplaceright 3
            & 
            \drawmarkedpersistentplace\ar@[mygray][ld]
            \\
            & \drawtrans {t_{bc}}\ar[d]\ar@[mygray][ld]
            & 
            & \drawtrans {t_d}\ar[d] 
            \\
            \drawpersistentplace
             \namepersright{6}
            &
            \drawplace
            \nameplaceright 7
            &
            &
            \drawplace
            \nameplaceright 5
            \\
          }
          $$
\subcaption{}\label{fig:bcatomic}
\end{subfigure}
\begin{subfigure}[b]{0.3\textwidth} 
          $$
          \xymatrix@R=.8pc@C=.8pc{
            \drawmarkedplace\ar[rd]
            \nameplaceright 1
            & 
            \drawmarkedplace\ar[d]
            \nameplaceright 2
            & 
            \drawmarkedpersistentplace\ar@[mygray][ld]
            &
            \drawmarkedplace\ar[d]
            \nameplaceright 3
            & 
            \drawmarkedpersistentplace\ar@[mygray][ld]
            \\
            & \drawtrans {t_{bc}}\ar[d]\ar[rd]\ar@[mygray][lddddd]
            & 
            & \drawtrans {t_d}\ar[d] 
            \\
            &
            \drawplace\ar[rddd]
            \nameplacerightm {1_{t_{bc}}}
            &
            \drawplace\ar[d]
            \nameplacerightm {2_{t_{bc}}}
             &
            \drawplace
            \nameplaceright 5
            \\
            & &
            \drawtrans {b_{t_{bc}}}\ar[d] 
            \\
            & &
            \drawplace\ar[d]
            \nameplacerightm {4_{t_{bc}}}
             \\
            & &
            \drawtrans {c_{t_{bc}}}\ar[d] 
            \\
            \drawpersistentplace
             \namepersright{6}
            & &
            \drawplace
            \nameplaceright 7
            \\
          }
          $$
\subcaption{}\label{fig:bcexpanded}
\end{subfigure}
\caption{Expanding transactions II}
\end{figure}
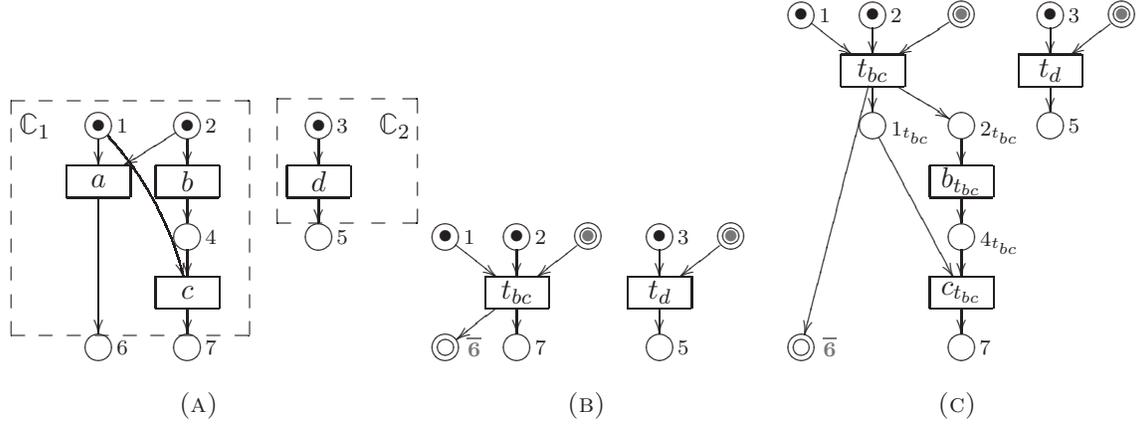

The improvement of the amount of concurrency that can be observed is even more evident if we consider the net $N$ in Fig.~\ref{fig:bcseq}. 
There are two s-cells $\bc_1$ and $\bc_2$: the former has two transactions $\theta_a=\{a\}$ and $\theta_{bc}=\{b,c\}$, the latter is trivial as it has only one transaction $\theta_d=\{d\}$. 
One process of $\dyntopt {\enc N}$ is in Fig.~\ref{fig:bcatomic}, where $t_{bc}$ and $t_d$ are executed concurrently. However it does not take into account the fact that the execution of $b$ and $c$ can be interleaved with that of $d$. If we expand $t_{bc}$ as discussed in the previous example, we get the process in Fig.~\ref{fig:bcexpanded}, where $d$ can be executed after $b$ and before $c$.

Formally, given a process $\theta$ and a transition name $t$, let $\theta_t$ be the process where any non final place/transition $n$ is renamed to $n_t$ and any final place is left unchanged. We say a positive transition $t_{\theta,\bc}$ of $\dyntopt {\enc N}$ is \emph{non-atomic} if the process $\theta$ involves more than one transition. Given a net $N$ and its uniformed net $\dyntopt {\enc N}$ we let $\dyntopt{ {\enc N}}_{\mathit{conc}}$ denote the persistent net obtained from $\dyntopt {\enc N}$ by removing each non-atomic transition $t_{\theta,\bc}$ and by adding, for each such transition, the places and transitions in $\theta_{t_{\theta,\bc}}$ together with a transition $t'_{\theta,\bc}$ such that $\preS t'_{\theta,\bc} = \preS t_{\theta,\bc}$ but whose postset consists of $\minp\theta_{t_{\theta,\bc}}$ together with the negative places in the postset of $t_{\theta,\bc}$.
%%% Local Variables:
%%% mode: latex
%%% TeX-master: "main.tex"
%%% End:

% !TEX root =  main.tex

\section{Probabilistic Nets}
\label{sec:probability}

%Building on the results from previous sections, 
We can now outline our methodology 
to assign probabilities to the concurrent runs of a Petri net, also in the presence of confusion.
Given a net $N$, we apply s-cell decomposition from Section~\ref{sec:structural-cells}, 
and then we assign probability distributions to the transactions available in each cell $\bc$ 
(and recursively to the s-cell decomposition of $N_{\bc}$).
Let $\mathcal{P}_\bc: \{ \theta \mid \theta:\bc\} \to [0,1]$ denote the probability 
distribution function of the s-cell $\bc$ (such that $\sum_{\theta:\bc} \mathcal{P}_\bc(\theta) = 1$).
Such probability distributions are defined locally and transferred automatically to the transitions 
in $\Tpos$ of the  dynamic p-net $\enc N$ defined in Section~\ref{sec:results}, in such a way
that $\mathcal{P}(t_{\theta,\bc}) = \mathcal{P}_\bc(\theta)$.
Each negative transition in $\Tneg$ has probability 1 because no choice is associated with it.
Since the uniformed net $\dyntopt{\enc N}$ has the same transitions of $\enc N$, the probability
distribution can be carried over $\dyntopt{\enc N}$ (thanks to Proposition~\ref{prop:dynvsflat}).

\paragraph{AB's probability distribution}
Building on the bijective correspondence in Theorem~\ref{theo:AB}, 
the distribution $\mathcal{P}_\bc$ can be chosen in such a way that it is consistent with the one
attached to the transitions of Abbes and Benveniste's branching cells (if any).

\paragraph{Purely local distribution}
Another simple way to define $\mathcal{P}_\bc$ is by assigning probability distributions to the arcs leaving the same place of the original net,
as if each place were able to decide autonomously which transition to fire.
Then, given a transaction $\theta:\bc$, we can set $\mathcal{Q}_\bc(\theta)$ be the product of the probability associated with the arcs of $N$ entering the transitions in $\theta$. Of course, in general it can happen that $\sum_{\theta:\bc} \mathcal{Q}_\bc(\theta) < 1$, as not all combinations are feasible.
%of token flows correspond to maximal processes. 
However, it is always possible to normalise the quantities of feasible assignments by setting $\mathcal{P}_\bc(\theta) = \frac{\mathcal{Q}_\bc(\theta)}{\sum_{\theta':\bc} \mathcal{Q}_\bc(\theta')}$ for any transaction $\theta:\bc$.

% !TEX root =  main.tex

\begin{exa}
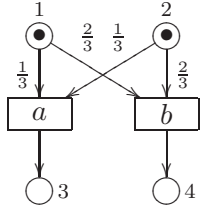
\begin{figure}
%\vspace{-20pt}
$$
 \xymatrix@R=1.5pc@C=2pc{
 \drawmarkedplace\ar[d]_{\frac{1}{3}} \ar[rd]^(.3){\frac{2}{3}}
 \nameplaceup 1
 &
 \drawmarkedplace\ar[d]^{\frac{2}{3}}\ar[ld]_(.3){\frac{1}{3}}
 \nameplaceup 2
 \\
 \drawtrans a\ar[d] 
 & \drawtrans b\ar[d] 
 \\
 \drawplace
 \nameplaceright 3
 &
 \drawplace
 \nameplaceright 4
 }
$$
\caption{A free-choice net}\label{fig:freechoice}
\end{figure} 
Take the free-choice net in Fig.~\ref{fig:freechoice}
and assume that decisions are local to each place.
Thus, place $1$ lends its token to $a$ with probability $p_1=\frac{1}{3}$ and to
$b$ with $q_1=\frac{2}{3}$.  Similarly, place $2$ lends its token to $a$
with probability $p_2=\frac{1}{3}$ and to $b$ with $q_2=\frac{2}{3}$.  
Then one can set $p_a = p_1 \cdot p_2 = \frac{1}{9}$ and $p_b = q_1
\cdot q_2 = \frac{4}{9}$.  However their sum is $\frac{5}{9} \neq 1$.  
This anomaly is due to the
existence of deadlocked choices with nonzero probabilities which disappear in the process semantics of nets. 
To some extent, the probabilities assigned to $a$ and $b$ should be conditional w.r.t. the fact that the local choices performed at places $1$ and $2$ are compatible, i.e., all non compatible choices are disregarded. This means that we need to normalize the values of $p_a$ and $p_b$ over their sum.
Of course, normalisation is possible only if there is at least one admissible alternative.  
In this simple example we get $\mathcal{Q}(a)=\frac{1}{9}$, $\mathcal{Q}(b)=\frac{4}{9}$, $\mathcal{P}(a)=\frac{1}{9}/\frac{5}{9} = \frac{1}{5}$ and $\mathcal{P}(b)=\frac{4}{9}/\frac{5}{9} = \frac{4}{5}$.
\end{exa}
%%% Local Variables:
%%% mode: latex
%%% TeX-master: "main.tex"
%%% End:

\begin{exa}
\label{ex:assignment-prob}
Suppose that in our running example we  assign uniform distributions to all arcs leaving a place. 
From simple calculation we have 
$\mathcal{P}_{\bc_1}(\theta_a) = \mathcal{P}_{\bc_1}(\theta_d) = \frac{1}{2}$ for the first cell,
$\mathcal{P}_{\bc_2}(\theta_e) = \mathcal{P}_{\bc_2}(\theta_f) = \frac{1}{2}$ for the second cell,
$\mathcal{P}_{\bc_3}(\theta_c) = \mathcal{P}_{\bc_{{3}}}(\theta_{bg}) = \frac{1}{2}$ for the third cell.
The transactions of nested cells are uniquely defined and thus have all probability $1$.
\end{exa}

Given a firing sequence $t_1;\cdots ; t_n$ we can set $\mathcal{P}(t_1;\cdots; t_n) = \prod_{i=1}^n \mathcal{P}(t_i)$.
Hence firing sequences that differ in the order in which  transitions are fired are assigned the
same probability.
%However, the sum of the probabilities of all maximal firing sequences is in general different than $1$, 
%because there can be different firing sequences that represent the same concurrent computation 
%(which is thus counted twice or more).
Thanks to Theorem~\ref{theo:concur},
%(and Corollary~\ref{cor:adm}), 
we can consider maximal persistent  processes
instead of firing sequences 
and set $\mathcal{P}(O) = \prod_{t\in O} \mathcal{P}(t)$.
In fact any maximal firing sequence in $O$ includes all transitions of $O$
and its probability is independent from the order of firing.
It follows from Theorem~\ref{theo:AB} that any maximal configuration has a corresponding 
maximal process (and viceversa) and since Abbes and Benveniste proved that the sum of the probabilities assigned to maximal configurations is $1$, the same holds for maximal persistent  processes.

\begin{exa}  
%Suppose that in our running example we are assigning uniform distributions to all choices available in an s-cell.
Suppose the distributions are assigned as in Example~\ref{ex:assignment-prob}.
Then, the persistent process in Fig.~\ref{fig:determ-persistent-process} has probability:
$
\mathcal{P}(O) =
\mathcal{P}(t_d) \cdot \mathcal{P}(t_f)\cdot \mathcal{P}(t_3)\cdot \mathcal{P}(t_8)\cdot \mathcal{P}(t_b)\cdot \mathcal{P}(t'_8) =
\frac{1}{2} \cdot \frac{1}{2} \cdot 1 \cdot 1 \cdot 1\cdot 1 = \frac{1}{4}
$.
There are other four maximal processes shown in Appendix~\ref{app:e} together with their probabilities.
We note that the sum of all probabilities assigned to maximal processes is indeed $1$.
\end{exa}
%%% Local Variables:
%%% mode: latex
%%% TeX-master: "main.tex"
%%% End:

% !TEX root =  main.tex

\section{Conclusion and Future Work}

%Confusion is long-standing problem in Petri net theory.
AB's branching cells are a sort of interpreter (or scheduler) 
for executing PESs in the presence of confusion. 
Our main results develop along two orthogonal axis. Firstly, 
%we show a tight correspondence between AB's 
%maximal configurations and  our maximal deterministic processes. Consequently, 
our approach
is an innovative construction with the following advantages:
%improves AB's construction in the following aspects:
%\begin{enumerate}
%\item Improvement of AB's construction.
\begin{enumerate}
\item Compositionality: s-cells are defined statically and locally, while AB's branching cells are defined dynamically and globally (by executing the whole event structure).
%(by manipulating the entire event structure).
\item Compilation vs interpretation:  AB's construction gives an interpreter that rules out some executions of an event structure. We instead compile a net into another one (with persistency) whose execution is driven by ordinary firing rules.
\item Complete concurrency: AB's recursively stopped configurations may include traces that cannot be executed  by the interpreter. 
%(i.e., in the presence of confusion)
Differently, 
 our notion of process captures all and only those executable traces of a concurrent computation.
%AB's concurrent computations are the configurations of an event structure. Thus, in the presence of confusion, they include traces that are not executable by AB's interpreter, while our notion of process captures all and only executable traces.
\item Simplicity: s-cells definition in terms of a closure relation takes a couple of lines (see Definition~\ref{def:scells}), while AB's branching cell definition 
%in terms of recursively stopped configurations in event structures requires 
is more involved. % (we have done at our best to concisely recall it in Section~\ref{sec:ABcells}).
%Then we prove that their construction and our construction assign the same probability distribution to corresponding configuration/processes.
\item Full matching: we define a behavioural correspondence 
%between AB's branching cells and our s-cells.
%, decorating related cells with the same probability distributions. 
%Moreover, we relate 
that relates AB's maximal configurations with our maximal deterministic processes,
preserving their probability assignment.
\end{enumerate}
%preserving their probability assignment.
%\item
 Secondly, we provide the following fully original perspectives:
\begin{enumerate}
\item Confusion removal: our target model is confusion-free.
\item Locally executable model:  probabilistic choices are confined to transitions with the same pre-set, and hence
can be resolved locally and concurrently. Besides, our target model relies on ordinary firing rules (with persistent places).
\item Processes: we define a novel notion of process for nets with persistency that conservatively extends the ordinary notion of process and captures the right amount of concurrency.
\item Goal satisfaction: our construction meets all requirements in the list of desiderata.
\end{enumerate}
%\end{enumerate} 

\noindent
This paper has extended the conference version~\cite{DBLP:conf/lics/BruniMM18} by including more examples and all detailed proofs of main results.
The idea of expanding transactions into the underlying processes is also original to this contribution.

Moreover, the construction presented here has opened the way to other interesting research directions.
First, it has led to the implementation of a tool, called {\tt RemConf}~\cite{maraschio} after ``removal of confusion'', that takes in input an acyclic net $N$ in the standard format {\tt .pnml} and returns the net $\dyntopt {\enc N}$.
As persistent places cannot be modeled in {\tt .pnml}, they are implemented using self-loops instead of input arcs when they are part of the preset of a transition.
This makes it possible to simulate the execution of $\dyntopt {\enc N}$ using any {\tt .pnml} compatible tool.
The tool is available at \url{http://remconf.di.unipi.it}.

Regarding OR causality, it can be accounted for by general event structures, where events can be enabled by distinct minimal sets of events. 
Several classes of event structures have been studied to disambiguate causality, in the sense that all alternative causes for an event must be somehow in conflict. This is the case, e.g., of stable event structures~\cite{DBLP:conf/rex/Winskel88} and bundle event structures~\cite{DBLP:conf/forte/Langerak92}. Five different classes of event structures that allow for causal ambiguity have been proposed in~\cite{DBLP:conf/concur/LangerakBK97}. In~\cite{DBLP:journals/lmcs/BaldanBCGMM18} we have extended the connection between p-nets and event structures in order to deal with OR causes. Generalising the work on ordinary nets, Petri nets with persistent places are related to a new subclass of general event structures, called locally connected, by means of a chain of coreflections relying on an unfolding construction, as for the original construction by Winskel.
The fact that a whole body of theory can be extended from Petri nets to p-nets witnesses that p-nets can be chosen as a general computational model and not just a convenient variant of an existing model.
The causal AND/OR-dependencies share some similarities also with the work on connectors and Petri nets with boundaries~\cite{DBLP:journals/corr/BruniMMS13}
that we would like to formalize.

We also want to investigate the connection between our s-cell structure and Bayesian networks, so to make forward and backward reasoning techniques available in our setting.
Some results in this direction can be found in~\cite{DBLP:journals/corr/abs-1807-06305}.

Our construction is potentially complex:
%The construction of s-cells is to some extent hierarchical: 
given a s-cell $\bc$ %and its corresponding net $N_{\bc}$ 
we recursively consider the nested s-cells 
%arising from the decomposition of  
in $N_{\bc} \ominus p$, for any initial place $p\in N_{\bc}$.
In the worst case, %when each s-cell is tightly connected, 
the number of nested s-cells can be exponential in the number of their initial places.
However s-cells are typically much smaller than the whole net and it can be the case that 
%, so that even if the net grows very large it can be the case that 
the size of all s-cells is bound by some fixed $k$.
In this case, the number of s-cells in our construction can still become exponential on the constant $k$, but linear w.r.t. the number of places of the net.

A limitation of our approach is that it applies to finite occurrence nets only (or, equivalently, to finite PESs).
As a future work, we plan %to extend the approach 
to deal with cycles and unfolding semantics.
This requires some efforts and we conjecture it is feasible only if the net is safe and its behaviour has some 
%form of 
regularity:
the same s-cell can be executed several times in a computation but every instance is restarted without tokens left from previous rounds.
%

%, event structures with conflicts, concurrency, and AND/OR causes (here only concurrency + AND/OR causality, but no conflict. TO CHECK: in flow event structures OR+conflict). To investigate connections with petri nets with boundaries and connectors.

%Remark: esecuzione atomica delle transazioni esprime massima concorrenza secondo AB, ma noi possiamo permettere una versione piu' generale usando commit ma permettendo interleaving.

%Advantages: statically defined, locally defined, compositional definition, complete concurrency is recovered by introducing deterministic nonsequential processes for persistent nets.

%MOVED FROM INTRO:
%Locally dynamic nets can be easily defined in JOIN calculus [], and programmed in JoCaml, Polyphonic C\#, C$\omega$ and Join Java []. The JOIN calculus can be considered as a Petri net-based mobile calculus analogous to pi-calculus. Its feature of limiting its nondeterministic choice process to fresh sites could be arguably combined with the branching cell approach to provide a confusion free, mobile version of nets.

%%% Local Variables:
%%% mode: latex
%%% TeX-master: "main.tex"
%%% End:

%% Acknowledgments
\paragraph{Acknowledgments}                            %% acks environment is optional
                                        %% contents suppressed with 'anonymous'
  %% Commands \grantsponsor{<sponsorID>}{<name>}{<url>} and
  %% \grantnum[<url>]{<sponsorID>}{<number>} should be used to
  %% acknowledge financial support and will be used by metadata
  %% extraction tools.
The idea of  structural cells emerged, with a different goal, in collaboration with Lorenzo Galeotti after his MSc thesis.
The tool RemConf has been developed as part of the MSc thesis of Gianluca Maraschio.
We thank Glynn Winskel, Holger Hermanns, {Joost-Pieter Katoen} for giving us insightful references.
The research has been partially supported by
  EU H2020 RISE programme under the Marie Marie Sk\l{}odowska-Curie grant agreement 778233,
    by the UBACyT projects 20020170100544BA and 20020170100086BA, 
    by the PIP project 11220130100148CO and
by Universit\`a di Pisa projects PRA\_2016\_64 \emph{Through the fog}
    and PRA\_2018\_66  \emph{DECLWARE: Metodologie dichiarative per la progettazione e il deployment di applicazioni}.
The third author carried on part of the work while attending a Program on Logical
Structures in Computation at Simons Institute, Berkeley, 2016.

%\paragraph{Acknowledgments}

%% Bibliography
\bibliographystyle{plain}
\bibliography{biblio}
%\input{editedmain}

%\newpage

%% Appendix
\appendix
% !TEX root =  main.tex

\section{Detailed proofs of results in Section 2.4}

\begin{prop}[Proposition~\ref{prop:dynvsflat}]
  Let $N = (T, b) \in\dn{\mathbb{S}}$. Then,
  \begin{enumerate}
  \item 
    $N\tr t N'$ implies $\dyntopt N\tr t \dyntopt {N'}$;
  \item Moreover,  $\dyntopt N \tr t N'$ implies there exists $N''$ such that $N\tr t N''$ and $N' = \dyntopt {N''}$.
  \end{enumerate}
\end{prop}

\begin{proof}
We start by showing that  $N\tr t N'$ implies $\dyntopt N\tr t \dyntopt {N'}$.
If $N\tr t N'$ then $t = S\pntrans (T', b') \in T$,  $S\subseteq b$  and 
 $N' =  (T\cup T', (b\setminus S)\cup b')$. 
 By definition of $\dyntopt{\_}$, it holds that  
 $\dyntopt N = (\mathbb{S} \cup {\bf P}_{\DT N}, \DT N,
  F, b\cup b_T)$
  where 
     ${\bf P}_{\DT N}=  \{ {\bf p}_{t'} \ |\ {t'}\in \DT{N}\}$.
  Note that  
\[
 \begin{array}{l@{\ =\ }ll}
 \DT N &  T \cup \bigcup_{t'\in T}  \DT {\postS {t'}} & \mbox{by def. of } \DT{\_} \\
 &  T \cup \bigcup_{t'\in T}  \DT {\postS {t'}}  \cup  \DT {\postS {t}}& t \in T \\
 &  T \cup \bigcup_{t'\in T}  \DT {\postS {t'}}  \cup  T'  \cup \bigcup_{t'\in T'}  \DT {\postS {t'}} & \mbox{by def. of } \DT{\_}\\
 &  (T \cup T')  \cup \bigcup_{t'\in T\cup T'}  \DT {\postS {t'}} & \mbox{by assoc. and}\\
 \multicolumn{3}{r}{ \mbox{comm. of } \cup} \\
 & \DT {N'}  & \mbox{by def. of } \DT{\_} 
  \end{array}
 \]

Hence,  $\dyntopt {N'} = (\mathbb{S} \cup {\bf P}_{\DT N}, \DT N,
  F, (b\setminus S)\cup b' \cup b_{T\cup T'})$. By the definition of $\dyntopt {\_}$, 
     $t \in \DT N$ and $F$ is  such that 
     $t:S \cup \{{\bf p}_t\} \to b' \cup b_{T'}$. 
     Hence, $t$ is enabled in  $b\cup b_T$ because $S\subseteq b$ and
     ${\bf p}_t\in b_T$. Consequently, 
     $b\cup b_T \tr t ((b\cup b_T)\setminus (S\cup \{{\bf p}_t\})) \cup b'\cup b_{T'}  
     = (b\setminus S) \cup b_T \cup b'\cup  b_{T'} = (b\setminus S) \cup b'  \cup b_{T\cup T'}$.
     
The proof for $\dyntopt N \tr t N'$ implies there exists $N''$ such that $N\tr t
    N''$ and $N' = \dyntopt {N''}$ follows by analogous arguments.
\end{proof}

%%% Local Variables:
%%% mode: latex
%%% TeX-master: "main.tex"
%%% End:

% !TEX root =  main.tex

\section{Proofs of results in Section 3}
This section presents the proofs of the results in Section 3. 
Note that we need some auxiliary lemmas that are not present in the main text of the paper.
They are marked by the keyword ``Aux'' to avoid ambiguities. 
For reviewer's convenience, the high-level proof sketches are separated from the proofs in full detail,
that are included in a separate section. 

We start by showing that the encoding of a net into a dynamic net does not add computations. 
We show that each reachable marking $b$ of the dynamic net can be associated with a reachable marking  $m$
of the original net, when disregarding negative information. 
We remark that in general the relation between such $b$ and $m$ is that 
$b \cap P \subseteq m$ and not necessarily $b \cap P = m$ (see, e.g., Lemma~\ref{lem:net-mimicks-encoding}). 
This is because the transitions $t_{\theta,\bc}$ generated by the 
encoding ($\Tpos$) always consume the tokens in all minimal places of the branching cell $\bc$.
%, even when the associated transaction does not.  
This choice is immaterial for the behaviour of the encoded net, as made explicit
by the main results in the paper.

\begin{lem}[Aux.]\label{lem:net-mimicks-encoding}
Let  $N = (P,T,F,m)$. If $\enc N \rightarrow^* (T,b)$ then $m \rightarrow^* m'$ and $b \cap P \subseteq m'$.
\end{lem}

\begin{proof} The proof follows by induction on the length of the reduction  $\enc N \rightarrow^n (T,b)$.
\begin{itemize}
  \item {\bf Base case (n=0)}. It follows immediately because $b = m$. 
  \item {\bf Inductive case (n = k+1)}. Then, $\enc N \rightarrow^k (T',b')\tr t (T,b)$. 
  By inductive hypothesis, $m \rightarrow^* m''$ and $b' \cap P \subseteq m''$. We now proceed by case analysis on the 
  shape of $t$. 
  \begin{itemize}
    \item $t = \minp \bc   \pntrans 
  	(\emptyset, \maxp\theta \cup \co{\maxp \bc\setminus\maxp \theta})$.  Then, $\minp \bc \subseteq b'$,
	$T = T'$ and $b = (b' \setminus \minp \bc) \cup \maxp\theta \cup \co{\maxp \bc\setminus\maxp \theta}$.
	Since $\minp \bc \subseteq P$, we have $\minp \bc \subseteq m''$. Moreover, $\theta : \bc $ implies 
	$\minp \theta \subseteq\minp \bc \subseteq m''$.  Since $\theta$ is a deterministic process,
	$m'' \rightarrow^* (m''\setminus\minp\theta) \cup \maxp\theta$. Then, take 
	$m' = (m''\setminus\minp\theta) \cup \maxp\theta$.
	
	Note that $b\cap P = ((b' \cap P) \setminus \minp \bc) \cup \maxp\theta$. We  use  $(b' \cap P) \subseteq m''$
	and $\minp \theta \subseteq \minp\bc$  to conclude that $b\cap P \subseteq m'$.

  \item $t =  \co{\bf p} \pntrans   (T', \co{\maxp \bc\setminus\maxp {(\remInitial {N_{\bc}} p)}})$. It follows immediately because
  $b' \cap P = b \cap P$. \qedhere
 \end{itemize}
\end{itemize}
\end{proof}

\begin{cor}[Aux.]\label{cor:one-safe-dyn} 
If $N$ is 1-safe then $\enc N$ is \OneInfSafe.
\end{cor}

\begin{lem}[Lemma~\ref{lemma:nested-disjoint-presets}] 
%Let $\enc N \in \dn{P\cup\co {\bf P}}$. If $\enc N \rightarrow^* (T,b)$ then for all $t,t'\in T$ such that
%$\preS t \neq \preS{t'}$ and  $\preS t \cap \preS{t'} \neq \emptyset$  it holds that 
% $\co {(\preS t \cup \preS{t'})}\cap b \neq \emptyset$.
Let $\enc N \in \dn{P\cup\co {\bf P}}$. If $\enc N \rightarrow^* (T,b)$ then for all $t,t'\in T$ such that
$\preS t \neq \preS{t'}$ and  $\preS t \cap \preS{t'} \cap P\neq \emptyset$  it holds that 
there is $p\in P \cap (\preS t \cup \preS{t'})$ such that $\co{\bf p} \in b$.

\end{lem}

\begin{proof} The proof follows by induction on the length of the firing sequence $\enc N \xrightarrow{t_1\cdots t_n} (T,b)$.
\begin{itemize}
 \item {\bf Base Case $n = 0$}. It holds trivially because any pair of different transitions in $T$ have either 
  the same preset (i.e., if they are taken from $\Tpos$  and originate from the same s-cell) or disjoint presets (i.e., if they are taken both 
 from  $\Tpos$ but originate from different s-cells, or  both from $\Tneg$, or one from $\Tneg$ and the other from $\Tpos$).
  \item {\bf Inductive step $n = k+1$}. Hence, $\enc N \xrightarrow{t_1\cdots t_k} (T',b') \tr {t_{k+1}} (T,b)$.   By inductive hypothesis, 
  for all $t,t'\in T'$ such that
and  $\preS t \cap \preS{t'} \cap P\neq \emptyset$,  it holds that 
there is $p\in P \cap (\preS t \cup \preS{t'})$ such that $\co{\bf p} \in b'$. 
Then, we proceed by case analysis on
$t_{k+1}$. 
  \begin{itemize}
      \item $t_{k+1} = \minp \bc \pntrans 
              (\emptyset, \maxp\theta \cup \co{\maxp \bc\setminus\maxp \theta})$. It holds trivially because
              $T=T'$ and $b'\cap\ \co {\bf P} \subseteq b$.%\textcolor{blue}{(RB: per\`o $b$ \`e diverso da $b'$)}.  
      \item $t_{k+1} =  \co{\bf p}  \pntrans   (T'', \co{\maxp \bc\setminus\maxp {(\remInitial {N_{\bc}} p)}})$ for 
      some  $\bc$,  $p \in \minp {\bc}$, and $(T'', \emptyset) = \enc{\remInitial {N_{\bc}} p}$. Then $T = T' \cup T''$. 
      By the definition of $\enc{\_}$,  we have that for all $t,t'\in T''$ either  (i) $\preS t = \preS{t'}$ or 
      (ii) $\preS t \cap \preS{t'} = \emptyset$
      (reasoning analogously to the Base Case). It remains to consider the cases in which $t$ and $t'$ are taken one  
      from $T'$ and the other from $T''$. W.l.o.g., we consider $t\in T''$ and $t'\in T'$
      and proceed as follows. 
      \changed{By the definition of $\enc {\_}$, $t\in T''$ implies either (i) $\preS t \subseteq \co{\bf P}$ or 
      (ii) $\preS t = \minp{\bc_1}$ for $\bc_1\in\BC{\remInitial {N_{\bc}} p}$. 
      Case (i), follows immediately because there does not exist $t'$ s.t. $\preS t \cap \preS{t'} \cap P\neq \emptyset$.
      For (ii), we note that $\preS t' = \minp{\bc_2}$ with  $\bc_1 \neq \bc_2$, $\minp{\bc_1}\cap \minp{\bc_2}\neq\emptyset$
      and  $\preS t\cup \preS{t'} \subseteq P$}.
      We proceed by contradiction and assume $\co {(\preS t \cup \preS{t'})}\cap b = \emptyset$. 
      There must exist \changed{an} s-cell $\bc_3$ such that $\bc_1\cup\bc_2 \subseteq \bc_3$ (because $\bc_1$ and $\bc_2$ are 
      closed under immediate conflict and their union introduces  immediate
      conflict between the transitions consuming from the shared places in $\minp{\bc_1}\cap\minp{\bc_2}$). 
      If $\bc_2 = \bc_3$, 
      then $\bc_1\subset \bc_2$. Hence $p\in \minp{\bc_2}$ and $p\in\preS{t'}$, which contradicts $\co {(\preS t \cup \preS{t'})}\cap b = \emptyset$ 
      because $t_{k+1}$  enabled at $b$ implies 
      $\co{\bf p}\in b$. Otherwise, $\bc_2\subset \bc_3$. Consequently, there exists (at least) a transition $t''\in T'$ such that $\preS{t''} = \minp{\bc_3}$ 
      and $\preS{t'} \neq \changed{\preS{t''}}$. 
      Since $t'\in T'$ and
      $t''\in T'$, we can use inductive hypothesis to conclude that 
      $\co{(\preS{t'}\cup\preS{t''})}\cap b \neq \emptyset$. The proof is completed by noting that this
       is in contradiction with the assumption $\co{(\preS{t}\cup\preS{t'})}\cap b = \emptyset$
      because \changed{ $\preS{t''} \supseteq \preS{t}\cup\preS{t'}$}. \qedhere
      
\end{itemize}
\end{itemize}
\end{proof}

\noindent
In what follows we write $p \prec q$ if $p \preceq q$ and $p\neq q$. The following auxiliary result 
provides some invariants about the configurations that can be reached by an encoded dynamic net.

\begin{lem}[Aux.]\label{lem:aux-invariant-mutex}
 If $\enc N \rightarrow^* (T,b)$ then 
  \begin{enumerate}
     \item $ {p}\in b$ implies $\co {\bf p}\not\in b$; 
     \item if $\co {\bf p}\in b$ and $p \preceq q$ then $q \not\in b$; 
     \item if $p\preceq q$, $p\in b$ and $\co{\bf q}\in b$ then there exists $r\prec q$ and $\co{\bf r}\in b$; and
     \item if $(T,b)\tr t $ and $\preS t = \minp \bc$ for some $\bc$ then $(\maxp \bc \cup \co{\maxp \bc}) \cap b =\emptyset$.
  \end{enumerate}
\end{lem}

\begin{proof} The proof follows by induction on the length of the firing sequence $\enc N \xrightarrow{t_1\cdots t_n} (T,b)$.
\begin{itemize}
  \item {\bf Base Case $n = 0$}. Hence $(T,b) = \enc N$.
  \begin{enumerate}
  \item It follows  from $b \subseteq P$. 
  \item Since $b \subseteq P$ there is no $\co {\bf p}\in b$. 
  \item Since $b \subseteq P$ there is no $\co {\bf q}\in b$.  
  \item It follows from the fact 
  that $N$ is an occurrence net,  $b\subseteq \minp N$, and hence there does not exist any
   $\bc\in\BC N$  such that \changed{$(\maxp\bc  \cup \co{\maxp \bc})\cap b\neq \emptyset$}.
  \end{enumerate}
  \item {\bf Inductive step $n = k+1$}.  Hence, $\enc N \xrightarrow{t_1\cdots t_k} (T',b') \tr {t_{k+1}} (T,b)$.   By inductive hypothesis, 
  (1) ${p}\in b'$ implies $\co {\bf p}\not\in b'$;
  (2) if $\co {\bf p}\in b'$ and $p \preceq q$ then $q \not\in b'$; 
  (3) if $p\preceq q$, $p\in b'$ and $\co{\bf q}\in b'$ then there exists $r\prec q$ and $\co{\bf r}\in b'$; and
  (4) if $(T',b')\tr t $ and $\preS t = \minp \bc$ for some ${\bc}$ then $(\maxp {\bc} \cup \co{\maxp {\bc}}) \cap b' =\emptyset$.
     We now proceed by case analysis on 
     $t_{k+1}$. 
  \begin{itemize}
      \item $t_{k+1} = \minp {\bc_{k+1}} \pntrans  (\emptyset, \maxp\theta \cup \co{\maxp {\bc_{k+1}}\setminus\maxp \theta})$
      for some s-cell $\bc_{k+1}$ and transaction $\theta : \bc_{k+1}$. 
      Hence, $b = (b' \setminus \minp {\bc_{k+1}}) \cup (\maxp\theta \cup \co{\maxp {\bc_{k+1}}\setminus\maxp \theta})$. 
      \begin{enumerate}
      \item
      We proceed by contradiction. Assume that there exists $p$ such that 
      $p\in b$ and  $\co{\bf p} \in b$. 
      Since $p\in b$ we have that either $p\in b' \setminus \minp {\bc_{k+1}}$ or $p\in \maxp\theta$.
      First, assume $p\in b' \setminus \minp {\bc_{k+1}}$.  By inductive hypothesis (1),  $\co{\bf p} \not\in b'$ and, hence, 
      $\co{\bf p} \not\in b' \setminus \minp {\bc_{k+1}}$. Therefore, it should be the case that 
      $\co{\bf p}\in (\maxp\theta \cup \co{\maxp {\bc_{k+1}}\setminus\maxp \theta})$.
      Hence, $\co{\bf p}\in\co{\maxp {\bc_{k+1}}}$ and  ${p}\in{\maxp {\bc_{k+1}}}$.
      Since $t_{k+1}$ is enabled at $(T',b')$, we can use inductive hypothesis (4) on ${t_{k+1}}$ to conclude 
       $(\maxp {\bc_{k+1}} \cup \co{\maxp {\bc_{k+1}}}) \cap b' =\emptyset$. Consequently, ${p}\in{\maxp {\bc_{k+1}}}$ 
       implies $p\not\in b'$. But this is in contradiction with the 
       assumption that $p\in b' \setminus \minp {\bc_{k+1}}$. Assume instead
       $p \in \maxp\theta$. 
       %It is immediate to notice that $p \in \maxp\theta$ implies
       Then
      $\co{\bf p}\not\in\co{\maxp {\bc_{k+1}}\setminus\maxp \theta}$. 
      Hence, it should be the case that $\co{\bf p}\in  b' \setminus \minp {\bc_{k+1}}$. But this is also in 
      contradiction with  the hypothesis (4) $(\maxp {\bc_{k+1}} \cup \co{\maxp {\bc_{k+1}}}) \cap b' =\emptyset$.

      \item We proceed by contradiction. 
      Assume  there exist $p$ and $q$ such that
      $\co {\bf p}\in b$, $p \preceq q$ and $q \in b$. 
      \begin{itemize}
       \item
      \changed{Firstly, consider}  $\co{\bf p}\in b'\setminus \minp {\bc_{k+1}}$, which implies $\co{\bf p}\in b'$. 
      By inductive hypothesis (2), for all $q$ s.t. $p\preceq q$ it holds that $q\not\in b'$.%\setminus \minp {\bc_{k+1}}$.
      \changed{Hence, it should be the case that} $q\in (\maxp\theta \cup \co{\maxp {\bc_{k+1}}\setminus\maxp \theta})$\changed{. Hence 
      either (i) $p = q$, (ii) $p\in\minp{\bc_{k+1}}$ or  (iii) $p\prec p'$ and $p'\prec q$ for some $p'\in\minp{\bc_{k+1}}$}. 
      \changed{For (i), note that $q = p \in (\maxp\theta \cup \co{\maxp {\bc_{k+1}}\setminus\maxp \theta})$. Hence,
      $q\in \maxp {\bc_{k+1}}$,  
      which is in contradiction with the assumption  $\co{\bf p} \in b'$ and the inductive hypothesis (4).
      For (ii), note that it implies $p\in b'$, which is in contradiction with the assumption  $\co {\bf p}\in b'$ and 
      inductive hypothesis (1).
      For (iii), note that $\co{\bf p} \in b'$ and $p\prec p'$ imply $p'\not\in b'$ by inductive hypothesis (2), which is 
      in contradiction with the fact that $p'\in\minp{\bc_{k+1}}$ and $t_{k+1}$ is enabled.}
      \item Assume instead
      $\co{\bf p} \in \maxp\theta \cup \co{\maxp {\bc_{k+1}}\setminus\maxp \theta}$. 
      Hence,  $p	 \in \maxp {\bc_{k+1}}\setminus\maxp \theta$ and therefore
       $p \in {\maxp {\bc_{k+1}}}$. Suppose there is $q\in b'$ and $p\preceq q$. Note that   $p'\preceq p$ for all $p'\in\minp{\bc_{k+1}}$ by definition of branching cells. 
       By transitivity of $\preceq$,  $p'\preceq q$ for all $p'\in\minp{\bc_{k+1}}$. Since \changed{$t_{k+1}$} is enabled at $b'$,  $\minp{\bc_{k+1}}\subseteq b'$.
       By using Lemma~\ref{lem:net-mimicks-encoding}, we can conclude that $q\not\in b'$ for all $q$ s.t. $p\preceq q$,
       which contradicts
       the hypothesis  $q\in b'$ and $p\preceq q$. 
       Assume instead $q\in \maxp\theta \cup \co{\maxp {\bc_{k+1}}\setminus\maxp \theta}$. Hence, $q\in \maxp\theta$. 
       Hence, $p \neq q$. Moreover, $p\in \maxp {\bc_{k+1}}$ and $q\in \maxp {\bc_{k+1}}$ contradict the assumption
       $p\preceq q$.       
      \end{itemize}
      
      \item We proceed by case analysis. 
      If $p\in b'$ and $\co{\bf q}\in b'$ then the proof follows by inductive hypothesis. 
      If $p\in b'$ and $\co{\bf q}\not\in b'$, then 
      $\co{\bf q}\in\maxp\theta \cup \co{\maxp {\bc_{k+1}}\setminus\maxp \theta}$. Therefore,
      $p\preceq q$ implies $p\in\minp \bc$, which contradicts the assumption $p\in b$.
      If $p\not\in b'$ and $\co{\bf q}\in b'$, then $p\in \maxp\theta$, which contradicts $p\preceq q$.
      
%  
%      \item If $\co{\bf q}\in b'$ the proof follows by inductive hypothesis and by noting that $p\in\maxp\theta$
%      and $p\preceq q$ imply there exists $r\in\minp \bc$ and $r\preceq q$ (by transitivity of $\preceq$). 
%      If $\co{\bf q}\in\maxp\theta \cup \co{\maxp {\bc_{k+1}}\setminus\maxp \theta}$, 
%      follows by contradiction  because $p\preceq q$ and $p\in b'$ implies $p\in\minp \bc$ by Lemma~\ref{lem:net-mimicks-encoding}. Therefore, there 
%      does not exist $p$ such that $p\preceq q$ and $p\in b$.
%      
      \item Let $t\in T$ such that $\preS t = \minp \bc \subseteq b$ for some $\bc$. 
      Since $t$ is enabled at $b$ and $\enc N$ is \OneInfSafe by Corollary~\ref{cor:one-safe-dyn}, then 
      $\bc_{k+1}\cap\bc = \emptyset$.
      If $t$ is enabled at $(b' \setminus \minp {\bc_{k+1}})$ then $t$ is enabled at $b'$. By inductive 
      hypothesis (4), we conclude that 
      $(\maxp {\bc} \cup \co{\maxp {\bc}}) \cap (b'\setminus\minp {\bc_{k+1}})  =\emptyset$. 
      If $t$ is not enabled at $(T',b')$, then it holds that for $x \in \minp\bc$ exists $y\in (\maxp{\bc_{k+1}}\cup \co{\maxp{\bc_{k+1}}})$ such 
      that $y \preceq x$. By inductive hypothesis  $(\maxp {\bc_{k+1}} \cup \co{\maxp {\bc_{k+1}}}) \cap b' =\emptyset$, 
      hence $(\maxp {\bc} \cup \co{\maxp {\bc}}) \cap b' =\emptyset$. Therefore, $(\maxp {\bc} \cup \co{\maxp {\bc}}) \cap b =\emptyset$.
    \end{enumerate}
   
 \item $t_{k+1} =  \co{\bf r} \pntrans   (T'', \co{\maxp {\bc_{k+1}}\setminus\maxp {(\remInitial {N_{{\bc_{k+1}}}} r)}})$ for some s-cell $\bc_{k+1}$ and place $r\in \minp{\bc_{k+1}}$. 
 Then, $T = T' \cup T''$ with $\enc{\remInitial {N_{{\bc_{k+1}}}} r} = (T'',\_)$ and $b = b' \cup \co{\maxp {\bc_{k+1}}\setminus\maxp {(\remInitial {N_{{\bc_{k+1}}}} r)}}$.
       \begin{enumerate}
      \item
      We proceed by contradiction. Assume that there exists $p$ such that 
      $p\in b$ and  $\co{\bf p} \in b$. Note that $p\in b$ implies $p\in b'$.
      By inductive hypothesis (1),  $\co{\bf p} \not\in b'$. 
      Therefore, it should be the case that 
      $\co{\bf p}\in\co{\maxp {\bc_{k+1}}\setminus\maxp {(\remInitial {N_{{\bc_{k+1}}}} r)}}$.
      Consequently $p\in\maxp {\bc_{k+1}}$ and $p\not\in\maxp {(\remInitial {N_{{\bc_{k+1}}}} r)}$. 
      Hence, $r\preceq p$. Since $t$ is enabled at $b'$, $\co{\bf r}\in b'$. By 
      inductive hypothesis (2), $p\not\in b'$ which contradicts the hypothesis $p\not\in b$.

       \item We proceed by contradiction. Assume  there exist $p$ and $q$ such that
       $\co {\bf p}\in b$, $p \preceq q$ and $q \in b$. Note that $q\in b$ implies $q\in b'$.
        Assume  $\co{\bf p}\in b'$. By inductive hypothesis, for all 
       $q$ s.t. $p\preceq q$ then $q\not\in b'$ and, hence it is in contradiction with assumption $q\in b$.
       Assume instead
      $\co{\bf p} \in \co{\maxp {\bc_{k+1}}\setminus\maxp {(\remInitial {N_{{\bc_{k+1}}}} r)}}$. As before, we conclude that 
      $r\preceq p$. By transitivity of $\preceq$, we have $r\preceq q$. By inductive hypothesis (2), $q\not\in b'$, which is 
      in contradiction with assumption $q\in b$.
 
      \item For $\co{\bf q} \in b'$, it follows immediately by inductive hypothesis. For   $\co{\bf q} \in \co{\maxp {\bc_{k+1}}\setminus\maxp {(\remInitial {N_{{\bc_{k+1}}}} r)}}$,
      it follows straightforwardly because $r \preceq q$ and $\co{\bf r}\in b$.
          
      \item Assume $\preS t = \minp \bc \subseteq b$ for some $\bc$. Hence,   $\preS t = \minp \bc \subseteq b'$. There are two cases:
      \begin{itemize}
      \item Suppose $t\in T'$.  By inductive hypothesis (4), 
      $(\maxp {\bc} \cup \co{\maxp {\bc}}) \cap b' = \emptyset$. We show that  the following holds
      $$
      (\maxp {\bc} \cup \co{\maxp {\bc}}) \cap 
      \co{\maxp {\bc_{k+1}}\setminus\maxp {(\remInitial {N_{{\bc_{k+1}}}} r)}} = \emptyset
      $$ 
      It is enough to show that
      $$ \co{\maxp {\bc}} \cap 
      \co{\maxp {\bc_{k+1}}\setminus\maxp {(\remInitial {N_{{\bc_{k+1}}}} r)}} = \emptyset$$
        We proceed by contradiction and assume
      there exists $q$ such that $q\in \maxp {\bc}$ and $q\in(\maxp {\bc_{k+1}}\setminus\maxp {(\remInitial {N_{{\bc_{k+1}}}} r)})$.
      Because $q\in(\maxp {\bc_{k+1}}\setminus\maxp {(\remInitial {N_{{\bc_{k+1}}}} r)})$,  $r \preceq q$. Since $q \in \maxp {\bc}$,
      $r\in \minp\bc$ (because $\bc$ is closed under causality). Hence $r\in b'$ because $t$ is enabled at $b'$. By the contrapositive of inductive hypothesis (1), $\co{\bf r}\not\in b'$,
      but this is in contradiction with the hypothesis that $t_{k+1}$ is enabled at $b'$.

     \item Suppose $t\in T''$.  Then, $\preS t \cap \minp {(\remInitial {N_{{\bc_{k+1}}}} r)} = \emptyset$.
      Hence, for all $q\in\maxp  \bc$ there exists $s\in\minp \bc$ s.t. $s\preceq q$. 
      Since $t$ is enabled at $b$, $\minp \bc \subseteq b$ holds. By Lemma~\ref{lem:net-mimicks-encoding}, $\maxp \bc \cap b = \emptyset$. 
      We show by contradiction that $\co{\maxp \bc} \cap b = \emptyset$ does not hold either. 
      Assume that there exists $\co{\bf q}\in \co{\maxp \bc}$ and $\co{\bf q} \in b$. Since there 
      exists $s\in \minp\bc \subseteq b$ and $s\preceq q$, we can use inductive hypothesis (3) 
      to conclude that  there exist   $o \preceq q$ s.t. $\co{\bf o}  \in  b$. By the
      inductive hypothesis (2) $q\not\in b$, and this is in contradiction with the assumption of $t$ enabled at $b$. \qedhere
      
    \end{itemize}
    
    \end{enumerate}

\end{itemize}
\end{itemize}
\end{proof}

\begin{lem}[Lemma~\ref{lemma:negpos}]
  If $\enc N \rightarrow^* (T,b)$ and $\co {\bf p}\in b$ then $(T,b) \rightarrow^* (T',b')$ implies that $p\not\in b'$.
\end{lem}

\begin{proof}
If $\co {\bf p}\in b$ then $\co {\bf p}\in b'$ because  $\co{\bf p}$ is persistent. Moreover, $\enc N \rightarrow^* (T',b')$.
By the contrapositive of Lemma~\ref{lem:aux-invariant-mutex}(1), $p\not\in b'$.
\end{proof}

\begin{thm}[Theorem~\ref{th:confusion-free}]
  Let $\enc N \in \dn{P\cup\co{\bf P}}$. If $\enc N \rightarrow^* (T,b) \tr
  t $ and $(T,b) \tr {t'} $ then either $\preS t = \preS {t'}$ or
  $\preS t \cap \preS {t'} = \emptyset$.
\end{thm}

\begin{proof}
By contradiction. Assume $t, t'$ such that  $(T,b) \tr  t $, $(T,b) \tr {t'}$, $\preS t \neq \preS {t'}$, and $\preS t \cap \preS {t'} \neq \emptyset$. 
By construction of the encoding, it must be the case that $\preS t \subseteq P$ and $\preS {t'} \subseteq P$. Hence, 
 $\preS t \cap \preS {t'} \cap P \neq \emptyset$. By Lemma~\ref{lemma:nested-disjoint-presets} 
there exists $p\in P \cap (\preS t \cup \preS{t'})$ such that $\co{\bf p} \in b$. By Lemma~\ref{lemma:negpos}, $p\not\in b$, 
which is in contradiction with the assumptions $(T,b) \tr  t $ and  $(T,b) \tr {t'}$.
\end{proof}

%%% Local Variables:
%%% mode: latex
%%% TeX-master: "main.tex"
%%% End:

% !TEX root =  main.tex

\subsection{Detailed proofs of results in Section 3}

\begin{lem}[Lemma~\ref{lem:net-mimicks-encoding}]
Let  $N = (P,T,F,m)$. If $\enc N \rightarrow^* (T,b)$ then $m \rightarrow^* m'$ and $b \cap P \subseteq m'$.
\end{lem}

\begin{proof} The proof follows by induction on the length of the reduction  $\enc N \rightarrow^n (T,b)$.
\begin{itemize}
  \item {\bf Base case (n=0)}. It follows immediately because $b = m$. 
  \item {\bf Inductive case (n = k+1)}. Then, $\enc N \rightarrow^k (T',b')\tr t (T,b)$. 
  By inductive hypothesis, $m \rightarrow^* m''$ and $b' \cap P \subseteq m''$. We now proceed by case analysis on the 
  shape of $t$. 
  \begin{itemize}
    \item $t = \minp \bc   \pntrans 
  	(\emptyset, \maxp\theta \cup \co{\maxp \bc\setminus\maxp \theta})$.  Then, $\minp \bc \subseteq b'$,
	$T = T'$ and $b = (b' \setminus \minp \bc) \cup \maxp\theta \cup \co{\maxp \bc\setminus\maxp \theta}$.
	Since $\minp \bc \subseteq P$, we have $\minp \bc \subseteq m''$. Moreover, $\theta : \bc $ implies 
	$\minp \theta \subseteq\minp \bc \subseteq m''$.  Since $\theta$ is a deterministic process,
	$m'' \rightarrow^* (m''\setminus\minp\theta) \cup \maxp\theta$. Then, take 
	$m' = (m''\setminus\minp\theta) \cup \maxp\theta$.
	
	Note that $b\cap P = ((b' \cap P) \setminus \minp \bc) \cup \maxp\theta$. We  use  $(b' \cap P) \subseteq m''$
	and $\minp \theta \subseteq \minp\bc$  to conclude that $b\cap P \subseteq m'$.

  \item $t =  \co{\bf p} \pntrans   (T', \co{\maxp \bc\setminus\maxp {(\remInitial {N_{\bc}} p)}})$. It follows immediately because
  $b' \cap P = b \cap P$. \qedhere
 \end{itemize}
\end{itemize}
\end{proof}

\begin{lem}[Lemma~\ref{lemma:nested-disjoint-presets} ]
%Let $\enc N \in \dn{P\cup\co {\bf P}}$. If $\enc N \rightarrow^* (T,b)$ then for all $t,t'\in T$ such that
%$\preS t \neq \preS{t'}$ and  $\preS t \cap \preS{t'} \neq \emptyset$  it holds that 
% $\co {(\preS t \cup \preS{t'})}\cap b \neq \emptyset$.
Let $\enc N \in \dn{P\cup\co {\bf P}}$. If $\enc N \rightarrow^* (T,b)$ then for all $t,t'\in T$ such that
$\preS t \neq \preS{t'}$ and  $\preS t \cap \preS{t'} \cap P\neq \emptyset$  it holds that 
there is $p\in P \cap (\preS t \cup \preS{t'})$ such that $\co{\bf p} \in b$.

\end{lem}

\begin{proof} The proof follows by induction on the length of the firing sequence $\enc N \xrightarrow{t_1\cdots t_n} (T,b)$.
\begin{itemize}
 \item {\bf Base Case $n = 0$}. It holds trivially because any pair of different transitions in $T$ have either 
  the same preset (i.e., if they are taken from $\Tpos$  and originate from the same s-cell) or disjoint presets (i.e., if they are taken both 
 from  $\Tpos$ but originate from different s-cells, or  both from $\Tneg$, or one from $\Tneg$ and the other from $\Tpos$).
  \item {\bf Inductive step $n = k+1$}. Hence, $\enc N \xrightarrow{t_1\cdots t_k} (T',b') \tr {t_{k+1}} (T,b)$.   By inductive hypothesis, 
  for all $t,t'\in T'$ such that
and  $\preS t \cap \preS{t'} \cap P\neq \emptyset$,  it holds that 
there is $p\in P \cap (\preS t \cup \preS{t'})$ such that $\co{\bf p} \in b'$. 
Then, we proceed by case analysis on
$t_{k+1}$. 
  \begin{itemize}
      \item $t_{k+1} = \minp \bc \pntrans 
              (\emptyset, \maxp\theta \cup \co{\maxp \bc\setminus\maxp \theta})$. It holds trivially because
              $T=T'$ and $b'\cap\ \co {\bf P} \subseteq b$.%\textcolor{blue}{(RB: per\`o $b$ \`e diverso da $b'$)}.  
      \item $t_{k+1} =  \co{\bf p}  \pntrans   (T'', \co{\maxp \bc\setminus\maxp {(\remInitial {N_{\bc}} p)}})$ for 
      some  $\bc$,  $p \in \minp {\bc}$, and $(T'', \emptyset) = \enc{\remInitial {N_{\bc}} p}$. Then $T = T' \cup T''$. 
      By the definition of $\enc{\_}$,  we have that for all $t,t'\in T''$ either  (i) $\preS t = \preS{t'}$ or 
      (ii) $\preS t \cap \preS{t'} = \emptyset$
      (reasoning analogously to the Base Case). It remains to consider the cases in which $t$ and $t'$ are taken one  
      from $T'$ and the other from $T''$. W.l.o.g., we consider $t\in T''$ and $t'\in T'$
      and proceed as follows. 
       Note that, by construction of $\enc N$, $\preS t \subseteq \co{\bf P}$ implies $|\preS t| = 1$ for any $t$. Hence, 
      the only possibility is $\preS t = \minp{\bc_1}$ with $\bc_1\in\BC{\remInitial {N_{\bc}} p}$  
      and $\preS t' = \minp{\bc_2}$ with  $\bc_1 \neq \bc_2$ and $\minp{\bc_1}\cap \minp{\bc_2}\neq\emptyset$. 
      Note that $\preS t\cup \preS{t'} \subseteq P$.
      We proceed by contradiction and assume $\co {(\preS t \cup \preS{t'})}\cap b = \emptyset$. 
      There must exist a s-cell $\bc_3$ such that $\bc_1\cup\bc_2 \subseteq \bc_3$ (because $\bc_1$ and $\bc_2$ are 
      closed under immediate conflict and their union introduces  immediate
      conflict between the transitions consuming from the shared places in $\minp{\bc_1}\cap\minp{\bc_2}$). 
      If $\bc_2 = \bc_3$, 
      then $\bc_1\subset \bc_2$ and hence $p\in \minp{\bc_2}$ and $p\in\preS{t'}$, which contradicts $\co {(\preS t \cup \preS{t'})}\cap b = \emptyset$ 
      because $t_{k+1}$  enabled at $b$ implies 
      $\co{\bf p}\in b$. Otherwise, $\bc_2\subset \bc_3$. Consequently, there exists (at least) a transition $t''\in T'$ such that $\preS{t''} = \minp{\bc_3}$ 
      and $\preS{t'} \neq \preS{t'''}$. 
      Since $t'\in T'$ and
      $t''\in T'$, we can use inductive hypothesis to conclude that 
      $\co{(\preS{t'}\cup\preS{t''})}\cap b \neq \emptyset$. The proof is completed by noting that this
       is in contradiction with the assumption $\co{(\preS{t}\cup\preS{t'})}\cap b \neq \emptyset$
      because $\preS{t''} = \preS{t}\cup\preS{t'}$. \qedhere
      
\end{itemize}
\end{itemize}
\end{proof}

%In what follows we write $p \prec q$ if $p \preceq q$ and $p\neq q$.

\begin{lem}[Lemma~\ref{lem:aux-invariant-mutex}]
 If $\enc N \rightarrow^* (T,b)$ then 
  \begin{enumerate}
     \item $ {p}\in b$ implies $\co {\bf p}\not\in b$; 
     \item if $\co {\bf p}\in b$ and $p \preceq q$ then $q \not\in b$; 
     \item if $p\preceq q$, $p\in b$ and $\co{\bf q}\in b$ then there exists $r\prec q$ and $\co{\bf r}\in b$; and
     \item if $(T,b)\tr t $ and $\preS t = \minp \bc$ for some $\bc$ then $(\maxp \bc \cup \co{\maxp \bc}) \cap b =\emptyset$.
  \end{enumerate}
\end{lem}

\begin{proof} The proof follows by induction on the length of the firing sequence $\enc N \xrightarrow{t_1\cdots t_n} (T,b)$.
\begin{itemize}
  \item {\bf Base Case $n = 0$}. Hence $(T,b) = \enc N$.
  \begin{enumerate}
  \item It follows  from $b \subseteq P$. 
  \item Since $b \subseteq P$ there is no $\co {\bf p}\in b$. 
  \item Since $b \subseteq P$ there is no $\co {\bf q}\in b$.  
  \item It follows from the fact 
  that $N$ is an occurrence net,  $b\subseteq \minp N$, and hence there does not exist any
   $\bc\in\BC N$  such that $\maxp\bc \cap b \neq \emptyset$.
  \end{enumerate}
  \item {\bf Inductive step $n = k+1$}.  Hence, $\enc N \xrightarrow{t_1\cdots t_k} (T',b') \tr {t_{k+1}} (T,b)$.   By inductive hypothesis, 
  (1) ${p}\in b'$ implies $\co {\bf p}\not\in b'$;
  (2) if $\co {\bf p}\in b'$ and $p \preceq q$ then $q \not\in b'$; 
  (3) if $p\preceq q$, $p\in b'$ and $\co{\bf q}\in b'$ then there exists $r\prec q$ and $\co{\bf r}\in b'$; and
  (4) if $(T',b')\tr t $ and $\preS t = \minp \bc$ for some ${\bc}$ then $(\maxp {\bc} \cup \co{\maxp {\bc}}) \cap b' =\emptyset$.
     We now proceed by case analysis on 
     $t_{k+1}$. 
  \begin{itemize}
      \item $t_{k+1} = \minp {\bc_{k+1}} \pntrans  (\emptyset, \maxp\theta \cup \co{\maxp {\bc_{k+1}}\setminus\maxp \theta})$
      for some s-cell $\bc_{k+1}$ and transaction $\theta : \bc_{k+1}$. 
      Hence, $b = (b' \setminus \minp {\bc_{k+1}}) \cup (\maxp\theta \cup \co{\maxp {\bc_{k+1}}\setminus\maxp \theta})$. 
      \begin{enumerate}
      \item
      We proceed by contradiction. Assume that there exists $p$ such that 
      $p\in b$ and  $\co{\bf p} \in b$. 
      Since $p\in b$ we have that either $p\in b' \setminus \minp {\bc_{k+1}}$ or $p\in \maxp\theta$.
      First, assume $p\in b' \setminus \minp {\bc_{k+1}}$.  By inductive hypothesis (1),  $\co{\bf p} \not\in b'$ and, hence, 
      $\co{\bf p} \not\in b' \setminus \minp {\bc_{k+1}}$. Therefore, it should be the case that 
      $\co{\bf p}\in (\maxp\theta \cup \co{\maxp {\bc_{k+1}}\setminus\maxp \theta})$.
      Hence, $\co{\bf p}\in\co{\maxp {\bc_{k+1}}}$ and  ${p}\in{\maxp {\bc_{k+1}}}$.
      Since $t_{k+1}$ is enabled at $(T',b')$, we can use inductive hypothesis (4) on ${t_{k+1}}$ to conclude 
       $(\maxp {\bc_{k+1}} \cup \co{\maxp {\bc_{k+1}}}) \cap b' =\emptyset$. Consequently, ${p}\in{\maxp {\bc_{k+1}}}$ 
       implies $p\not\in b'$. But this is in contradiction with the 
       assumption that $p\in b' \setminus \minp {\bc_{k+1}}$. Assume instead
       $p \in \maxp\theta$. 
       %It is immediate to notice that $p \in \maxp\theta$ implies
       Then
      $\co{\bf p}\not\in\co{\maxp {\bc_{k+1}}\setminus\maxp \theta}$. 
      Hence, it should be the case that $\co{\bf p}\in  b' \setminus \minp {\bc_{k+1}}$. But this is also in 
      contradiction with the the hypothesis (4) $(\maxp {\bc_{k+1}} \cup \co{\maxp {\bc_{k+1}}}) \cap b' =\emptyset$.

      \item We proceed by contradiction. 
      Assume  there exist $p$ and $q$ such that
      $\co {\bf p}\in b$, $p \preceq q$ and $q \in b$. 
      Assume  $\co{\bf p}\in b'$. 
      By inductive hypothesis (2), for all $q$ s.t. $p\preceq q$ it holds that $q\not\in b'\setminus \minp {\bc_{k+1}}$.
      Moreover,  if $q\in (\maxp\theta \cup \co{\maxp {\bc_{k+1}}\setminus\maxp \theta})$ implies  
      $p'\preceq q$ for all $p'\in\minp{\bc_{k+1}}$ by definition of branching cells. 
      Since $t$ is enabled at $b'$, $\minp{\bc_{k+1}}\subseteq b'$ and hence $p\in b'$, but this is 
      in contradiction with inductive hypothesis (1), i.e., $\co {\bf p}\in b'$ implies $ p\not\in b'$.
      Assume instead
      $\co{\bf p} \in \maxp\theta \cup \co{\maxp {\bc_{k+1}}\setminus\maxp \theta}$. 
      Hence,  $p	 \in \maxp {\bc_{k+1}}\setminus\maxp \theta$ and
       $p \in {\maxp {\bc_{k+1}}}$. Suppose there is $q\in b'$ and $p\preceq q$. Note that   $p'\preceq p$ for all $p'\in\minp{\bc_{k+1}}$ by definition of branching cells. 
       By transitivity of $\preceq$,  $p'\preceq q$ for all $p'\in\minp{\bc_{k+1}}$. Since $t$ is enabled at $b'$,  $\minp{\bc_{k+1}}\subseteq b'$.
       By using Lemma~\ref{lem:net-mimicks-encoding}, we can conclude that $q\not\in b'$ for all $q$ s.t. $p\preceq q$,
       which contradicts
       the hypothesis  $q\in b'$ and $p\preceq q$. 
       Assume instead $q\in \maxp\theta \cup \co{\maxp {\bc_{k+1}}\setminus\maxp \theta}$. Hence, $q\in \maxp\theta$. 
       Hence, $p \neq q$. Moreover, $p\in \maxp {\bc_{k+1}}$ and $q\in \maxp {\bc_{k+1}}$ contradict the hypothesis
       $p\preceq q$.       
      
      \item If $\co{\bf q}\in b'$ the proof follows by inductive hypothesis and by noting that $p\in\maxp\theta$
      and $p\preceq q$ imply there exists $r\in\minp \bc$ and $r\preceq q$ (by transitivity of $\preceq$). 
      If $\co{\bf q}\in\maxp\theta \cup \co{\maxp {\bc_{k+1}}\setminus\maxp \theta}$, 
      follows by contradiction  because $p\preceq q$ and $p\in b'$ implies $p\in\minp \bc$ by Lemma~\ref{lem:net-mimicks-encoding}. Therefore, there 
      does not exist $p$ such that $p\preceq q$ and $p\in b$.
      
      \item Let $t\in T$ such that $\preS t = \minp \bc \subseteq b$ for some $\bc$. 
      Since $t$ is enabled at $b$ and $\enc N$ is \OneInfSafe by Corollary~\ref{cor:one-safe-dyn}, then 
      $\bc_{k+1}\cap\bc = \emptyset$.
      If $t$ is enabled at $(b' \setminus \minp {\bc_{k+1}})$ then $t$ is enabled at $b'$. By inductive 
      hypothesis (2), we conclude that 
      $(\maxp {\bc} \cup \co{\maxp {\bc}}) \cap (b'\setminus\minp {\bc_{k+1}})  =\emptyset$. 
      If $t$ is not enabled at $(T',b')$, then it holds that for $x \in \bc$ exists $y\in (\maxp{\bc_{k+1}}\cup \co{\maxp{\bc_{k+1}}})$ such 
      that $y \preceq x$. By inductive hypothesis  $(\maxp {\bc_{k+1}} \cup \co{\maxp {\bc_{k+1}}}) \cap b' =\emptyset$, 
      hence $(\maxp {\bc} \cup \co{\maxp {\bc}}) \cap b' =\emptyset$. Therefore, $(\maxp {\bc} \cup \co{\maxp {\bc}}) \cap b =\emptyset$.
    \end{enumerate}
   
 \item $t_{k+1} =  \co{\bf r} \pntrans   (T'', \co{\maxp {\bc_{k+1}}\setminus\maxp {(\remInitial {N_{{\bc_{k+1}}}} r)}})$ for some s-cell $\bc_{k+1}$ and place $r\in \minp{\bc_{k+1}}$. 
 Then, $T = T' \cup T''$ with $\enc{\remInitial {N_{{\bc_{k+1}}}} r} = (T'',\_)$ and $b = b' \cup \co{\maxp {\bc_{k+1}}\setminus\maxp {(\remInitial {N_{{\bc_{k+1}}}} r)}}$.
       \begin{enumerate}
      \item
      We proceed by contradiction. Assume that there exists $p$ such that 
      $p\in b$ and  $\co{\bf p} \in b$. Note that $p\in b$ implies $p\in b'$.
      By inductive hypothesis (1),  $\co{\bf p} \not\in b'$. 
      Therefore, it should be the case that 
      $\co{\bf p}\in\co{\maxp {\bc_{k+1}}\setminus\maxp {(\remInitial {N_{{\bc_{k+1}}}} r)}}$.
      Consequently $p\in\maxp {\bc_{k+1}}$ and $p\not\in\maxp {(\remInitial {N_{{\bc_{k+1}}}} r)}$. 
      Hence, $r\preceq p$. Since $t$ is enabled at $b'$, $\co{\bf r}\in b'$. By 
      inductive hypothesis (2), $p\not\in b'$ which contradicts the hypothesis $p\not\in b$.

       \item We proceed by contradiction. Assume  there exist $p$ and $q$ such that
       $\co {\bf p}\in b$, $p \preceq q$ and $q \in b$. Note that $q\in b$ implies $q\in b'$.
        Assume  $\co{\bf p}\in b'$. By inductive hypothesis, for all 
       $q$ s.t. $p\preceq q$ then $q\not\in b'$ and, hence it is in contradiction with assumption $q\in b$.
       Assume instead
      $\co{\bf p} \in \co{\maxp {\bc_{k+1}}\setminus\maxp {(\remInitial {N_{{\bc_{k+1}}}} r)}}$. As before, we conclude that 
      $r\preceq p$. By transitivity of $\preceq$, we have $r\preceq q$. By inductive hypothesis (2), $q\not\in b'$, which is 
      in contradiction with assumption $q\in b$.
 
      \item For $\co{\bf q} \in b$, it follows immediately by inductive hypothesis. For   $\co{\bf q} \in \co{\maxp {\bc_{k+1}}\setminus\maxp {(\remInitial {N_{{\bc_{k+1}}}} r)}}$,
      it follows straightforwardly because $r \preceq q$ and $\co{\bf r}\in b$.
          
      \item Assume $\preS t = \minp \bc \subseteq b$ for some $\bc$. Hence,   $\preS t = \minp \bc \subseteq b'$. There are two cases:
      \begin{itemize}
      \item Suppose $t\in T'$.  By inductive hypothesis (4), 
      $(\maxp {\bc} \cup \co{\maxp {\bc}}) \cap b' = \emptyset$. We show that  the following holds
      $$
      (\maxp {\bc} \cup \co{\maxp {\bc}}) \cap 
      \co{\maxp {\bc_{k+1}}\setminus\maxp {(\remInitial {N_{{\bc_{k+1}}}} r)}} = \emptyset
      $$ 
      It is enough to show that
      $$ \co{\maxp {\bc}} \cap 
      \co{\maxp {\bc_{k+1}}\setminus\maxp {(\remInitial {N_{{\bc_{k+1}}}} r)}} = \emptyset$$
        We proceed by contradiction and assume
      there exists $q$ such that $q\in \maxp {\bc}$ and $q\in(\maxp {\bc_{k+1}}\setminus\maxp {(\remInitial {N_{{\bc_{k+1}}}} r)})$.
      Because $q\in(\maxp {\bc_{k+1}}\setminus\maxp {(\remInitial {N_{{\bc_{k+1}}}} r)})$,  $r \preceq q$. Since $q \in \maxp {\bc}$,
      $r\in \minp\bc$ (because $\bc$ is closed under causality). Hence $r\in b'$ because $t$ is enabled at $b'$. By the contrapositive of inductive hypothesis (1), $\co{\bf r}\not\in b'$,
      but this is in contradiction with the hypothesis that $t_{k+1}$ is enabled at $b'$.

     \item Suppose $t\in T''$.  Then, $\preS t \cap \minp {(\remInitial {N_{{\bc_{k+1}}}} r)} = \emptyset$.
      for some $\bc$. Hence, for all $q\in\maxp  \bc$ there exists $s\in\minp \bc$ s.t. $s\preceq q$. 
      Since $t$ is enabled at $b$, $\minp \bc \subseteq b$ holds. By Lemma~\ref{lem:net-mimicks-encoding}, $\maxp \bc \cap b = \emptyset$. 
      We show by contradiction that $\co{\maxp \bc} \cap b = \emptyset$ does not hold either. 
      Assume that there exists $\co{\bf q}\in \co{\maxp \bc}$ and $\co{\bf q} \in b$. Since there 
      exists $s\in \bc \subseteq b$ and $s\preceq q$, we can use inductive hypothesis (3) 
      to conclude that  there exist $\co{\bf s'}  \subseteq  b$ and $s'\preceq q$ and $s' \in\minp \bc$. By the
      inductive hypothesis (1) $s'\not\in b$, and this is in contradiction with the assumption of $t$ enabled at $b$. qedhere
      
    \end{itemize}
    
    \end{enumerate}

\end{itemize}
\end{itemize}
\end{proof}

%\begin{lem}\label{lemma:negpos} 
%  If $\enc N \rightarrow^* (T,b)$ and $\co {\bf p}\in b$ then $(T,b) \rightarrow^* (T',b')$ implies that $p\not\in b'$.
%\end{lem}
%
%\begin{proof}
%If $\co {\bf p}\in b$ then $\co {\bf p}\in b'$ because  $\co{\bf p}$ is persistent. Moreover, $\enc N \rightarrow^* (T',b')$.
%By the contrapositive of Lemma~\ref{lem:aux-invariant-mutex}(1), $p\not\in b'$.
%\end{proof}
%
%
%\begin{reptheorem}{th:confusion-free}
%  Let $\enc N \in \dn{P\cup\co P}$. If $\enc N \rightarrow^* (T,b) \tr
%  t $ and $(T,b) \tr {t'} $ then either $\preS t = \preS {t'}$ or
%  $\preS t \cap \preS {t'} = \emptyset$.
%\end{reptheorem}
%
%\begin{proof}
%By contradiction. Assume $t, t'$ such that  $(T,b) \tr  t $, $(T,b) \tr {t'}$, $\preS t \neq \preS {t'}$, and $\preS t \cap \preS {t'} \neq \emptyset$. 
%By construction of the encoding, it must be the case that $\preS t \subseteq P$ and $\preS {t'} \subseteq P$. Hence, 
% $\preS t \cap \preS {t'} \cap P \neq \emptyset$. By Lemma~\ref{lemma:nested-disjoint-presets} 
%there exists $p\in P \cap (\preS t \cup \preS{t'})$ such that $\co{\bf p} \in b$. By Lemma~\ref{lemma:negpos}, $p\not\in b$, 
%which is in contradiction with the assumptions $(T,b) \tr  t $ and  $(T,b) \tr {t'}$.
%\end{proof}
%

%%% Local Variables:
%%% mode: latex
%%% TeX-master: "main.tex"
%%% End:

% !TEX root =  main.tex

\section{Proofs of results in Section 4} 
This section presents the proof sketches of the results in Section 4. 
As in Appendix~A, we exploit some auxiliary lemmas marked by the keyword ``Aux'' and full proofs are 
provided separately. 
%
%We start by providing some auxiliary results. 
We start by showing that reductions of a encoded net correspond to recursively stopped configurations of 
the event structure.
\begin{lem}[Aux.]\label{lem:aux-red-rstopped}
Let $N = (P, T, F, m)$ and $\mathcal{E}$ the event structure of $N$. If 
$\enc N \xrightarrow{t_1\cdots t_n}  (T,b)$ and $v = \bigcup_{1\leq i\leq n} \dec {t_i}$, then
\begin{enumerate}
\item $b\cap P = \minp \{ e \ | \ e\in \mathcal{E}^v  \mbox{ and } \down{e}=\{e\}\}$; and
\item If $(T,b)\tr t$ then $\dec t \neq \emptyset$ implies  $\dec t$ is a stopped 
configuration of $\mathcal{E}^v$.
\end{enumerate}
\end{lem}

\begin{proof} If follows by induction on the length of  $\enc N \xrightarrow{t_1\cdots t_n} (T,b)$.
\begin{itemize}
  \item {\bf Base case (n=0)}. Then, $v = \emptyset$ and 
  $\mathcal{E}^v = \mathcal{E}$. Moreover,  $b = m$. 
  \begin{enumerate}
  \item
  It is immediate to notice that $m$ corresponds to the preset of all minimal events of $\mathcal{E}$.
  \item Since $t$ is enabled, $\preS t\subseteq m$. Hence, $\preS t = \minp \bc$ with $\bc\in\BC N$. 
  Therefore, $\bc$ corresponds to 
  a branching cell of  $\mathcal{E}$.  By the definition of $\enc{\_}$, $t$ is associated with some $\theta : \bc$,
  which is  a maximal, conflict-free set of transitions in $\bc$. Hence, $\dec t$ is a stopped 
  configuration of $\mathcal{E}$. 
  \end{enumerate}
  \item {\bf Inductive case (n = k+1)}. Then, $\enc N \xrightarrow{t_1\cdots t_k} (T_k,b_k)\xrightarrow{t_{k+1}} (T,b)$. 
  By inductive hypothesis, letting $v_k= \bigcup_{1\leq i\leq k} \dec {t_i}$, we assume 
  (1) $b_k\cap P = \minp \{ e \ | \ e\in \mathcal{E}^{v_k}  \mbox{ and } \down{e}=\{e\}\}$, and
 (2) If $(T_k,b_k)\tr t$ then $\dec t \neq \emptyset$ implies  $\dec t$ is a stopped 
configuration of $\mathcal{E}^{v_k}$.
  
  We now proceed by case analysis on the shape of the applied rule: 
  \begin{itemize}
        \item $t_{k+1} = \minp \bc \pntrans 
              (\emptyset, \maxp\theta \cup \co{\maxp \bc\setminus\maxp \theta})$. Hence, $v = v_k \cup \dec{\theta}$ and 
              $b \cap P = (b_k \cap P \setminus \minp \bc) \cup \maxp \theta$.
              
         \begin{enumerate}
	 \item Then:
	\[ 
	\begin{array}{ll}
	 & \{ e \ | \ e\in \mathcal{E}^v  \mbox{ and } \down{e}=\{e\}\} \\
	 = &  \{ e \ | \ e\in \mathcal{E}^{v_k}  \mbox{ and } e \not \in \bc \mbox{ and } \down{e}=\{e\}\}\qquad
	 \\ & \hfill   \cup \{ e \ | \ e\in \mathcal{E}^{v_k}  \mbox{ and }  \down{e}\subseteq\{e\} \cup \dec{\theta}\} \\
	 \end{array}
	 \]
	 The proof is completed by noting that 
	 $$\begin{array}{l}
	 \minp {\{ e \ | \ e\in \mathcal{E}^{v_k}  \mbox{ and } e \not \in \bc \mbox{ and } \down{e}=\{e\}\}} \ = \qquad\\ \hfill  (b_k \cap P\setminus \minp \bc)
	 \end{array}$$
	 and $$ \minp {\{ e \ | \ e\in \mathcal{E}^{v_k}  \mbox{ and }  \down{e}\subseteq\{e\} \cup \dec{\theta}\}}  = \minp \theta$$
  	\item Take $t$ such that $\preS t = \bc_t$. Then,  $ \bc_t \subseteq b\cap P$.  By Theorem~\ref{th:confusion-free}, there cannot be 
	$t'$ enabled at $b$ and $\preS {t'} \neq \minp \bc_t$ and  $\minp \bc_t \cap \preS {t'} \neq \emptyset$. 
	By using inductive hypothesis (1), we conclude that all events in direct conflict with $\bc_t$ in $\mathcal{E}^v$ are in $\bc$. Hence, 
	$\dec{\theta}$ is a stopped configuration of $\mathcal{E}^v$.
	\end{enumerate}
	\item $t_{k+1} =  \co{\bf p}  \pntrans   (T'', \co{\maxp \bc\setminus\maxp {(\remInitial {N_{\bc}} p)}})$ for 
      some  $\bc$,  $p \in \minp {\bc}$, and $(T'', \emptyset) = \enc{\remInitial {N_{\bc}} p}$. Then $T = T' \cup T''$. 
         \begin{enumerate}
	 \item	
  	 Immediate because $b_k \cap P = b \cap P$.
	  \item It follows analogously to the previous case. 
	\end{enumerate} \qedhere
      
  \end{itemize}

\end{itemize}
\end{proof}

\begin{lem}[Aux.]\label{lem:aux-propagation} 
Let $\enc N \in \dn{P\cup\co{\bf P}}$. If $\enc N \rightarrow^* (T,b)$ 
then there exists $(T',b')$ such that $(T,b) \Tr {} (T',b')$ and
\begin{enumerate}
  \item $b'\cap P = b\cap P$;
  \item for all $p,q$, if  $\co{\bf p}\in b$ and  $p\preceq q$, then  $\co{\bf q}\in b'$;
  \item for all $\bc\in \BC N$ and $\co{\bf Q}\subseteq \co{\bf P}$, if $\co{\bf Q} \subseteq b'$ then 
  for all $\bc'\in \BC{\remInitial {N_{\bc}} Q}$ and $\theta : \bc'$ there exists $t\in T'$ such that 
  $t = \minp{\bc'} \tr{}  (\emptyset, \maxp\theta \cup \co{\maxp {\bc'}\setminus\maxp \theta})$.
\end{enumerate}
\end{lem}

\begin{proof} 
\begin{enumerate}
\item It follows straightforwardly by analysis of the applied rules. They are of the form
$ \co{\bf p} \pntrans  (T', \co{\maxp \bc\setminus\maxp {(\remInitial {N_{\bc}} p)}})$, which 
does not consume nor produce tokens in regular places. 
\item
By induction on the length of the chain $p = p_0\prec \ldots \prec p_n = q$ (this is a finite chain because $N$ is
a finite occurrence net). The inductive step follows by straightforward inspection of the shape of the transitions 
with negative premises.
\item By straightforward induction on the number  $n$ of elements in $\co{\bf Q}$, i.e., $n = |\co{\bf Q}|$.\qedhere
\end{enumerate}
\end{proof}

\begin{lem}[Aux]\label{lem:aux-r-stopped-reduction} 
Let $N = (P, T, F, m)$ and $\mathcal{E}$ the event structure of $N$. If 
$v$ is recursively stopped configuration and $v = \bigcup_{1\leq i\leq n}  {v_i}$ is 
a valid decomposition, then 
\begin{enumerate}
\item $\enc N \Tr {t_1\cdots t_n}  (T,b)$ and $v = \bigcup_{1\leq i\leq n} \dec {t_i}$;
\item $b\cap P = \minp \{ e \ | \ e\in \mathcal{E}^v  \mbox{ and } \down{e}=\{e\}\}$;
\item If $v'$ is a stopped 
configuration of $\mathcal{E}^v$, then there exists $t\in T$ s.t.  $(T,b)\tr t$ and $\dec t = v'$; 
\item For all $e\in \mathcal{E}$, if $e\not\in(\mathcal{E}^v \cup v)$ implies $\co{\preS e} \cap b = 0$.
\end{enumerate}
\end{lem}

\begin{proof}If follows by induction on the length $n$ of the decomposition $v = \bigcup_{1\leq i\leq n}  {v_i}$.
\begin{itemize}
  \item {\bf Base case (n=0)}. Then, $v = \emptyset$ and 
  $\mathcal{E}^v = \mathcal{E}$. Moreover,  $b = m$. Then 
  \begin{enumerate}
  \item
  It is immediate because $(T,b)=\enc N$ and $m = b$.
  \item
  Since $b = m$, $b$ corresponds to the preset of all minimal events of $\mathcal{E}^\emptyset =  \mathcal{E}$.
  \item If $v'$ is a stopped configuration of $\mathcal{E}$, then there exists $\bc\in\BC N$ such that 
  $v'\subseteq\bc$. Since $v'$ is a maximal configuration, there exists $\theta : \bc$ such 
  that $\events{\theta} = v'$. Hence, there exists $t\in T$ such that $\dec t = v'$.  Since, $v'$ is part of an initial 
  prefix, $\preS t = \minp \bc \subseteq m$. Hence, $t$ is enabled.
    \item
  It trivially holds because there does not exist $e\in \mathcal{E}$ and $e\not\in(\mathcal{E}^v \cup v)$.
  \end{enumerate}
  \item {\bf Inductive case (n = k+1)}. Take $v'= \bigcup_{1\leq i\leq k+1}  {v_i}$ and  $v = v_{k+1}\cup v'$. Then, 
  (1) $\enc N \Tr{t_1\cdots t_k} (T_k,b_k)$ and  $v' = \bigcup_{1\leq i\leq k} \dec {t_i}$; and
  (2) $b_k\cap P = \minp \{ e \ | \ e\in \mathcal{E}^{v'}  \mbox{ and } \down{e}=\{e\}\}$; and
  (3) If $v''$ is a stopped 
configuration of $\mathcal{E}^{v_k}$, then there exists $t\in T$ s.t.  $(T_k,b_k)\tr t$ and $\dec t = v''$; and 
  (4)  For all $e\in \mathcal{E}$, if $e\not\in(\mathcal{E}^{v'} \cup {v'})$ implies $\co{\preS e} \cap b = 0$.

 By inductive hypothesis (3), there exists $t_{k+1}$ such that  $\dec {t_{k+1}}= v_{t_{k+1}}$  and $(T_k,b_k)\tr {t_{k+1}}$.
 Then, take $(T_k,b_k)\tr t (T',b')$. By using Lemma~\ref{lem:aux-propagation}, we conclude that there 
 exists $(T_{k+1},b_{k+1})$ such that
 $(T_k,b_k)\Tr {t_{k+1}} (T_{k+1},b_{k+1})$ where:
 \begin{description}
  \item[(a)] $b_{k+1}\cap P = b_k\cap P$;
  \item[(b)] for all $p,q$, if  $\co{\bf p}\in b_k$ and  $p\preceq q$, then  $\co{\bf q}\in b_{k+1}$;
  \item[(c)] for all $\bc\in \BC N$ and $\co{\bf Q}\subseteq \co{\bf P}$, if $\co{\bf Q} \subseteq b_{k+1}$ then 
  for all $\bc'\in \BC{\remInitial {N_{\bc}} Q}$ and $\theta : \bc'$ there exists $t\in T_{k+1}$ such that 
  $t = \minp{\bc'} \tr{}  (\emptyset, \maxp\theta \cup \co{\maxp {\bc'}\setminus\maxp \theta})$.
 \end{description}
 
 Then,
 \begin{enumerate}
   \item It follows immediately because $\dec {t_{k+1}} = v_{t_{k+1}}$;
   \item Then, $t_{k+1} = \minp \bc \pntrans 
              (\emptyset, \maxp\theta \cup \co{\maxp \bc\setminus\maxp \theta})$.  Moreover, 
              $b' \cap P = (b_k \cap P\setminus \minp \bc) \cup \maxp \theta$. Hence, 
       	\[ 
	\begin{array}{ll}
	 & \{ e \ | \ e\in \mathcal{E}^v  \mbox{ and } \down{e}=\{e\}\} \\
	 = &  \{ e \ | \ e\in \mathcal{E}^{v'}  \mbox{ and } e \not \in \bc \mbox{ and } \down{e}=\{e\}\} \qquad\\
	    & \hfill\cup\ \{ e \ | \ e\in \mathcal{E}^{v'}  \mbox{ and }  \down{e}\subseteq\{e\} \cup \dec{\theta}\} \\
	 \end{array}
	 \]
	 The proof is completed by noting that 
	 $$\minp {\{ e \ | \ e\in \mathcal{E}^{v_k}  \mbox{ and } e \not \in \bc \mbox{ and } \down{e}=\{e\}\}} = (b_k \cap P\setminus \minp \bc)$$
	 and $ \minp {\{ e \ | \ e\in \mathcal{E}^{v_k}  \mbox{ and }  \down{e}\subseteq\{e\} \cup \dec{\theta}\}}  = \minp \theta$ and  by using (a) above. 
 \item It follows from (c).
 \item It follows from (b).\qedhere
 \end{enumerate}
\end{itemize}
\end{proof}

\begin{lem}[Lemma~\ref{theo:AB}]
 Let $N$ be an occurrence net.  
  \begin{enumerate}
    
  \item
    If $\enc N \Tr {t_0\cdots t_n}$, then $v = \bigcup_{1\leq
      i\leq n} \dec {t_i}$ is recursively-stopped in $\mathcal{E}_N$
    and $(\dec {t_i})_{1\leq i\leq n}$ is a valid decomposition of
    $v$.
  \item
    If $v$ is recursively-stopped in $\mathcal{E}_N$, then for any
    valid decomposition $(v_i)_{1\leq i\leq n}$ there exists $\enc N
    \Tr {t_0 \cdots t_n}$ such that $\dec {t_i} = v_i$.
  \end{enumerate}
\end{lem}

\begin{proof} 1). It follows from Lemma~\ref{lem:aux-red-rstopped}. 2). It follows from Lemma~\ref{lem:aux-r-stopped-reduction}(1). 
\end{proof}

\begin{thm}[Theorem~\ref{theo:corr-ab}] Let $N$ be an occurrence net. 
   \begin{enumerate}
    
  \item
    If $\dyntopt {\enc N} \Tr {t_1\cdots t_n}$, then $v = \bigcup_{1\leq
      i\leq n} \dec {t_i}$ is recursively-stopped in $\mathcal{E}_N$
    and $(\dec {t_i})_{1\leq i\leq n}$ is a valid decomposition of
    $v$.
  \item
    If $v$ is recursively-stopped in $\mathcal{E}_N$, then for any
    valid decomposition $(v_i)_{1\leq i\leq n}$ there exists $\dyntopt {\enc N}
    \Tr {t_1 \cdots t_n}$ such that $\dec {t_i} = v_i$.
  \end{enumerate}
% For every maximal configuration $v$ of $N$ 
% there exists a firing sequence $\dyntopt {\enc N} \Tr {t_0 \cdots t_n}$ such that $v = \bigcup_{1\leq
%      i\leq n} \dec {t_i}$.
\end{thm}

\begin{proof} It follows from Lemma~\ref{theo:AB} and Proposition~\ref{prop:dynvsflat}.
\end{proof}

%%% Local Variables:
%%% mode: latex
%%% TeX-master: "main.tex"
%%% End:

% !TEX root =  main.tex

\subsection{Detailed proofs of results in Section 4} 
This section is devoted to prove the main results in Section 4. We start by providing some auxiliary results. 

\begin{lem}[Lemma~\ref{lem:aux-red-rstopped}]
Let $N = (P, T, F, m)$ and $\mathcal{E}$ the event structure of $N$. If 
$\enc N \xrightarrow{t_1\cdots t_n}  (T,b)$ and $v = \bigcup_{1\leq i\leq n} \dec {t_i}$, then
\begin{enumerate}
\item $b\cap P = \minp \{ e \ | \ e\in \mathcal{E}^v  \mbox{ and } \down{e}=\{e\}\}$; and
\item If $(T,b)\tr t$ then $\dec t \neq \emptyset$ implies  $\dec t$ is a stopped 
configuration of $\mathcal{E}^v$.
\end{enumerate}
\end{lem}

\begin{proof} If follows by induction on the length of the reduction $\enc N \xrightarrow{t_1\cdots t_n} (T,b)$.
\begin{itemize}
  \item {\bf Base case (n=0)}. Then, $v = \emptyset$ and 
  $\mathcal{E}^v = \mathcal{E}$. Moreover,  $b = m$. 
  \begin{enumerate}
  \item
  It is immediate to notice that $m$ corresponds to the preset of all minimal events of $\mathcal{E}$.
  \item Since $t$ is enabled, $\preS t\subseteq m$. Hence, $\preS t = \minp \bc$ with $\bc\in\BC N$. 
  Therefore, $\bc$ corresponds to 
  a branching cell of  $\mathcal{E}$.  By the definition of $\enc{\_}$, $t$ is associated with some $\theta : \bc$,
  which is  a maximal, conflict-free set of transitions in $\bc$. Hence, $\dec t$ is a stopped 
  configuration of $\mathcal{E}$. 
  \end{enumerate}
  \item {\bf Inductive case (n = k+1)}. Then, $\enc N \xrightarrow{t_1\cdots t_k} (T_k,b_k)\xrightarrow{t_{k+1}} (T,b)$. 
  By inductive hypothesis, letting $v_k= \bigcup_{1\leq i\leq k} \dec {t_i}$, we assume 
  (1) $b_k\cap P = \minp \{ e \ | \ e\in \mathcal{E}^{v_k}  \mbox{ and } \down{e}=\{e\}\}$, and
 (2) If $(T_k,b_k)\tr t$ then $\dec t \neq \emptyset$ implies  $\dec t$ is a stopped 
configuration of $\mathcal{E}^{v_k}$.
  
  We now proceed by case analysis on the shape of the applied rule: 
  \begin{itemize}
        \item $t_{k+1} = \minp \bc \pntrans 
              (\emptyset, \maxp\theta \cup \co{\maxp \bc\setminus\maxp \theta})$. Hence, $v = v_k \cup \dec{\theta}$ and 
              $b \cap P = (b_k \cap P \setminus \minp \bc) \cup \maxp \theta$.
              
         \begin{enumerate}
	 \item Then:
	\[ 
	\begin{array}{ll}
	 & \{ e \ | \ e\in \mathcal{E}^v  \mbox{ and } \down{e}=\{e\}\} \\
	 = &  \{ e \ | \ e\in \mathcal{E}^{v_k}  \mbox{ and } e \not \in \bc \mbox{ and } \down{e}=\{e\}\}\qquad
	 \\ & \hfill   \cup \{ e \ | \ e\in \mathcal{E}^{v_k}  \mbox{ and }  \down{e}\subseteq\{e\} \cup \dec{\theta}\} \\
	 \end{array}
	 \]
	 The proof is completed by noting that 
	 $$\begin{array}{l}
	 \minp {\{ e \ | \ e\in \mathcal{E}^{v_k}  \mbox{ and } e \not \in \bc \mbox{ and } \down{e}=\{e\}\}} \ = \qquad\\ \hfill  (b_k \cap P\setminus \minp \bc)
	 \end{array}$$
	 and $$ \minp {\{ e \ | \ e\in \mathcal{E}^{v_k}  \mbox{ and }  \down{e}\subseteq\{e\} \cup \dec{\theta}\}}  = \minp \theta$$
  	\item Take $t$ such that $\preS t = \bc_t$. Then,  $ \bc_t \subseteq b\cap P$.  By Theorem~\ref{th:confusion-free}, there cannot be 
	$t'$ enabled at $b$ and $\preS {t'} \neq \minp \bc_t$ and  $\minp \bc_t \cap \preS {t'} \neq \emptyset$. 
	By using inductive hypothesis (1), we conclude that all events in direct conflict with $\bc_t$ in $\mathcal{E}^v$ are in $\bc$. Hence, 
	$\dec{\theta}$ is a stopped configuration of $\mathcal{E}^v$.
	\end{enumerate}
	\item $t_{k+1} =  \co{\bf p}  \pntrans   (T'', \co{\maxp \bc\setminus\maxp {(\remInitial {N_{\bc}} p)}})$ for 
      some  $\bc$,  $p \in \minp {\bc}$, and $(T'', \emptyset) = \enc{\remInitial {N_{\bc}} p}$. Then $T = T' \cup T''$. 
         \begin{enumerate}
	 \item	
  	 Immediate because $b_k \cap P = b \cap P$.
	  \item It follows analogously to the previous case. 
	\end{enumerate} 
      
  \end{itemize}

\end{itemize}
\end{proof}

%\begin{lem}\label{lem:aux-propagation} 
%Let $\enc N \in \dn{P\cup\co{\bf P}}$. If $\enc N \rightarrow^* (T,b)$ 
%then there exists $(T',b')$ such that $(T,b) \Tr {} (T',b')$ and
%\begin{enumerate}
%  \item $b'\cap P = b\cap P$;
%  \item for all $p,q$, if  $\co{\bf p}\in b$ and  $p\preceq q$, then  $\co{\bf q}\in b'$;
%  \item for all $\bc\in \BC N$ and $\co{\bf Q}\subseteq \co{\bf P}$, if $\co{\bf Q} \subseteq b'$ then 
%  for all $\bc'\in \BC{\remInitial {N_{\bc}} Q}$ and $\theta : \bc'$ there exists $t\in T'$ such that 
%  $t = \minp{\bc'} \tr{}  (\emptyset, \maxp\theta \cup \co{\maxp {\bc'}\setminus\maxp \theta})$.
%\end{enumerate}
%\end{lem}
%
%\begin{proof} 
%\begin{enumerate}
%\item It follows straightforwardly by analysis of the applied rules. They are of the form
%$ \co{\bf p} \pntrans  (T', \co{\maxp \bc\setminus\maxp {(\remInitial {N_{\bc}} p)}})$, which 
%does not consume nor produce tokens in regular places. 
%\item
%By induction on the length $p = p_0\prec \ldots \prec p_n = q$ (this is a finite chain because $N$ is
%a finite occurrence net). The inductive step follows by straightforward inspection of the shape of the transitions 
%with negative premises.
%\item By straightforward induction on the number  $n$ of elements in $\co{\bf Q}$, i.e., $n = |\co{\bf Q}|$.
%\end{enumerate}
%\end{proof}
%

\begin{lem}[Lemma~\ref{lem:aux-r-stopped-reduction} ]
Let $N = (P, T, F, m)$ and $\mathcal{E}$ the event structure of $N$. If 
$v$ is recursively stopped configuration and $v = \bigcup_{1\leq i\leq n}  {v_i}$ is 
a valid decomposition, then 
\begin{enumerate}
\item $\enc N \Tr {t_1\cdots t_n}  (T,b)$ and $v = \bigcup_{1\leq i\leq n} \dec {t_i}$;
\item $b\cap P = \minp \{ e \ | \ e\in \mathcal{E}^v  \mbox{ and } \down{e}=\{e\}\}$;
\item If $v'$ is a stopped 
configuration of $\mathcal{E}^v$, then there exists $t\in T$ s.t.  $(T,b)\tr t$ and $\dec t = v'$; 
\item For all $e\in \mathcal{E}$, if $e\not\in(\mathcal{E}^v \cup v)$ implies $\co{\preS e} \cap b = 0$.
\end{enumerate}
\end{lem}

\begin{proof}If follows by induction on the length $n$ of the decomposition $v = \bigcup_{1\leq i\leq n}  {v_i}$.
\begin{itemize}
  \item {\bf Base case (n=0)}. Then, $v = \emptyset$ and 
  $\mathcal{E}^v = \mathcal{E}$. Moreover,  $b = m$. Then 
  \begin{enumerate}
  \item
  It is immediate because $(T,b)=\enc N$ and $m = b$.
  \item
  Since $b = m$, $b$ corresponds to the preset of all minimal events of $\mathcal{E}^\emptyset =  \mathcal{E}$.
  \item If $v'$ is a stopped configuration of $\mathcal{E}$, then there exists $\bc\in\BC N$ such that 
  $v'\subseteq\bc$. Since $v'$ is a maximal configuration, there exists $\theta : \bc$ such 
  that $\events{\theta} = v'$. Hence, there exists $t\in T$ such that $\dec t = v'$.  Since, $v'$ is part of an initial 
  prefix, $\preS t = \minp \bc \subseteq m$. Hence, $t$ is enabled.
    \item
  It trivially holds because there does not exist $e\in \mathcal{E}$ and $e\not\in(\mathcal{E}^v \cup v)$.
  \end{enumerate}
  \item {\bf Inductive case (n = k+1)}. Take $v'= \bigcup_{1\leq i\leq k+1}  {v_i}$ and  $v = v_{k+1}\cup v'$. Then, 
  (1) $\enc N \Tr{t_1\cdots t_k} (T_k,b_k)$ and  $v' = \bigcup_{1\leq i\leq k} \dec {t_i}$; and
  (2) $b_k\cap P = \minp \{ e \ | \ e\in \mathcal{E}^{v'}  \mbox{ and } \down{e}=\{e\}\}$; and
  (3) If $v''$ is a stopped 
configuration of $\mathcal{E}^{v_k}$, then there exists $t\in T$ s.t.  $(T_k,b_k)\tr t$ and $\dec t = v''$; and 
  (4)  For all $e\in \mathcal{E}$, if $e\not\in(\mathcal{E}^{v'} \cup {v'})$ implies $\co{\preS e} \cap b = 0$.

 By inductive hypothesis (3), there exists $t_{k+1}$ such that  $\dec {t_{k+1}}= v_{t_{k+1}}$  and $(T_k,b_k)\tr {t_{k+1}}$.
 Then, take $(T_k,b_k)\tr t (T',b')$. By using Lemma~\ref{lem:aux-propagation}, we conclude that there 
 exists $(T_{k+1},b_{k+1})$ such that
 $(T_k,b_k)\Tr {t_{k+1}} (T_{k+1},b_{k+1})$ where:
 \begin{description}
  \item[(a)] $b_{k+1}\cap P = b_k\cap P$;
  \item[(b)] for all $p,q$, if  $\co{\bf p}\in b_k$ and  $p\preceq q$, then  $\co{\bf q}\in b_{k+1}$;
  \item[(c)] for all $\bc\in \BC N$ and $\co{\bf Q}\subseteq \co{\bf P}$, if $\co{\bf Q} \subseteq b_{k+1}$ then 
  for all $\bc'\in \BC{\remInitial {N_{\bc}} Q}$ and $\theta : \bc'$ there exists $t\in T_{k+1}$ such that 
  $t = \minp{\bc'} \tr{}  (\emptyset, \maxp\theta \cup \co{\maxp {\bc'}\setminus\maxp \theta})$.
 \end{description}
 
 Then,
 \begin{enumerate}
   \item It follows immediately because $\dec {t_{k+1}} = v_{t_{k+1}}$;
   \item Then, $t_{k+1} = \minp \bc \pntrans 
              (\emptyset, \maxp\theta \cup \co{\maxp \bc\setminus\maxp \theta})$.  Moreover, 
              $b' \cap P = (b_k \cap P\setminus \minp \bc) \cup \maxp \theta$. Hence, 
       	\[ 
	\begin{array}{ll}
	 & \{ e \ | \ e\in \mathcal{E}^v  \mbox{ and } \down{e}=\{e\}\} \\
	 = &  \{ e \ | \ e\in \mathcal{E}^{v'}  \mbox{ and } e \not \in \bc \mbox{ and } \down{e}=\{e\}\} \qquad\\
	    & \hfill\cup\ \{ e \ | \ e\in \mathcal{E}^{v'}  \mbox{ and }  \down{e}\subseteq\{e\} \cup \dec{\theta}\} \\
	 \end{array}
	 \]
	 The proof is completed by noting that 
	 $$\minp {\{ e \ | \ e\in \mathcal{E}^{v_k}  \mbox{ and } e \not \in \bc \mbox{ and } \down{e}=\{e\}\}} = (b_k \cap P\setminus \minp \bc)$$
	 and $ \minp {\{ e \ | \ e\in \mathcal{E}^{v_k}  \mbox{ and }  \down{e}\subseteq\{e\} \cup \dec{\theta}\}}  = \minp \theta$ and  by using (a) above. 
 \item It follows from (c).
 \item It follows from (b).\qedhere
 \end{enumerate}
\end{itemize}
\end{proof}

%%% Local Variables:
%%% mode: latex
%%% TeX-master: "main.tex"
%%% End:

% !TEX root =  main.tex

\section{Proofs of results in Section 5}

\begin{thm}[Theorem~\ref{theo:concur}]
Let $\sigma= t_1; \cdots; t_n$ with $n\geq 0$ be a, possibly empty, firing sequence of a persistent process, and $t$ a transition not in $\sigma$.
The following conditions are all equivalent:
(i)~%the transition 
$t$ is enabled after $\sigma$;
(ii)~there is a collection of causes of $t$ which appears in $\sigma$; 
(iii)~$\bigwedge_{i=1}^n t_i$ implies $\Phi(t)$.
%(iv)~$\bigwedge_{i=1}^n t_i$ implies $\Psi(t)$.
\end{thm}

\begin{proof} 
%We first prove that (ii) and (iii) are equivalent, then 
%that (i) implies (iii) and at last that (iii) implies (i).
\begin{description}
\item[ii) $\Leftrightarrow$ (iii):]
We have that $\bigwedge_{i=1}^n t_i$ implies $\Phi(t)$ iff there is a prime implicant $\bigwedge_{j=1}^m t_{i_j}$ of $\Phi(t)$ that is implied by $\bigwedge_{i=1}^n t_i$. This is the case iff the collection of causes $\{t_{i_1},...,t_{i_m}\}$ appears in $\sigma$.

\item[(i) $\Rightarrow$ (iii):] The proof is by induction on the length $n$ of the sequence. 

For the base case, if $n=0$ it means that $t$ is enabled in the initial marking, i.e., that its pre-set only contains initial places of the process and thus $\Phi(t) = true$.

For the inductive case, assume the property holds for any shorter sequence $t_1; \cdots ; t_k$ with $0\leq k < n+1$ and let us prove that it holds for $\sigma= t_1 ; \cdots ; t_{n+1}$. Let $b_0$ the initial bag of the process. As $t$ is enabled after $\sigma$ we have $b_0\xrightarrow{\sigma}b\xrightarrow{t}$ for some bag $b$. Since $t$ is enabled in $b$, we have $\preS{t} \subseteq b$, i.e., for any $s\in \preS{t}$ we have $b(s)\in \{1,\infty\}$ (by definition of p-net, $\preS{t}$ is not empty). We need to prove that $\Phi(t) = \wedge_{s\in\preS{t}} \Phi(s)$ is implied by $\bigwedge_{i=1}^{n+1} t_i$, i.e., that for any $s\in\preS{t}$ the formula  $\Phi(s)$ is implied by $\bigwedge_{i=1}^{n+1} t_i$.
Take a generic $s\in \preS{t}$. Either $\preS{s} = \emptyset$, in which case $s$ is initial and $\Phi(s) = true$, or $\preS{s} \neq \emptyset$ and $\Phi(s) = \bigvee_{t'\in\preS{s}} (t' \wedge \Phi(t'))$. Since $b(s)\in \{1,\infty\}$, there must exist an index $j\in [1,n+1]$ such that $t_j\in \preS{s}$.
Take $t'=t_j$. Since $\sigma$ is a firing sequence, the transition $t_j$ is enabled after $\sigma' = t_1 ; \cdots ; t_{j-1}$. As $k = j-1 < n+1$, by inductive hypothesis $\Phi(t_j)$ is implied by $\bigwedge_{i=1}^{j-1} t_i$ and thus also by $\bigwedge_{i=1}^{n+1} t_i$. Since $\bigwedge_{i=1}^{n+1} t_i$ clearly implies $t_j$ we have that $\bigwedge_{i=1}^{n+1} t_i$ implies $\Phi(s) = t_j \wedge \Phi(t_j)$.

\item[(iii) $\Rightarrow$ (i):] Suppose $\bigwedge_{i=1}^n t_i$ implies $\Phi(t) = \bigwedge_{s\in\preS{t}} \Phi(s)$. If for all $s\in\preS{t}$ we have $\preS{s} = \emptyset$, then $t$ is enabled in the initial marking and as the process is deterministic no transition can steal tokens from $\preS{t}$ and $t$ remains enabled after the firing of any $\sigma= t_1 ; \cdots ; t_n$. Otherwise, $\Phi(t) = \bigwedge_{s\in\preS{t}} \bigvee_{t'\in\preS{s}\neq \emptyset} (t' \wedge \Phi(t'))$. Thus, for any $s\in\preS{t}$ with $\preS{s}\neq \emptyset$ there exists some $t'\in\preS{s}$ such that $\bigwedge_{i=1}^n t_i$ implies $t' \wedge \Phi(t')$. Since $\bigwedge_{i=1}^n t_i$ implies $t'$ then there exists some index $k\in [1,n]$ such that $t'=t_k$ and $s$ becomes marked during the firing of $\sigma$. As the process is deterministic, no transition can steal tokens from $s$. Since all the places in the pre-set of $t$ becomes marked during the firing of $\sigma$, then $t$ is enabled after $\sigma$.\qedhere
\end{description}
\end{proof}
%%% Local Variables:
%%% mode: latex
%%% TeX-master: "main.tex"
%%% End:

 %!TEX root =  main.tex

\newpage
\section{Additional processes of the running example}\label{app:e}
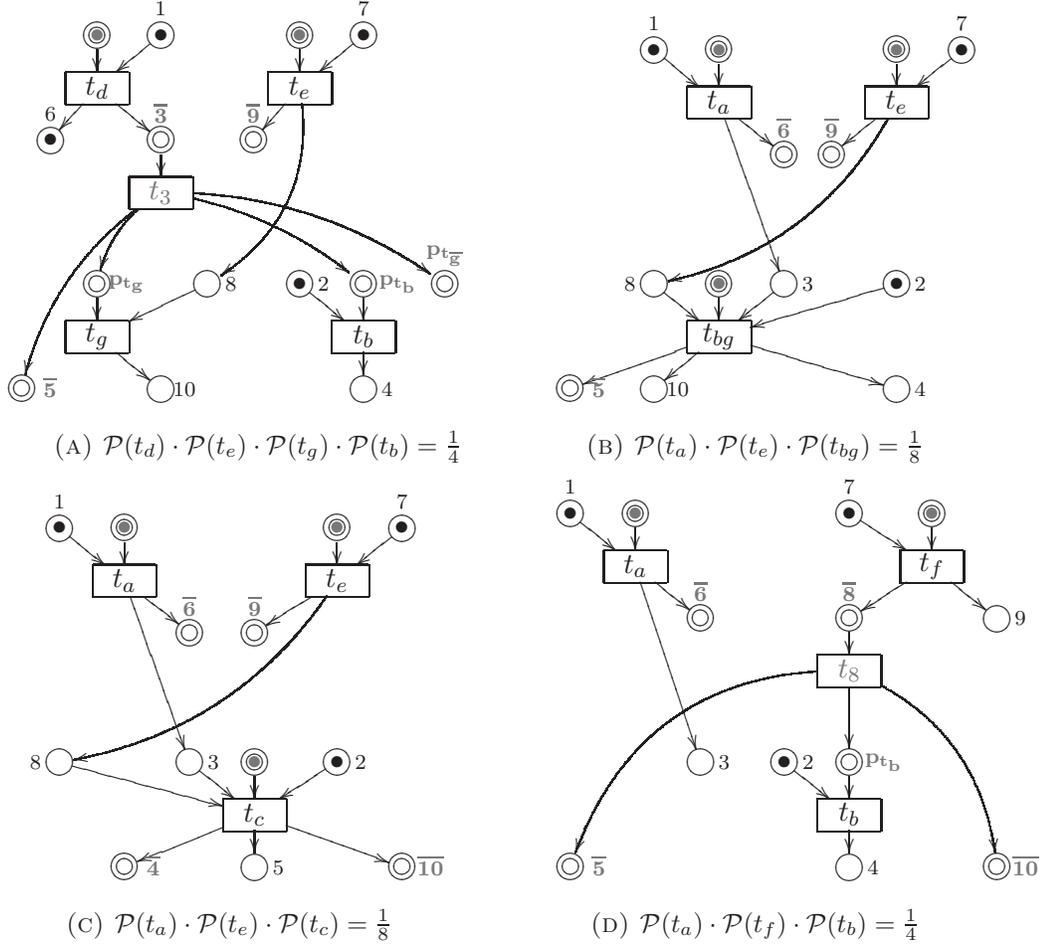
\begin{figure}[h]
%%%%LL
\begin{subfigure}[b]{0.5\textwidth}
$$
 \xymatrix@R=.7pc@C=.01pc{
&
&
\drawmarkedpersistentplace\ar@[mygray][d]
& 
\drawmarkedplace\ar[dl]
\nameplaceup 1
 &
 &
 &
 &
 \drawmarkedpersistentplace\ar@[mygray][d]
 &
 \drawmarkedplace\ar[dl]
 \nameplaceup 7
 &
 \\
&
 & 
 \drawtrans {t_d}\ar[dl]\ar@[mygray][dr] 
 &
 &
  &
  &
  &
  \drawtrans {t_e}\ar@[mygray][dl]\ar@/^3ex/[ddddlll] 
  &
  &
 \\
&
 \drawmarkedplace
 \nameplaceup 6
 &
 &
\drawpersistentplace\ar@[mygray][d] 
 \namepersup{ 3}
 &
 &
 &
 \drawpersistentplace
 \namepersup{ 9}
 &
 &
 &
 &
  \\
 &&&
 \drawtransu {\color{mygray}t_3}\ar@[mygray]@/_2ex/[ddddlll] \ar@[mygray]@/^/[ddrrrrr]\ar@[mygray]@/^2ex/[ddrrrrrrrr]\ar@[mygray]@/_1ex/[ddl]
 &&&&&
 \\
 &&&&&&&&&&
   \\
 &
&
  \drawpersistentplace\ar@[mygray][d] 
 \namepersrightD{p_{t_g}\ }
&&
 \drawplace\ar[dll]
 \nameplaceright {8}
 &
&
&
 \drawmarkedplace\ar[dr]
 \nameplaceright {2}
 &
  \drawpersistentplace\ar@[mygray][d] 
 \namepersrightD{p_{t_b}}
 &&&
 \drawpersistentplace
 \namepersupD{p_{t_{\co g}}}
  \\
 &
 &
 \drawtrans {t_g}\ar[dr] 
 &
 &
 &
 &
  &
  &
 \drawtrans {t_b}\ar[d]
  &
 &
 &&&&
 \\
 \drawpersistentplace
 \namepersright{5}
 &
 &
 &
 \drawplace
 \nameplaceright {10}
 &
 &
 &
 &
&
 \drawplace
 \nameplaceright {4}
 &
 &
 &&&
  }
$$
\subcaption{$\mathcal{P}(t_d) \cdot \mathcal{P}(t_e)\cdot \mathcal{P}(t_g)\cdot \mathcal{P}(t_b) = \frac{1}{4}$}\label{fig:process1}
\end{subfigure}
 %%%% RL1
 \begin{subfigure}[b]{.35\textwidth}
$$
 \xymatrix@R=.7pc@C=.6pc{
&
& 
\drawmarkedplace\ar[dr] 
\nameplaceup 1
 &
 \drawmarkedpersistentplace\ar@[mygray][d]
 &
 &
 &
 \drawmarkedpersistentplace\ar@[mygray][d]
 &
 \drawmarkedplace\ar[dl]
 \nameplaceup 7
 \\
 & 
 &
 &
  \drawtrans {t_a}\ar@[mygray][dr]\ar[ddddr] 
  &
  &
  &
  \drawtrans {t_e}\ar@[mygray][dl]\ar@/^3ex/[ddddllll] 
  &
  &
 \\
 &
 &
 &
 &
 \drawpersistentplace
 \namepersup{ 6}
 &
 \drawpersistentplace
 \namepersup{ 9}
 &
 &
 &
 &
  \\
 &&
 &&&&&
 \\
 &&&&&&&&&
   \\
&
&
 \drawplace\ar[dr]
  \nameplaceleft {8}
 &
 \drawmarkedpersistentplace\ar@[mygray][d]
 &
 \drawplace\ar[dl]
 \nameplaceright {3}
&
&
 \drawmarkedplace\ar[dlll]
 \nameplaceright {2}
 &
  \\
 &
 &
 &
 \drawtrans {t_{bg}}\ar[drrr]\ar@[mygray][dlll]\ar[dl]
 &
 &
  &
  &
  &
 &
 \\
  \drawpersistentplace
 \namepersright{5}
 &
 &
 \drawplace
 \nameplaceright {10}
 &
 &
 &
 &
 \drawplace
 \nameplaceright {4}
 &
 &
    }
$$
\subcaption{$\mathcal{P}(t_a) \cdot \mathcal{P}(t_e)\cdot \mathcal{P}(t_{bg}) = \frac{1}{8}$}
\end{subfigure}
%%%%RL2
 \begin{subfigure}[b]{.45\textwidth}
$$
 \xymatrix@R=.7pc@C=.6pc{
&
& 
\drawmarkedplace\ar[dr] 
\nameplaceup 1
 &
 \drawmarkedpersistentplace\ar@[mygray][d]
 &
 &
 &
 \drawmarkedpersistentplace\ar@[mygray][d]
 &
 \drawmarkedplace\ar[dl]
 \nameplaceup 7
 \\
 & 
 &
 &
  \drawtrans {t_a}\ar@[mygray][dr]\ar[ddddr] 
  &
  &
  &
  \drawtrans {t_e}\ar@[mygray][dl]\ar@/^3ex/[ddddllll] 
  &
  &
 \\
 &
 &
 &
 &
 \drawpersistentplace
 \namepersup{ 6}
 &
 \drawpersistentplace
 \namepersup{ 9}
 &
 &
 &
 &
  \\
 &&
 &&&&&
 \\
 &&&&&&&&&
   \\
&
&
 \drawplace\ar[drrr]
 \nameplaceleft {8}
 &
 &
 \drawplace\ar[dr]
 \nameplaceright {3}
&
\drawmarkedpersistentplace\ar@[mygray][d]
&
 \drawmarkedplace\ar[dl]
 \nameplaceright {2}
 &
  \\
 &
 &
 &
 &
 &
 \drawtrans {t_c}\ar[d]\ar@[mygray][drr]\ar@[mygray][dll]
  &
  &
  &
 &
 \\
 &
 &
 &
  \drawpersistentplace
 \namepersright{4}
 &
 &
 \drawplace
 \nameplaceright {5}
&
 &
  \drawpersistentplace
 \namepersright{10}
  }
$$
\subcaption{$\mathcal{P}(t_a) \cdot \mathcal{P}(t_e)\cdot \mathcal{P}(t_c) = \frac{1}{8}$}
\end{subfigure}
 %%% RR
\qquad
 \begin{subfigure}[b]{.35\textwidth}
$$
 \xymatrix@R=.7pc@C=.6pc{
\drawmarkedplace\ar[dr] 
\nameplaceup 1
 &
 \drawmarkedpersistentplace\ar@[mygray][d]
 &
 &
 &
 &
 \drawmarkedplace\ar[dr]
 \nameplaceup 7
 &
 \drawmarkedpersistentplace\ar@[mygray][d]
 \\
 &
  \drawtrans {t_a}\ar@[mygray][dr]\ar[ddddr] 
  &
  &
  &
  &
  &
   \drawtrans {t_f}\ar@[mygray][dl]\ar[dr] 
 \\
 &
 &
 \drawpersistentplace
 \namepersup{ 6}
 &
 &
 &
\drawpersistentplace\ar@[mygray][d]
 \namepersup{ 8}
 &
 &
 \drawplace
 \nameplaceright 9
  \\
 &&&&&
 \drawtransu {\color{mygray}t_8}\ar@[mygray][dd]\ar@[mygray]@/^3ex/[ddddrr] \ar@[mygray]@/_5ex/[ddddlllll] 
 \\
&&&&&&&
   \\
 &
 &
 \drawplace
 \nameplaceright {3}
&
&
 \drawmarkedplace\ar[dr]
 \nameplaceright {2}
 &
  \drawpersistentplace\ar@[mygray][d] 
 \namepersrightD{p_{t_b}}
 &&
   \\
 &
 &
 &
  &
  &
 \drawtrans {t_b}\ar[d]
  &
 &
 &&&&
 \\
 \drawpersistentplace
 \namepersright{5}
  &
 &
 &
 &
&
 \drawplace
 \nameplaceright {4}
 &
 &
  \drawpersistentplace
 \namepersright{10}
  }
$$
\subcaption{$\mathcal{P}(t_a) \cdot \mathcal{P}(t_f)\cdot \mathcal{P}(t_b) = \frac{1}{4}$}
\end{subfigure}

\caption{Processes of the net $\dyntopt {\enc N}$ (running example)}\label{fig:allproc}
%\label{fig:persistent-encoding}
%\end{subfigure}
\end{figure}

We show in Fig.~\ref{fig:allproc} the additional processes of the net  $\dyntopt {\enc N}$ of the running example and 
their probabilities.

%%% Local Variables:
%%% mode: latex
%%% TeX-master: "main.tex"
%%% End:

\end{document}